\documentclass[preprint]{aastex}

\slugcomment{accepted by AJ}

\shorttitle{QSO Hosts on Fundamental Plane}
\shortauthors{Wolf \& Sheinis}

\begin{document}


\title{Host Galaxies of Luminous Quasars: Structural Properties and \\
the Fundamental Plane}


\author{Marsha J. Wolf\altaffilmark{1}, Andrew I. Sheinis\altaffilmark{2} } 
\affil{University of Wisconsin - Madison, Department of Astronomy, \\
475 N. Charter Street, Madison, WI 53706}

\altaffiltext{1}{mwolf@astro.wisc.edu}
\altaffiltext{2}{sheinis@astro.wisc.edu}


\begin{abstract}

  We present stellar velocity dispersion measurements in the host
  galaxies of 10 luminous quasars (M$_{V}<-$23) using the Ca H\&K
  lines in off-nuclear spectra. We combine these data with effective
  radii and magnitudes from the literature to place the host galaxies
  on the Fundamental Plane (FP) where their properties are compared to
  other types of galaxies. We find that the radio-loud (RL) QSO hosts
  have similar properties to massive elliptical galaxies, while the
  radio-quiet (RQ) hosts are more similar to intermediate mass
  galaxies. The RL hosts lie at the upper extreme of the FP due to
  their large velocity dispersions ($\langle \sigma_{*} \rangle$ =
  321~km~s$^{-1}$), low surface brightness ($\langle \mu_{e}(r)
  \rangle$ = 20.8~mag~arcsec$^{-2}$), and large effective radii
  ($\langle R_{e} \rangle$ = 11.4~kpc), and have $\langle M_{*}
  \rangle$ = 1.5 x 10$^{12}$ M$_{\sun}$ and $\langle M/L \rangle$ =
  12.4. In contrast, properties of the RQ hosts are $\langle
  \sigma_{*} \rangle$ = 241~km~s$^{-1}$, $\langle M_{*} \rangle \sim$ 4.4 x
  10$^{11}$ M$_{\sun}$, and M/L $\sim$ 5.3. The distinction between
  these galaxies occurs at $\sigma_{*}\sim$~300~km~s$^{-1}$, R$_{e}
  \sim$ 6~kpc, and corresponding M$_{*} \sim$ 5.9 $\pm$ 3.5 x
  10$^{11}$ M$_{\sun}$. Our data support previous results that PG QSOs
  are related to gas-rich galaxy mergers that form intermediate-mass
  galaxies, while RL QSOs reside in massive early-type galaxies, most
  of which also show signs of recent mergers or interactions. Most
  previous work has drawn these conclusions by using estimates of the
  black hole mass and inferring host galaxy properties from that,
  while here we have relied purely on directly measured host galaxy
  properties.

\end{abstract}



\section{Introduction}

A growing understanding of the connection between galaxies and their
central black holes has emerged over the last decade. We now know that
all galaxies with a bulge contain supermassive black holes
\citep{kormendy04} and that black hole mass is correlated with host
galaxy stellar velocity dispersion
\citep{gebhardt00a,ferrarese00,tremaine02}. Furthermore, the inclusion
of an amount of energy equal to that expected from AGN feedback to
quench star formation above a critical halo mass in semi-analytic
galaxy formation models \citep{cattaneo06,dekel06} reproduces the
galaxy demographics and bimodality of properties observed in large
surveys (SDSS:
\citet{kauffmann03a,kauffmann03b,hogg03,baldry04,heavens04,cidfernandes05};
GOODS: \citet{giavalisco04}; COMBO-17: \citet{bell04}; DEEP/DEEP2:
\citet{koo03,koo05,faber07}; MUNICS: \citet{drory01}; FIRES:
\citet{labbe03}; K20: \citet{cimatti02}; GDDS:
\citet{mccarthy04}). These facts suggest that the growth
mechanisms of the black hole and galaxy must be connected. However,
details of the physical processes that make this connection, such as
how AGN energy interacts with and is dissipated by surrounding halo
gas, are not yet understood.

One way to investigate these processes is to understand the nature of
the host galaxies. Does something in the galaxy trigger AGN activity?
Do active quasars exist in galaxies with similar properties? Studies
of AGN host galaxies have reached different conclusions. One group of
collaborators, \citep{MKDBOH99,HKDB00,NDKHBJ00} believes these objects
to be predominantly normal massive ellipticals, including Nolan et
al. (2001) who found that most quasar host galaxies had evolved
stellar populations, 10~Gyr old, with only a very small amount of
recent star formation.  However, \citet{mil81}, in the first
spectroscopic investigation of a sample of these objects, concluded
that they are not normal luminous ellipticals, a result which was
later confirmed with deeper spectroscopy from the Keck telescope
\citep{mts96,she01,miller03}. Moreover, \citet{CS00,CS01} have seen
evidence of star formation within the past 100~Myr in quasars hosts
with far-infrared excesses using deep spectra from Keck.  There is
still debate about whether the different results are due to better
quality spectra taken closer to the nucleus using 8-10~m class
telescopes, or whether real differences exist in the host galaxy
properties of the different quasar samples studied
\citep{lacy06}. This question will no doubt be answered as more data
are analyzed from the larger extra-galactic surveys.

In this work we study luminous quasars (M$_{V}<-$23) in which the
galaxy is actively feeding the central supermassive black hole. It is
here that we should be able to investigate connections between the
black hole and its surrounding galaxy from which we can draw
conclusions about how the black hole may or may not affect galaxy
formation and evolution. We begin in this first paper by analyzing the
structural properties of the QSO host galaxies with the use of
directly measured stellar velocity dispersions, previously
unobtainable for quasars this luminous. We use these data to place the
host galaxies on the Fundamental Plane and ascertain their structural
properties relative to other types of galaxies. In future papers we
will investigate whether these objects follow the M$_{BH}$-$\sigma$
relation and analyze the host galaxy stellar populations to look for
indications of star formation activity relative to quasar activity.

This paper is organized in the following manner. In \S
\ref{data_section} we describe the sample selection and data analysis,
including our removal of scattered quasar light from the observed
spectra and the measurement of stellar velocity dispersions. We also
present the comparison objects from the literature that are used in
our analysis. In \S \ref{result_section} we use the fundamental
parameters derived in \S \ref{data_section} to analyze the Fundamental
Plane locations and mass-to-light ratios of these objects. In \S
\ref{discussion} we discuss the properties of our QSO host galaxies
relative to the comparison objects and their implication that two
different classes of objects are present in our sample. Finally, \S
\ref{summary} summarizes our work and presents the main
conclusions. Further details about our stellar velocity dispersion
measurement limitations and potential biases can be found in Appendix
\ref{bias}.

\section{Data and Analysis \label{data_section}}

\subsection{Sample Selection}

Our full sample is primarily drawn from the 20 nearby luminous quasars
of \citet{bahcall97}, with the addition of a few objects from
\citet{dunlop03} and \citet{guyon06}. The analyses presented here
include ten QSOs from this larger sample of 28 whose spectra show
sufficient signal-to-noise ratio (S/N) to allow measurement of the
velocity dispersion via the techniques described below.  These ten
objects are PG~0052+251, PHL~909, PKS~0736+017, 3C~273, PKS~1302-102,
PG~1309+355, PG~1444+407, PKS~2135-147, 4C~31.63, and PKS~2349-014.

The full sample from \citet{bahcall97} were selected solely on the
basis of luminosity (M$_{V} < -$22.9), redshift (z $\leq$ 0.20), and
galactic latitude ($|b| >$ 35$\degr$). All the quasars in the
\citet{veron91} catalog that satisfied the redshift, luminosity, and
galactic latitude criteria were included, amounting to 14 objects.

These 14 quasars with z $\leq$ 0.20 have an average (median) absolute
magnitude $\langle M_{V} \rangle = -$23.4 (23.2) and an average
redshift $\langle z \rangle =$ 0.17.  Only three radio-loud quasars
are present in the original sample of 14 objects.  By combining the
time available from GTO and GO programs \citet{bahcall97} added an
additional six quasars with redshifts in the range 0.20 $<$ z $<$
0.30; these additional objects satisfied the same luminosity and
galactic latitude constraints as the original sample. The additional
objects contained three radio-loud quasars, bringing the total number
of radio-loud objects in the sample to 6 of 20 quasars.

Bahcall's final sample of 20 objects shows an average (median) absolute
magnitude of -23.6 (-23.2); an average redshift, z = 0.19.  Nearly all
(18) of the quasars have 15.1 $<$ V $<$ 16.7, but two are much
brighter in the optical band: 3C 273 (V = 12.8) and HE 1029-140 (V =
13.9).

\subsection{Spectra}

Off-nuclear spectra of seven of the host galaxies presented here were
obtained with the Low Resolution Imaging Spectrograph \citep{oke94} on
the Keck telescope during 1996-1997.  Typical offsets for the
long-slit observations were 2-4\arcsec. Observed wavelength ranges
covered $\sim$~4500-7000~\AA~at a spectral resolution of
$\Delta\lambda\sim$~11~\AA~(300~km~s$^{-1}$). The average
S/N~\AA$^{-1}$ for the galaxies in this study ranges from 1.8 to 13.9,
with a mean of 5.3, over the wavelength range of
3850-4200~\AA. Further observing details for these Keck data can be
found in \citet{sheinis02} and \citet{miller03}.

We are currently obtaining spectra of additional host galaxies using
an integral field unit (IFU), Sparsepak \citep{bershady04,bershady05},
which feeds the Bench Spectrograph on the 3.5-m WIYN Telescope. The
central core of Sparsepak fibers are 4.7\arcsec~in diameter on
5.6\arcsec~spacings. Though not optimally sized for the host galaxies
in this study that have typical effective radii of 1.3-6.5\arcsec, we
are finding this IFU and the excellent image quality of WIYN to be
quite effective in isolating host galaxy light from the central quasar
light. We use a configuration of the Bench Spectrograph that provides
an observed wavelength coverage of $\sim$4270-7130 \AA~at a resolution
of $\Delta\lambda\sim$~5~\AA~(110~km~s$^{-1}$). In the work presented
here PG~1309+355 was observed on WIYN in June 2007; PHL~909 and
PKS~0736+017 were observed in December 2007. These spectra have S/N
spanning 2.4-8.9~\AA$^{-1}$, with an average of 5.7~\AA$^{-1}$, for
exposure times of 2-4 hours.

\subsection{Imaging}

These objects all have HST or ground-based adaptive optics (AO) images
in the archive for analysis of host galaxy properties. We take galaxy
magnitudes and effective radii from the \citet{bahcall97} 2-D de
Vaucouleurs profile fits for eight of the objects. These data for
4C~31.63 are provided by \citet{hamilton02} and \citet{guyon06} and
for PKS~0736+017 by \citet{dunlop03}. We use color corrections from
\citet{fukugita95} and \citet{poggianti97} to convert all to r-band
and adopt a cosmology of $\Omega_{M}=$~0.3, $\Omega_{\Lambda}=$~0.7,
and H$_{0}=$~70~km~s$^{-1}$~Mpc$^{-1}$. Relevant object data are
given in Table \ref{object_table}.

\subsection{Comparison Galaxies}

To aid in the interpretation of the nature of our QSO host galaxies, we
will compare their structural properties to other types of
galaxies. This comparison sample includes Sloan Digital Sky Survey
(SDSS) early-type galaxies from
\citet{bernardi03a,bernardi03b,bernardi06}. The \citet{bernardi03a}
sample consists of nearly 9000 early-type galaxies that are placed on
the Fundamental Plane in \citet{bernardi03b}. \citet{bernardi06}
search for the most massive galaxies by selecting SDSS objects with
high stellar velocity dispersions, $\sigma_{*}>$~350~km~s$^{-1}$,
resulting in approximately 70 galaxies massive enough to be considered
giant ellipticals. They find that these galaxies lie on the
Fundamental Plane, but at its outer extreme.

For further comparison we add the host galaxies of PG QSOs from
\citet{dasyra07} who were investigating the possibility that these
QSOs were triggered during mergers of galaxies that contain gas, as
are ultraluminous infrared galaxies (ULIRGs). They conclude that some,
but not all, ULIRGs may undergo a QSO phase as the merger evolves and
that the progenitors of both PG QSOs and ULIRGs are likely
similar. \citet{dasyra07} compare their QSOs to a sample of galaxy
merger remnants from \citet{rothberg06} that we add to our comparison
as well. These objects were selected to be late stage mergers in which
the cores have coalesced into single nuclei.

\begin{figure}
 \epsscale{0.5}
 \plotone{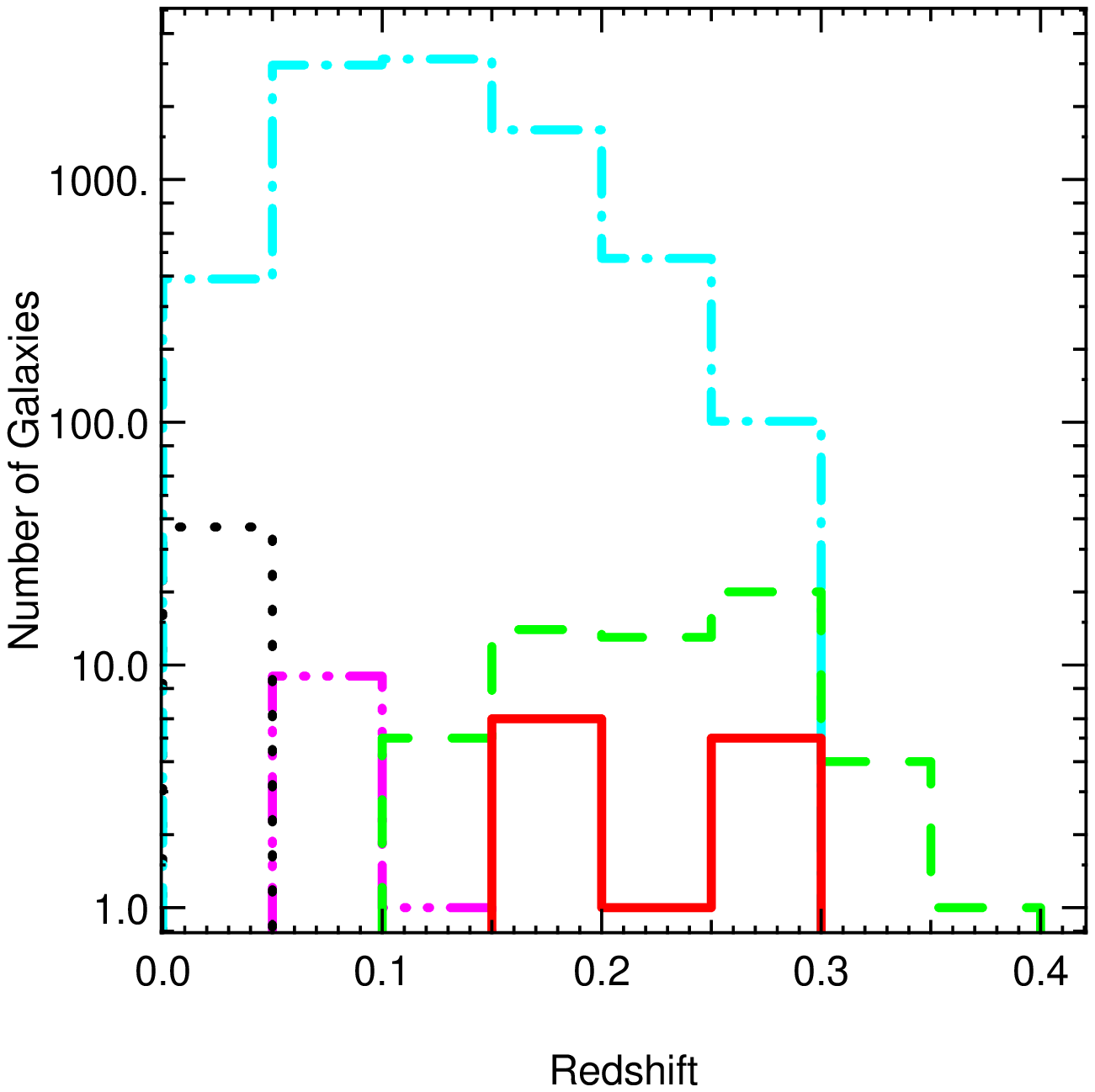}
 \caption{Redshift distributions of the galaxies. The solid red line
 is our sample of QSO host galaxies, the dot-dash cyan line is the
 large sample of early-type SDSS galaxies \citep{bernardi03a}, the
 dashed green line is sample of massive early-type SDSS galaxies with
 $\sigma_{*}>$~350~km~s$^{-1}$ \citep{bernardi06}, the dash-dot-dot
 magenta line is the sample of PG QSO hosts \citep{dasyra07}, and the
 dotted black line is the sample of merger remnant galaxies
 \citep{rothberg06}.  \label{redshift_fig} }
\end{figure}

The redshift distributions of our
QSO host galaxies and the comparison samples are shown in Figure
\ref{redshift_fig}.

\subsection{Surface Brightness and K-corrections}

\begin{figure}
 \epsscale{0.5}
 \plotone{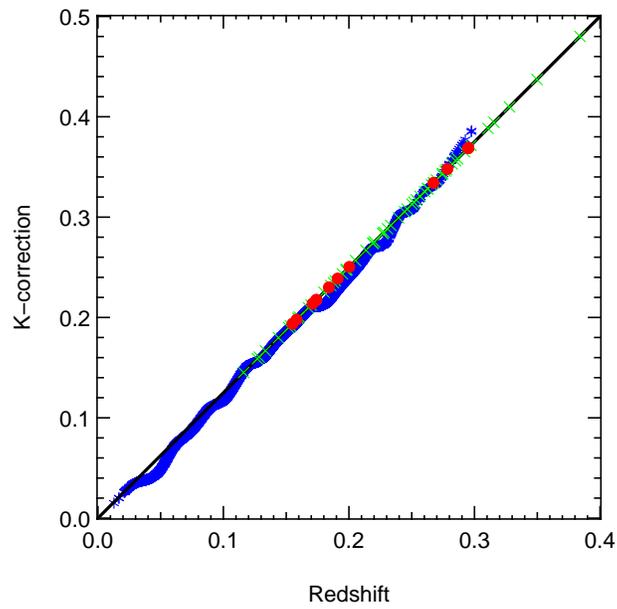}
 \caption{K-corrections for r-band.  The line is a linear fit to the
 K-corrections of the \citet{bernardi03a} galaxies ({\it blue
 asterisks}). This fit of slope 1.25 is applied to the redshifts of
 the \citet{bernardi06} galaxies ({\it green crosses}) and our QSO
 host galaxies ({\it red circles}) to derive their
 K-corrections. \label{kcor_fig} }
\end{figure}

Host galaxy magnitudes in \citet{bahcall97} are given for the F606W
HST filter. We convert these to r-band using their F606-V colors for
individual galaxies, and using the average colors for early-type
galaxies from \citet{faber89} and \citet{gebhardt03}, B$-$V$=$0.95,
and B$-$r$=$1.25. We take the magnitudes of 4C~31.63 from
\citet{hamilton02} in the F702W HST filter and of PKS~0736+017 from
\citet{dunlop03} in R-band, both of which we convert to r-band using
average colors for E and S0 galaxies from \citet{fukugita95}. With
m$_{r}$ we calculate host galaxy surface brightness from
\begin{equation}
\mu_{e}(r) = m_{r} + 2.5~log(2\pi R_{e}^{2}) - K(z) - 10~log(1+z) ,
\end{equation}
as in \citet{bernardi03a}, where R$_{e}$ is in arcsec. K-corrections
are given in \citet{bernardi03a} for the large sample of early-type
galaxies and we use their values. To derive K(z) for the other
galaxies, we linearly fit the \citet{bernardi03a} K-corrections that
are derived from a combination of \citet{bc03} models and early-type
galaxy templates from \citet{coleman80}, and apply them as
$K(z)=1.25z$ to the \citet{bernardi06} comparison galaxies and the QSO
host galaxies (Figure \ref{kcor_fig}). Below z=0.28 this fit gives a
median difference for the \citet{bernardi03a} galaxies of
$\Delta$K(z)=0.0040 with a standard deviation of
$\sigma_{K(z)}$=0.0038. Above z=0.28, affecting 16 of the \citet{bernardi06}
galaxies, these values change to $\Delta$K(z)=$-$0.0080 and
$\sigma_{K(z)}$=0.0028. Our K(z) relation is extrapolated to redshifts
higher than the fitted range for 5 of the \citet{bernardi06} galaxies.

\subsection{Scattered QSO Light Removal}

The light from the quasars in this study is as much as 3.5 magnitudes
brighter than the entire host galaxy. The regions of the host galaxies
that we observed were typically 5-8 mag fainter than the quasar. As a
result of the blurring produced by the Earth's atmosphere, plus a
small contribution from the intrinsic optical point-spread function
and diffraction produced by the telescope, some light from the quasar
entered the spectrograph in the off-nucleus observations, either
long-slit or fiber.

\begin{figure}
 \epsscale{0.7}
 \includegraphics[angle=-90,scale=0.65]{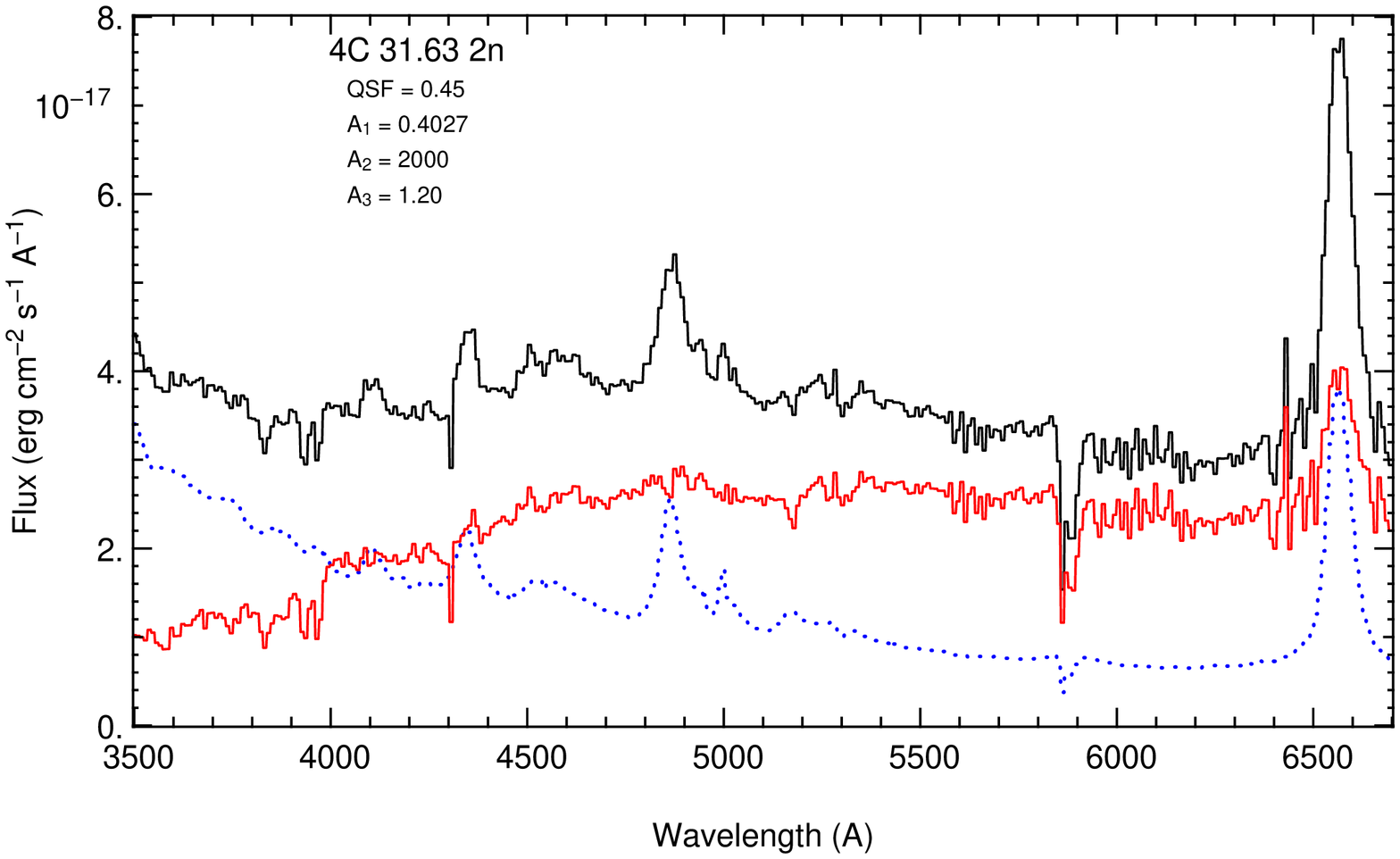}
 \includegraphics[angle=-90,scale=0.65]{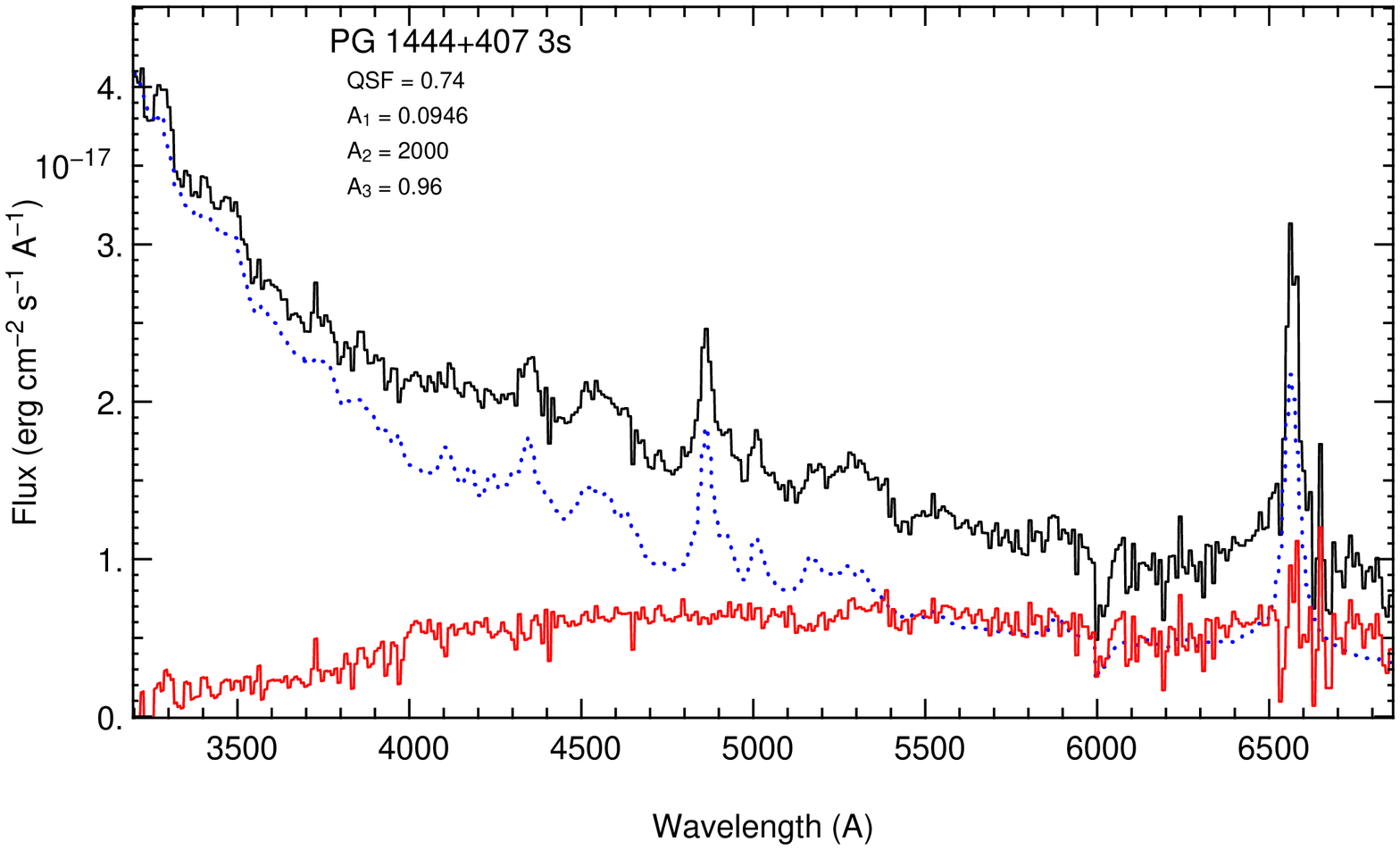}
 \caption{Examples of scattered quasar light removal on 4C~31.63 and
 PG 1444+407. The upper solid black lines are the extracted off-nuclear
 spectra, the dotted blue lines are the products of the scatter
 efficiency curves (with given best-fit parameters) and the nuclear
 quasar spectra, and the lower solid red lines are the final scatter
 subtracted host galaxies. All are binned to the instrumental
 resolution. QSF is the total quasar scattered light fraction in the
 extracted spectrum.
 \label{scatter_fig} }
\end{figure}

To spectrally remove the scattered light, we must both determine the
fraction of light in the extracted spectrum that is from the quasar
and also account for wavelength-dependent scattering efficiency due to
atmospheric seeing (in the sense that as seeing gets worse more
scattering into a fixed offset distance from the quasar occurs with
shorter wavelength). We approximate these two factors by deriving a
scattering efficiency curve of the form, $\psi$ = A$_{1}$ + A$_{2}
\lambda^{-A_{3}}$. We subtract the product of the scattering
efficiency curve and the observed central quasar spectrum from the
off-nuclear spectrum and iteratively fit that result by two-population
stellar synthesis galaxy model spectra from \citet{bc03}. Examples of
this approach are shown for 4C~31.63 and PG~1444+407 in Figure
\ref{scatter_fig}, where the upper solid black lines are the observed
off-nuclear spectra, the dotted blue lines are the product of the
nuclear quasar spectrum and the derived scattering efficiency curve,
and the lower solid red lines are the resulting scatter-subtracted host
galaxy spectra. For this work we are only concerned with a small
spectral region around the 4000~\AA~break to measure stellar velocity
dispersion. This scatter-subtraction approximation is sufficient for
such a small region. However, in future work, to accurately determine
constituent stellar populations of the host galaxies using the entire
observed spectrum, we plan to implement a more rigorous scattering model
based on the geometry of the slit/fiber, its distance from the central
quasar, approximate seeing conditions, and selective weighting of more
important spectral features, such as QSO broad line regions.

The basic model spectra fitting techniques used here are a modified
version of \citet{wolf07} and \citet{liu07}. We use 6 fitting
parameters: old age, young age, luminosity-weighed fraction of the
young age, and the scattering efficiency coefficients A$_{1}$,
A$_{2}$, and A$_{3}$. Given the relatively low S/N of the spectra we
are not able to distinguish metallicity, so only [Fe/H]=0 models are
used. Age ranges for the old stellar population run from 10 Myr up to
the age of the universe at the redshift of the quasar.  Young ages are
allowed to be any in the model grid that are younger than the current
old age.  $\chi^{2}$ is calculated at each wavelength and weighted by
the observed spectrum's noise at that point, $\chi_{i}^{2} =
[(F_{gal_{i}} - \alpha F_{model_{i}})/\sigma_{i}]^{2}$, where
$F_{gal}$ is the flux of the scatter-subtracted galaxy spectrum,
$F_{model}$ is the flux of the model galaxy spectrum, $\alpha$ is a
model spectrum normalization parameter, $\sigma$ is the observed
spectral noise, and the index $i$ refers to each wavelength point. For
the Keck data, noise spectra are generated by measuring the rms of the
restframe spectrum at three locations devoid of prominent spectral
features: $\sim$3600~\AA, $\sim$5400~\AA, and $\sim$6100~\AA. We fit a
polynomial to the measured rms values and extrapolate across the
entire wavelength range to generate the galaxy's noise spectrum.  For
the WIYN data, noise spectra are calculated directly using the
observed sky spectra. In both data, spectral features not included in
the the model spectra, such as sky residuals or galaxy emission lines,
are masked and do not affect the $\chi^{2}$ calculation. The
best-fitting combination of parameters is selected as the one that
provides the minimum $\chi^{2}$ value when summed over wavelength.

\subsection{Stellar Velocity Dispersions \label{vd_section} }

We measure velocity dispersions from stellar absorption lines in the
host galaxies by fitting a stellar template that has been convolved
with a gaussian profile to the off-nuclear, scatter-subtracted galaxy
spectrum. The continuum is normalized on both galaxy and template
spectra and a direct fit is done in pixel space using the code of Karl
Gebhardt \citep{gebhardt00b,gebhardt03}. The template is compiled from
21 weighted stellar spectra from the Coud\'{e} Feed Spectral Library
at KPNO \citep{valdes04}, that were selected to provide a good
representation of main sequence A stars through K giants. We use the
same group of stars as \citet{gebhardt03}, which are listed with their
parameters in Table \ref{star_table}. We are able to use this library
of stellar spectra taken with a different instrument because the
spectral resolution ($\Delta\lambda$ = 1.8~\AA) is much higher than
our host galaxy data.  Fitting parameters include the weights of the
individual stellar spectra and the gaussian width to which the
combined template is convolved. We select the best fit as the
combination that provides the lowest rms differences from the galaxy
spectrum. 

Velocity dispersion uncertainties are calculated through Monte Carlo
simulations by adding gaussian noise to each pixel in the final
template, which has very high S/N, at a level such that the mean
matches the noise in the initial galaxy spectrum and the standard
deviation is given by the rms of the initial fit. The velocity
dispersion is measured for 100 noise realizations and the mean and
standard deviation of these results provide the measured velocity
dispersion and its 1$\sigma$ uncertainty. The distribution of measured
values is also used to determine the sigma bias of the measurement. As
long as the bias is much less than the uncertainty then the
measurement is considered valid (see Appendix \ref{bias}).

We use the approximate wavelength range of 3850-4200~\AA~for our
velocity dispersion fits, which contains the Ca~II~H\&K (3968,
3934~\AA) absorption lines. Even though these lines were historically
avoided because of their intrinsic broadening, they have been shown to
work very well in the presence of mixed stellar populations in
galaxies \citep{kobulnicky00,gebhardt03}, as long as the mix of
stellar spectra in the fitting template is individually determined for
each galaxy. \citet{greene06} also point out that the Ca H\&K lines
should be good for measuring $\sigma_{*}$ in powerful AGNs because
they may be the only stellar absorption lines with sufficient
equivalent width to persist in high luminosity AGNs. We mask a few
features in the region including H$\delta$ since Balmer line widths
may be dominated by rotation and pressure broadening of A stars; the
H8 line just blueward of Ca K; and the Ca~H line in some cases that
show contamination. We also tested the Mg I{\it b} triplet (5167,
5173, 5184~\AA) region and found that it did not produce reliable
results in these data (see Appendix \ref{bias}).

The velocity dispersion fits for our ten host galaxies are shown in
Figure \ref{vd_fig}. The black solid lines are the galaxy spectra and
the dashed red overlays are the best-fitting templates. Yellow shaded
regions mark features that were masked and disregarded during the fit.
PKS~2135-147 contained particularly strong broad QSO emission at the
Ca~H line. 3C~273 and PKS~2349-014 also show contamination from narrow
emission near this line. Our measured dispersions are given in Table
\ref{vd_table}, along with the radii at which the galaxies were
observed, the relative fractions of stars in the best-fitting template
(A V, F-G V, G-K III), the spectral S/N over the fitted region, and
the amount of scattered quasar light that was removed from the
spectrum. Measured velocity dispersions range from 150 to
346~km~s$^{-1}$ with a mean of 272~km~s$^{-1}$. We tested the
possibility of the A stars in the template inflating the velocity
dispersions measured from the Ca II lines by removing them from the
available stellar templates and got the same results within the
uncertainties. In fact, the measured $\sigma$'s with the A stars in
the mix were systematically lower by $\sim$ 7\%.

\begin{figure}
 \epsscale{0.7}
 \includegraphics[angle=-90,scale=0.65]{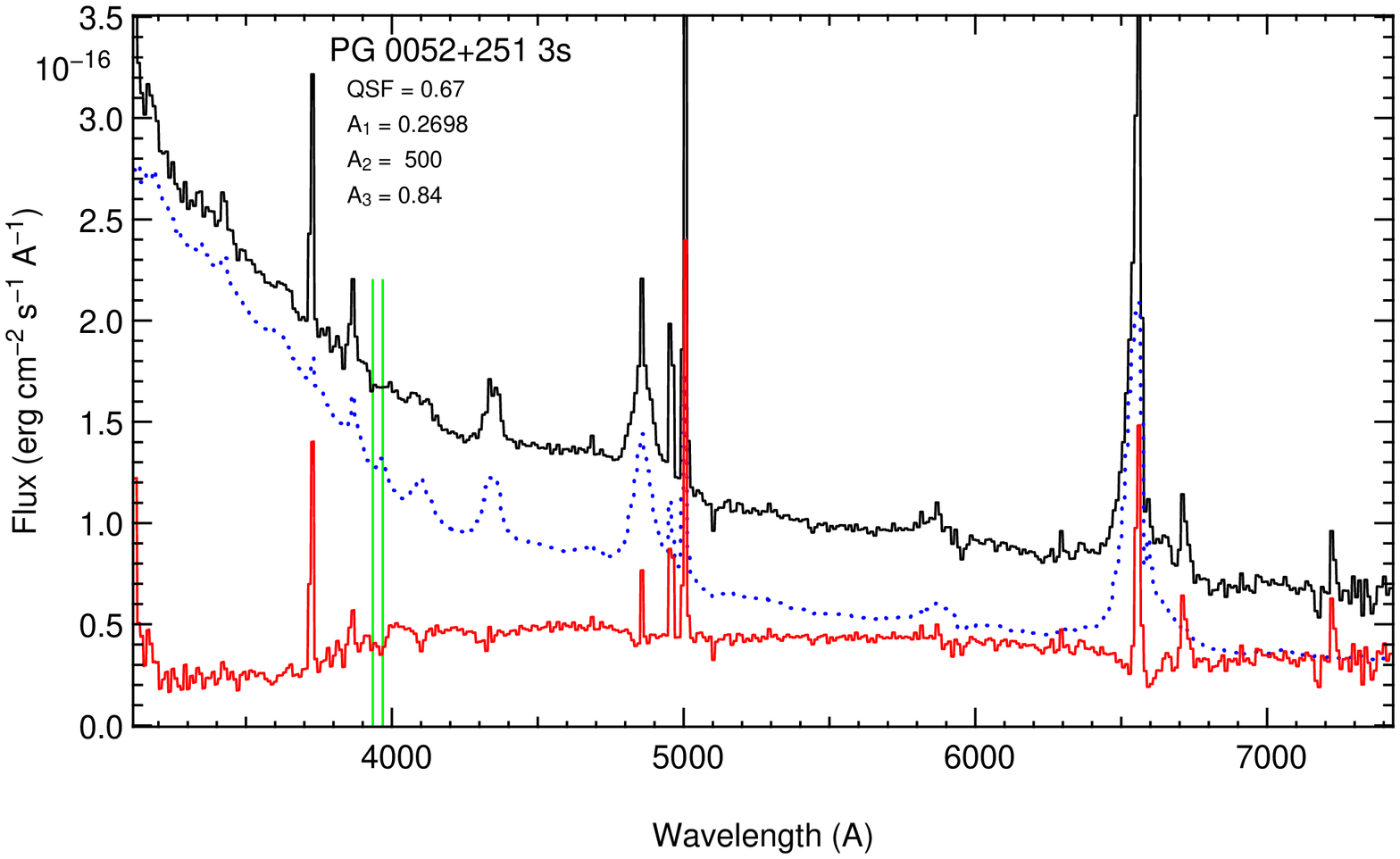}
 \includegraphics[angle=-90,scale=0.65]{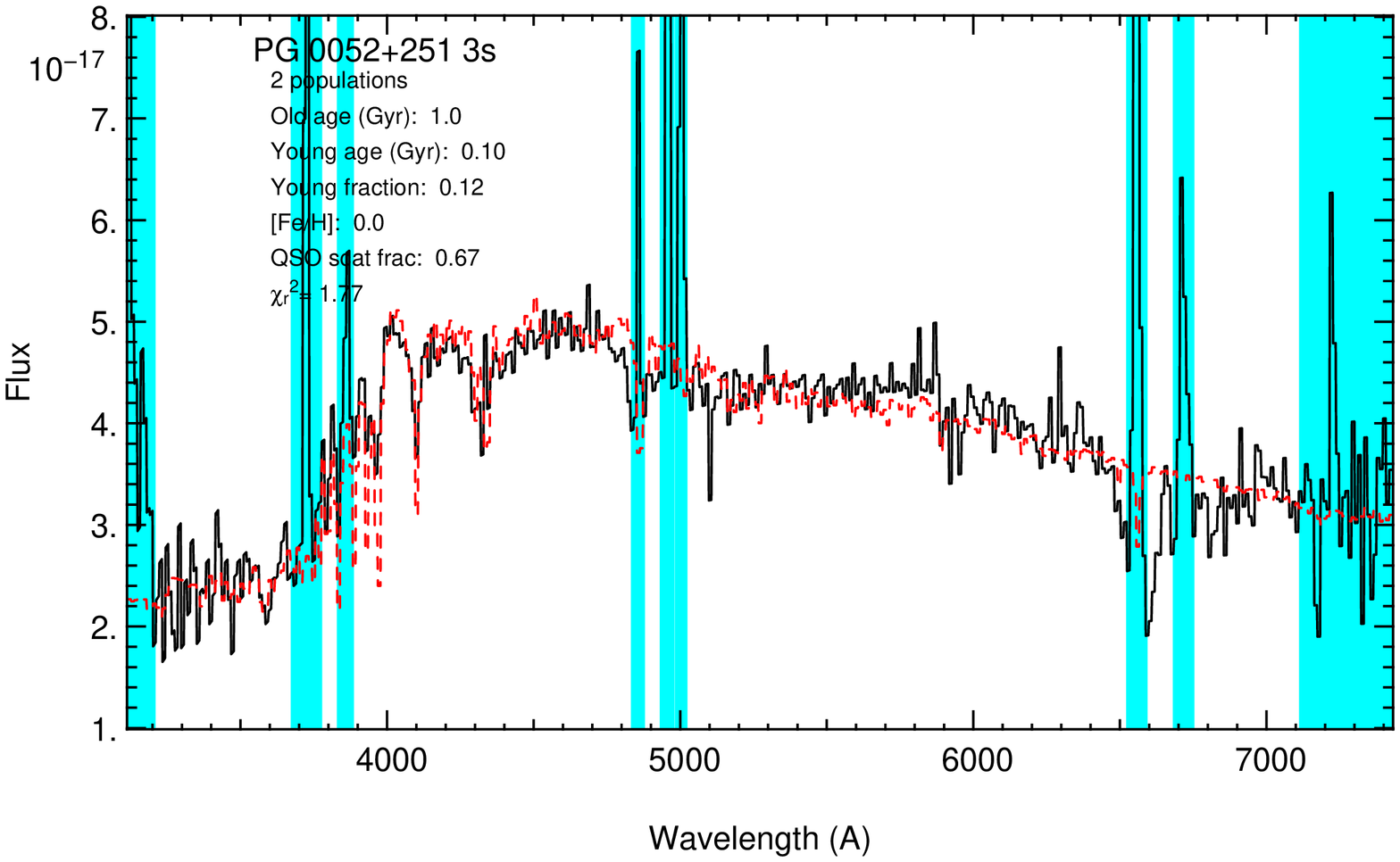}
 \caption{ Problematic quasar emission line from Ne III at
 3969~\AA~affecting the Ca H\&K region  for PG~0052+251. The QSO 
 scattered light subtraction is shown on top (legend same as in 
 Figure \ref{scatter_fig}) and the simultaneously  best fitting 
 stellar population model on the bottom. The Ca H\&K lines (marked 
 by green vertical lines in the top plot) appear affected.
 \label{bad_sub} }
\end{figure}

Note that three velocity dispersions are marked as upper limits. Tests
on velocity dispersion standard galaxies of \citet{mcelroy95} show
that we should be able to measure down to about 20\% below the
instrumental resolution of 300 km s$^{-1}$, or 240~km~s$^{-1}$ (see
Appendix \ref{bias}). We mark any measurements $<$250~km~s$^{-1}$ as
upper limits. For PG~0052+251, even though its fit looks reasonable in
Figure \ref{vd_fig}, the measured $\sigma_{*}$ exhibits higher sigma
bias than the rest of our objects over the range of input $\sigma$'s
relevant for our spectral resolution (see Appendix \ref{bias}). The
data seem to indicate that the true $\sigma_{*}$ is below our
resolution, therefore, we can only give an upper limit on the velocity
dispersion of PG~0052+251. On close inspection of its spectrum, this
measurement may be affected by insufficient scatter subtraction of a
Ne III emission line at 3969~\AA~in the quasar spectrum
\citep{hewitt93} that falls in the vicinity of the Ca H\&K lines,
causing a strange shape to the lines (Figure \ref{vd_fig}). Despite
the fact that this galaxy has relatively high S/N, the amount of
scattered quasar light and location of this feature prevent accurate
scatter subtraction around these lines. Figure \ref{bad_sub} shows our
scatter subtraction of PG~0052+251 with the Ca H\&K lines marked and
the stellar population model fit for this galaxy. Although the
continuum and many other features of the host galaxy are fit very well
by the model, the line depths between 3800 and 4100~\AA, which is
riddled with quasar emission, are very discrepant, signaling that the
galaxy absorption lines may be contaminated. It is possible that this
could contribute to the problematic sigma bias in the velocity
dispersion measurements on this galaxy. A further discussion of tests
of $\sigma_{*}$ measurement on galaxy spectra diluted by residual
scattered quasar light can be found in Appendix \ref{bias}.

\subsubsection{Aperture Correction}

Because velocity dispersion varies with galaxy radius, to match the
comparison data we aperture-correct all host galaxy velocity
dispersions from the radius at which $\sigma_{*}$ was measured to the 
radius of R$_{e}$/8, as in \citet{bernardi03a} (c.f. \citet{jorgensen95}),
\begin{equation}
\sigma_{cor} = \sigma_{meas} \left( \frac {R_{obs}} {R_{e}/8}
\right)^{0.04} .
\end{equation}
In the Keck data, the host galaxy spectra are off-nuclear long-slit
observations for which the extracted spectra are summed along the
slit, therefore, the weighted average radius along the summed portion
of the slit is used as the effective observed radius. For the WIYN
data the approximate radius at which the fiber was centered is used.
Effective observed radii are given in Table \ref{vd_table}.
Aperture-corrected velocity dispersions of the host galaxies are used
in the following Fundamental Plane analysis. Average aperture
corrections are 9\% for these galaxies.

\section{Results \label{result_section}}

\subsection{The Fundamental Plane}

Previous work has shown that host galaxies of AGN lie on the
Fundamental Plane (FP) occupied by quiescent early-type galaxies
(e.g. \citet{woo04,treu07,dasyra07}). Here we explore the properties of
the host galaxies of higher luminosity quasars. We place the
QSO host galaxies on the Fundamental Plane in Figure \ref{fp_fig},
using the r-band projection against effective radius, R$_{e}$ from
\citet{bernardi06},
\begin{equation}
log~\sigma_{*} + 0.2 \left[ \mu_{e}(r) - 19.64 \right] ,
\end{equation}
with R$_{e}$ in kpc, stellar velocity dispersion, $\sigma_{*}$, in
km~s$^{-1}$, and surface brightness, $\mu_{e}$(r), in
mag~arcsec$^{-2}$. Our QSO host galaxies are shown as large circles
(green are radio-loud and yellow are radio-quiet, using the division
defined in \citet{kellermann94} at
L$_{5~GHz}\sim$~10$^{26}$~W~Hz$^{-1}$).  The numbers inside these
symbols correspond to object numbers in the tables. The PG QSO host
galaxies of \citet{dasyra07} are shown as triangles. These objects are
all radio-quiet and some have known morphologies
\citep{guyon06}: large filled triangles are ellipticals,
crosses on triangles denote ellipticals with signs of interaction,
large open triangles are early-types, large open triangles with
asterisks are spirals, small filled triangles are unknown. Comparison
SDSS early-type galaxies are the small cyan points \citep{bernardi03a}
and early-types with $\sigma_{*}>$~350~km~s$^{-1}$ \citep{bernardi06}
are the open diamonds. Merger remnant galaxies from \citet{rothberg06}
are denoted as squares: normal mergers as filled red squares,
LIRG/ULIRGs as filled black squares, and shell ellipticals as open
squares.

The QSO hosts occupy both regions covered by the Bernardi et
al.~early-type galaxy samples, while the merger remnants lie within or
below the main early-type galaxy distribution. All of our radio-loud
QSO hosts lie near the upper envelope of the FP, while the radio-quiet
objects (both ours and those of Dasyra et al.) mostly lie within the
main locus. The implications of a location in the upper envelope of
the FP is higher stellar velocity dispersion, lower surface
brightness, or larger effective radius than normal for typical
early-type galaxies. The \citet{bernardi06} galaxies occupy this
region as well. They concluded that these massive galaxies were not
outliers, but merely at the outer extreme of the FP
distribution. Figure \ref{parameters_fig} plots the fundamental galaxy
parameters against each other. It is clear from these plots that our
host galaxies at the upper extreme of the FP have systematically
higher velocity dispersions, lower surface brightnesses, and larger
effective radii than the normal early-type galaxies.

A caveat to keep in mind is that we have used surface brightnesses and
effective radii derived from 2-D de Vaucouleurs profile fits to all
galaxies, regardless of morphology. This was an attempt to analyze all
objects in the same way for comparison, however some objects have
obvious morphological disturbances not typical of early-type
galaxies. Three host galaxies of radio-quiet QSOs show obvious
spiral structure.  The consequences of fitting disk galaxies with an
r$^{1/4}$ model, which is more characteristic of their bulges, is
discussed in \S \ref{profile_section}.

\subsection{Mass-to-Light Ratios}

Now we explore the behavior of the mass-to-light ratios (M/L) of the
galaxies. For this analysis we calculate the galaxy mass from:
\begin{equation}
M_{*} = \frac {5~\sigma_{*}^{2}~R_{e}} {G}~~[M_{\sun}],
\end{equation}
where 5 is a structural constant corresponding to $R_{tidal}/R_{core}
\approx 100$, where $R_{core}$ is the radius at which surface
brightness has dropped to half its central value, for King surface
brightness models of giant to intermediate mass ellipticals
\citep{bender92}. The luminosity is calculated from:
\begin{equation}
L = 2\pi~\langle I_{e} \rangle R_{e}^{2}~~[L_{\sun}],
\end{equation}
with $\langle I_{e} \rangle$, the mean surface brightness within
$R_{e}$, calculated from:
\begin{equation}
\mathrm{log}~\langle I_{e} \rangle = -0.4(\langle \mu_{e}
\rangle-c_{1} - c_{2}(z))~~[L_{\sun}~\mathrm{pc}^{-2}],
\end{equation}
where $c_{1}$=26.4 for the Gunn r-band \citep{jorgensen96}, $c_{2}(z)$
is a redshift-dependent correction to our cosmology, and $\mu_{e}$ is
the surface brightness in units of mag~arcsec$^{-2}$. 

Table \ref{mass_table} gives the values of M/L and stellar mass for
our host galaxies and the lower right panel of Figure
\ref{parameters_fig} shows the calculated M/L as a function of
velocity dispersion for our hosts and the comparison objects. As noted
in \citet{bernardi03b}, the slope of the (M/L)-$\sigma_{*}$ relation
can be quite different when comparing the M/L calculated from
observables and that predicted from the Fundamental Plane. Therefore,
we use the M/L calculated from observables. The host galaxies of
radio-loud quasars show systematically higher M/L than the radio-quiet
objects.

\subsection{Individual Objects \label{individuals} }

We now compile all information on our individual QSOs and their host
galaxies. The magnitudes quoted in this section are from
\citet{bahcall97} except for 4C~31.63, which is from
\citet{hamilton02}, and PKS~0736+017, which is from
\citet{dunlop03}. All quoted velocity dispersions are
aperture-corrected values. Stellar populations are qualitatively mentioned,
based on the model fitting performed during our scatter subtraction
process. More accurate stellar populations will be analyzed in future
work. Where relevant, implied star formation rates from [O II] line
emission in the entire host galaxies are taken from \citet{ho05}. For
reference, mean galaxy parameters for the different types of objects
are given in Table \ref{avg_table}.

\subsubsection{PG 0052+251}

PG~0052+251 (labeled 1 in our plots) is a radio-quiet quasar with
M$_{V}$=$-$24.1 and L$_{5 GHz}$=2.4x10$^{39}$ erg~s$^{-1}$. The host
galaxy, at M$_{V}$=$-$22.5, is obviously a spiral in the HST images
and was classified as such by both Bahcall et al.~and Hamilton et
al.~(also studied by \citet{bog82}, \citet{BPO85}, \citet{SM87},
\citet{HJN89}, \citet{dunlop93}, \citet{MR94}, \citet{BKS96}). It has a
companion at a distance of 14.1\arcsec. Bahcall et al.~suggest that
this galaxy is an Sb spiral with a fairly large bulge component, best
fit by an exponential disk 2-D profile. 

This host galaxy is the smallest (R$_{e}$=4.8 kpc) and brightest
($\mu_{e}$=19.35 mag~arcsec$^{-1}$) of our sample. It lies within the
early-type galaxy Fundamental Plane and has properties consistent with
the Dasyra et al.~PG QSO hosts. Its velocity dispersion has an upper
limit of 279~km~sec$^{-1}$. The galaxy spectrum contains many emission
lines indicating ongoing star formation and no significant old stellar
population. Furthermore, Bahcall et al.~observed 11 bright HII regions
in the HST image. \citet{ho05} estimates a star formation rate of 3.7
M$_{\sun}$~yr$^{-1}$ from the [OII] luminosity. We estimate a young
effective age for this galaxy.

\subsubsection{PHL 909 (0054+144)}

PH~909 (labeled 2 in our plots) is a radio-quiet quasar with
M$_{V}$=$-$24.1 and L$_{5 GHz}$=1.1x10$^{40}$ erg~s$^{-1}$. Unlike our
other radio-quiet hosts, this one is a normal E4 elliptical galaxy
with M$_{V}$=$-$22.2 and has a companion $\sim$12 arcsec to the
west. Extended emission toward this companion was noted by
\citet{dunlop93} and \citet{gehren84}, but not by \citet{bahcall97}.

This host galaxy has our smallest measured velocity dispersion
(167~km~s$^{-1}$) that is also at the low end of the Dasyra et al.~PG
QSO sample.  Its effective radius (6.7~kpc) lies in the transition
region between our radio-quiet and radio-loud objects. Its surface
brightness (20.4~mag~arcsec$^{-1}$) is lower than the rest of our
radio-quiet hosts and near the mean of our radio-loud hosts. The
stellar mass (2.2~x~10$^{11}$~M$_{\sun}$) and M/L (3.8) are the lowest
of our sample. Its spectrum shows [OII] emission and looks to be of
intermediate age.

\subsubsection{PKS 0736+017}

PKS~0736+017 (labeled 3 in our plots) is a radio-loud, compact,
flat-spectrum quasar \citep{gower84,romney84} with M$_{V}$=$-$23.2 and
L$_{5 GHz}$=9.0x10$^{42}$ erg~s$^{-1}$. The host galaxy
(M$_{V}$=$-$22.7) shows a disturbed morphology with nebulosity
extending to nearby companions \citep{dunlop93}.

This host galaxy's velocity dispersion (342~km~s$^{-1}$) is above
average for the radio-loud hosts. The effective radius (10.4~kpc) and
surface brightness (21.0~mag~arcsec$^{-1}$) are slightly below
average. It sits high on the Fundamental Plane for its size. The
stellar mass (14.1~x~10$^{11}$~M$_{\sun}$) is near the mean of the
radio-loud hosts, while its M/L (17.8) is above average. The spectrum
shows no emission lines and can be fit by an intermediate aged
population.

\subsubsection{3C 273 (PG 1226+023)}

3C~273 (labeled 4 in our plots) is the most luminous quasar in our
sample, both in the optical (M$_{V}$=$-$26.7) and the radio
(L$_{5 GHz}$=1.3x10$^{44}$ erg~s$^{-1}$). It has a large radio jet
that is also weakly detected in the optical HST images
\citep{bahcall97}. The host galaxy is classified as an elliptical by
\citet{bahcall97} and \citet{hamilton02} with
M$_{V}$=$-$23.2. There are no nearby companion galaxies.

This galaxy has a velocity dispersion of 334 km~sec$^{-1}$, falling at
the low end of the distribution of \citet{bernardi06} massive
elliptical galaxies, but slightly above average for our radio-loud
hosts. Its surface brightness (20.3 mag~arcsec$^{-1}$) is slightly
fainter than the mean of the \citet{bernardi03a} normal early-type
galaxies. Its effective radius is above average (10.1 kpc), but well
within the distribution of normal early-type galaxies and near the
mean of our six radio-loud objects and of the massive
early-types. The M/L (9.1) is above the normal and lower than the
massive early-type galaxies. Its spectrum shows [OII] emission, but
can be fit by a relatively old single population.

\subsubsection{PKS 1302-102}

PKS~1302-102 (labeled 5 in our plots) is a radio-loud, flat-spectrum,
quasar at M$_{V}$=$-$25.9 and L$_{5 GHz}$=9.5x10$^{42}$
erg~s$^{-1}$. The host galaxy, at M$_{V}$=$-$22.9, is classified as
an elliptical that may be slightly disturbed by two companions within
2.3\arcsec~\citep{hutchings92,bahcall97, hamilton02, guyon06}. Bahcall
et al.~visually classify this galaxy as an E4, however they find that
an exponential disk profile best fits the PSF-subtracted
image. \citet{guyon06} comment on the difficulty of decoupling the
host and its closer companion, but find that their 2-D fit favors an
r$^{1/4}$ model.

On the Fundamental Plane this host occupies a transition region
between our radio-loud and radio-quiet objects. Its velocity
dispersion (388~km~s$^{-1}$) is on the high end of our radio-loud
hosts, however the surface brightness (19.56~mag~arcsec$^{-1}$) is
higher than all other radio-loud hosts, and near that of our
radio-quiet QSO hosts. Its size (R$_{e}$=5.9~kpc) is comparable to the
radio-quiet hosts as well. The M/L (10.3) and stellar mass (10.3 x
10$^{11}$~M$_{\sun}$) are on the low end of the radio-loud objects,
but larger than the radio-quiet objects. This stellar mass is similar
to the massive early-type galaxies, however, the M/L is slightly
lower. The spectrum shows [OII] in emission, but the 4000~\AA~break
region suggests a fairly old stellar population. Current data quality
does not allow a secure fit over the rest of the spectrum.

\subsubsection{PG 1309+355}

PG~1309+355 (labeled 6 in our plots) could be called a
radio-intermediate quasar. It falls into the category of radio-quiet
using the \citet{kellermann94} cut at
L$_{5~GHz}\sim$~10$^{26}$~W~Hz$^{-1}$, but is classified as radio-loud
using the criterion of radio-to-optical flux density ratio greater
than 10 \citep{brinkmann97,yuan98,hamilton02}. The absolute magnitude
of the quasar is M$_{V}$=$-$24.4 and radio luminosity is L$_{5
GHz}$=1.8x10$^{41}$ erg~s$^{-1}$. The host galaxy, at
M$_{V}$=$-$22.8, is classified as an early-type spiral by
\citet{bahcall97} and \citet{hamilton02} and as a moderately elongated
elliptical by \citet{guyon06}. The HST images show tightly wrapped
spiral arms. There are no nearby companions.

This galaxy has a high effective radius (6.2~kpc) for our radio-quiet
hosts, which is equal to the highest of the Dasyra et al. PG QSO
hosts, lies well below the average of our radio-loud hosts, and is
between the means of the normal and massive quiescent early-type
galaxies. The velocity dispersion (265~km~sec$^{-1}$) falls at the
high end of the distribution of Dasyra et al.~PG QSO hosts and above
the mean of the normal early-type galaxies. The surface brightness
(19.62~mag~arcsec$^{-2}$) is near the mean of our radio-quiet hosts
and of the Dasyra et al.~PG QSO hosts. The M/L (5.0) and stellar mass
(5.0 x 10$^{11}$~M$_{\sun}$) are the lowest of our host galaxy
sample. Its M/L is very close to the Dasyra et al.~PG QSO hosts, while
its stellar mass is higher than the means of the PG QSO hosts, the
merger remnants, and the normal early-type galaxies. The host galaxy
spectrum shows [OII] emission, but is fit well by an old population.

\subsubsection{PG 1444+407}

PG~1444+407 (labeled 7 in our plots) is a radio-quiet quasar with
M$_{V}$=$-$25.3 and L$_{5 GHz}$=1.8 x 10$^{39}$ erg~s$^{-1}$. The QSO
spectrum is similar to that of 4C~31.63, showing strong Fe II
emission, broad Balmer lines and no detectable forbidden
emission-lines. The host galaxy, at M$_{V}$=$-$22.7, appears smooth
and elongated in the north-south direction in the HST images of
Bahcall. They suggest the host galaxy has the appearance of an E1,
although the light profile is better fit by an exponential disk. It is
classified as a spiral galaxy by Hamilton et al.~and a bar may be
present \citep{hutchings92,bahcall97}.

The size (R$_{e}$=5.3~kpc) of this host galaxy is near the average of
the normal early-type galaxies and slightly higher than the Dasyra et
al.~PG QSO hosts. Its velocity dispersion (316~km~s$^{-1}$) is the
highest of our radio-quiet hosts and near the mean of our radio-loud
hosts, though its stellar mass (6.4 x 10$^{11}$~M$_{\sun}$) is much
lower than the radio-loud hosts. Its surface brightness (19.51
mag~arcsec$^{-1}$) is near that of our other radio-quiet QSO hosts,
brighter than our radio-loud hosts, and about average compared to the
Dasyra et al.~PG QSO hosts. The spectrum can be fit with about 3/4 of
the light from a moderately young population and the rest old. [OII]
is in emission and \citet{ho05} estimates a star formation rate of
19.4~M$_{\sun}$~yr$^{-1}$ from the [OII] luminosity.

\subsubsection{PKS 2135-147}

PKS~2135-147 (labeled 8 in our plots) is a radio-loud quasar with
M$_{V}$=$-$24.7 and L$_{5 GHz}$=7.7x10$^{42}$ erg~s$^{-1}$. The host
galaxy (M$_{V}$=$-$22.4) is classified as an elliptical with two
companions within 5.5\arcsec~\citep{bahcall97,hamilton02}.

The size of this host (R$_{e}$=8.6~kpc) is average for our sample, but
slightly above the average of the high mass early-type galaxies. Its
surface brightness (20.83~mag~arcsec$^{-1}$) is average for our
radio-loud hosts, but lower than all other groups of objects. The
velocity dispersion (307~km~s$^{-1}$) is slightly below average for
the radio-loud hosts. The M/L (14.6) is average for our radio-loud
sample and stellar mass (9.4 x 10$^{11}$~M$_{\sun}$) is on the low end
of the radio-loud hosts. The spectrum shows many emission lines,
suggesting ongoing star formation, however, we are not able to make an
estimate of this galaxy's stellar populations with our current data
quality.

\subsubsection{4C 31.63 (2201+315)}

4C~31.63 (labeled 9 in our plots) is a radio-loud quasar at
M$_{V}$=$-$25.1 and L$_{5 GHz}$=1.9x10$^{43}$ erg~s$^{-1}$. It has
the spectrum of a typical strong Fe~II object. The host galaxy is one
of the few objects in our larger sample, and the only object in this
subsample, that is clearly a first-rank elliptical at
M$_{V}$=$-$23.8.  This galaxy is classified as a smooth elongated
elliptical from both optical HST images \citep{bahcall97,hamilton02}
and ground-based near infrared AO images \citep{guyon06} with three
apparent companions within 5\arcsec of the QSO.

This is our most extreme host galaxy on the Fundamental Plane, lying
in the upper right corner. It has the largest effective radius of our
objects (28.8 kpc) and, correspondingly, the lowest surface brightness
(22.18 mag~arcsec$^{-1}$). Its velocity dispersion measured in three
locations ranges from 259 to 339 km~sec$^{-1}$, but is consistent
within the error bars. The average measured $\sigma_{*}$ gives a mass
of 30.1 x 10$^{12}$ M$_{\sun}$, the highest of our sample.  The M/L
(14.3) is typical of the massive early-type galaxies.  The host galaxy
spectrum clearly reveals an old galaxy with no emission lines. We
estimate an old population with no apparent signs of a younger
one. This is the only host galaxy in our sample with the purely old
spectrum typical of a giant elliptical.

\subsubsection{PKS 2349-014}

PKS~2349-014 (labeled 10 in our plots) is a radio-loud quasar at
M$_{V}$=$-$24.5 and L$_{5 GHz}$=2.9x10$^{42}$ erg~s$^{-1}$. The QSO
has a typical spectrum with strong broad permitted lines and and much
narrower forbidden lines with a modest amount of Fe II emission. The
host galaxy (M$_{V}$=$-$23.2) appears very disturbed, although the
radial profile is well described by an r$^{1/4}$ model
\citep{bahcall97,hamilton02}. This galaxy shows obvious morphological
signs of gravitational interaction, such as large tidal arms and
extensive diffuse, off-center nebulosity. \cite{BKS295} identify
several distinct regions for which we have obtained spectra: the
diffuse nebulosity 3\arcsec~south of the nucleus and the ``wisps''
(possible tidal feature) 4\arcsec~north of the nucleus.

The velocity dispersions of these two components are the same within
error bars, at $<$250 and 302~km~s$^{-1}$, respectively, and are near
the average of our QSO host galaxy sample. The host galaxy surface
brightness (21.02~mag~arcsec$^{-1}$) is below the average of our
radio-loud objects and of the massive early-type galaxies. Its
effective radius (14.1~kpc) is the second highest in our sample and
larger than the average massive early-type galaxy. The M/L (8.3) and
stellar mass (12.2 x 10$^{11}$~M$_{\sun}$) are slightly below the
averages of the radio-loud hosts. The spectra of the two regions show
a number of emission lines. They can be fit by an underlying old
population with 20-50\% moderately young. \citet{ho05} estimates a
star formation rate for this host galaxy of 8.3~M$_{\sun}$~yr$^{-1}$
from the [OII] luminosity.

\section{Discussion \label{discussion} }

\subsection{ The Exponential Disk Galaxies \label{profile_section} }

Although we have thus far utilized the parameters of our QSO host
galaxies that were derived from 2-D r$^{1/4}$ profile fits,
\citet{bahcall97} actually found that four galaxies were better fit
by an exponential disk model: PG~0052+251, PG~1444+407, PG~1309+355, and
PKS~1302-102. The first three of these show obvious spiral structure and
the fourth one is described as a disturbed elliptical. We now analyze
the properties derived from exponential disk fits to these four
galaxies. Placement of disk galaxies on the Fundamental Plane is not
technically correct since this relation holds only for early-type
galaxies and bulges. Nonetheless, we do it here to investigate the
range in which these galaxies should fall, given that they likely have a
bulge component and that our velocity dispersion measurements may have 
contained light from both the disk and the bulge. Therefore, these
galaxies should exist somewhere between the positions of pure bulge
and pure disk. 

Table \ref{expfit_table} gives the exponential disk parameters derived
by \citet{bahcall97} and our galaxy properties calculated from
them. We convert Bahcall's disk scale lengths, R$_{s}$, to effective
radii, R$_{e}$, using R$_{e}$ = 1.6785 R$_{s}$ \citep{dV78}. The new
galaxy positions (for objects 1, 5, 6, and 7) are shown on the
Fundamental Plane in Figure \ref{fpexp_fig} and on galaxy parameter
plots in Figure \ref{expparam_fig}.  The lines connected to the
new points illustrate the range between the new exponential disk model
locations and the previous r$^{1/4}$ model locations for these
galaxies. It can be seen in Figure \ref{fpexp_fig} that all four host
galaxies move up on the Fundamental Plane. Interestingly,
PKS~1302-102, the radio-loud QSO, now resides among the locus of the
radio-loud objects, rather than appearing to exist in a transition
region between the radio-loud and radio-quiet groups. The radio-quiet
QSOs have also moved up into the region occupied by
massive early-type galaxies. Their masses derived from the new
radii, also given in Table \ref{expfit_table}, are indeed all well
above 10$^{11}$~M$_{\sun}$. Regardless of assumed morphology, these
are massive galaxies. It can be seen in Figure \ref{expparam_fig} that
these four radio-quiet galaxies now lie among the largest, lowest
surface brightness, and highest M/L QSO hosts from
\citet{dasyra07}. However, it is expected that disk galaxies in that
sample would move similarly if exponential disk models were used to
derive their parameters.

\subsection{Is Location on the Fundamental Plane Related to Host Galaxy Morphology?}

Of the six radio-loud objects in our sample, all are morphologically
classified (Table \ref{object_table}) as elliptical galaxies except
for PKS~2349-014 and PKS~0736+017, which are complex and interacting
with clear tidal tails, nevertheless well fit by a r$^{1/4}$ light
distribution. \citet{bahcall97} and \citet{schweizer96} suggest that
PKS~2349-014 may be a proto-elliptical galaxy resulting from a merger
of two spirals.  PKS~1302-102, PKS~2135-147, PKS~2349-014, and
4C~31.63 all have companions within 5\arcsec of the QSO. Of the four
radio-quiet objects, PG~0052+251 is classified as an Sb galaxy and
shows HII regions; PG~1444+407 is classified as an E1 by
\citet{bahcall97}, as a spiral by \citet{hamilton02}, and may contain
a bar \citep{hutchings92,bahcall97}; PG~1309+355 is an Sab with
tightly wound spiral arms; and PHL~909 is clearly an elliptical
galaxy.  PHL~909 is our only radio quiet host with a nearby companion.

Our QSO hosts at the upper envelope of the FP are either elliptical or
interacting galaxies. Three radio-quiet spiral galaxies and one
elliptical lie on the FP. If we also consider the \citet{dasyra07} QSO
hosts, ellipticals, interacting ellipticals, early-types, and spirals
all lie on the FP close to our four radio-quiet hosts. An early-type
and spiral also lie well below the FP among the merger remnants. This
lack of morphological consistency suggests that host galaxy morphology
alone is not the driver of position on the Fundamental Plane or in the
galaxy property plots of Figure~\ref{parameters_fig} for this small
sample.

\subsection{Evolution or Different Populations?
  \label{evolution_section} }

\begin{figure}
 \epsscale{0.7}
 \plotone{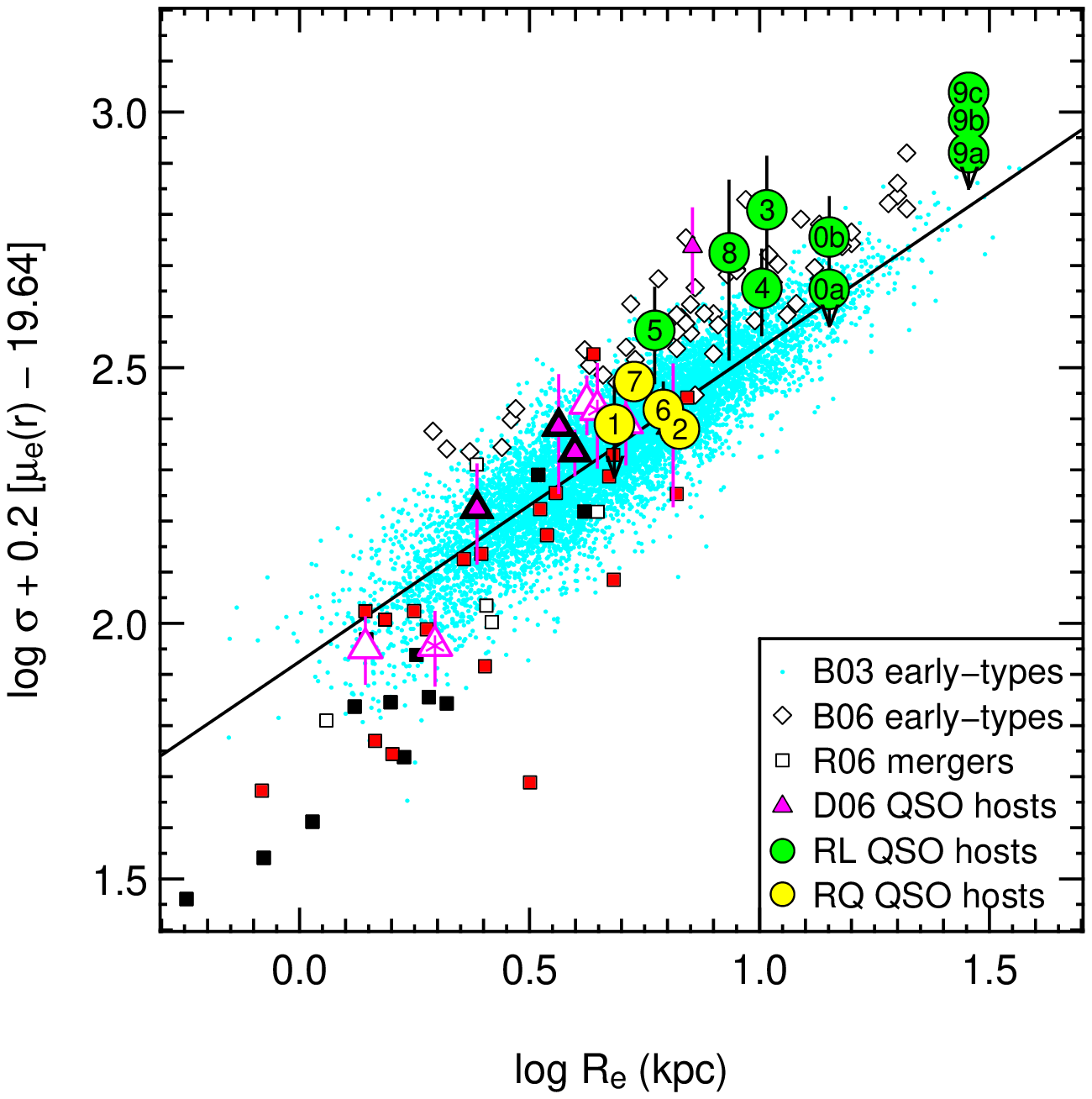}
 \caption{The Fundamental Plane using the projection from
   \citet{bernardi06}. QSO host galaxies from this work are shown as
   large circles (radio-loud are green, radio-quiet are yellow,
   numbers correspond to object numbers in the tables, with 10a and
   10b marked as 0a and 0b), radio-quiet PG QSO hosts from
   \citet{dasyra07} are triangles coded by morphology (large filled
   triangles are ellipticals, crosses denote ellipticals with signs of
   interaction [underneath other symbols in this plot; can be seen in
   Figure \ref{parameters_fig}], large open triangles are early-types,
   large open triangles with asterisks are spirals, small filled
   triangles are unknown), merger remnant galaxies from
   \citet{rothberg06} are squares (normal mergers are filled red
   squares, LIRG/ULIRGs are filled black squares, shell ellipticals
   are open squares), early-type galaxies from \citet{bernardi03a} are
   small dots, and early-types with $\sigma_{*}>$~350~km~s$^{-1}$ from
   \citet{bernardi06} are diamonds.
 \label{fp_fig} }
\end{figure}

\begin{figure}
 \includegraphics[scale=0.6]{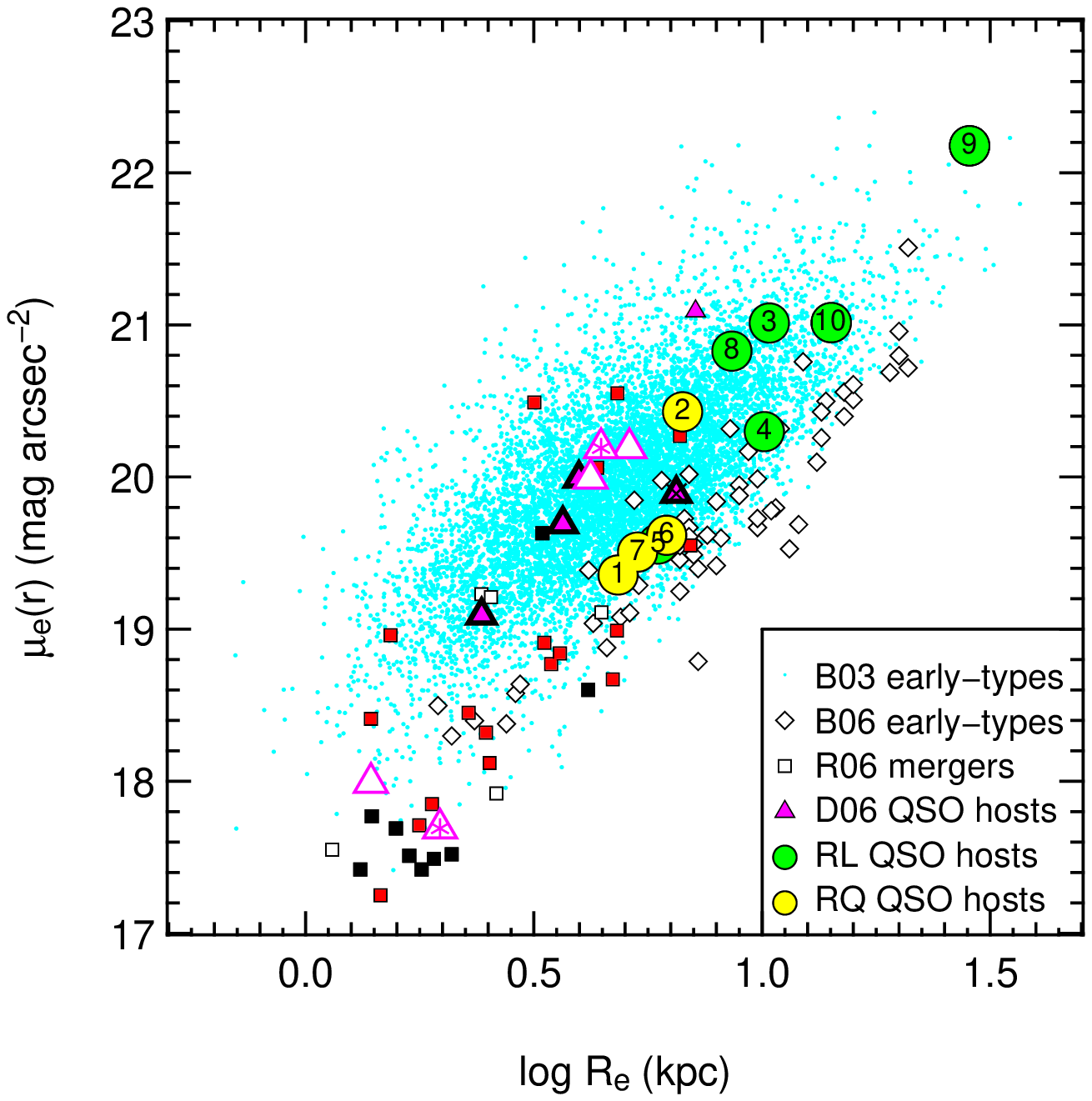}
 \includegraphics[scale=0.6]{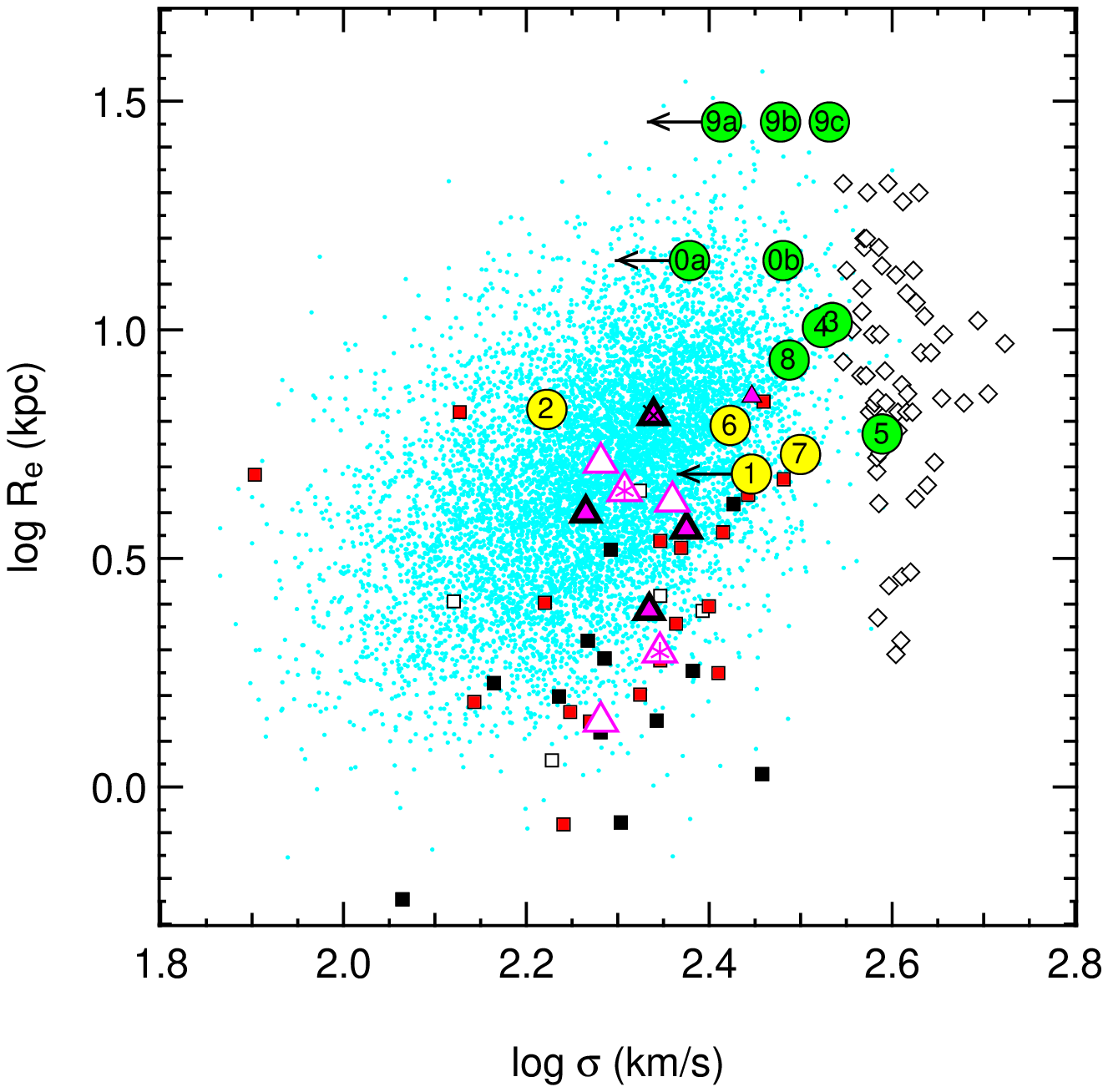}
 \includegraphics[scale=0.6]{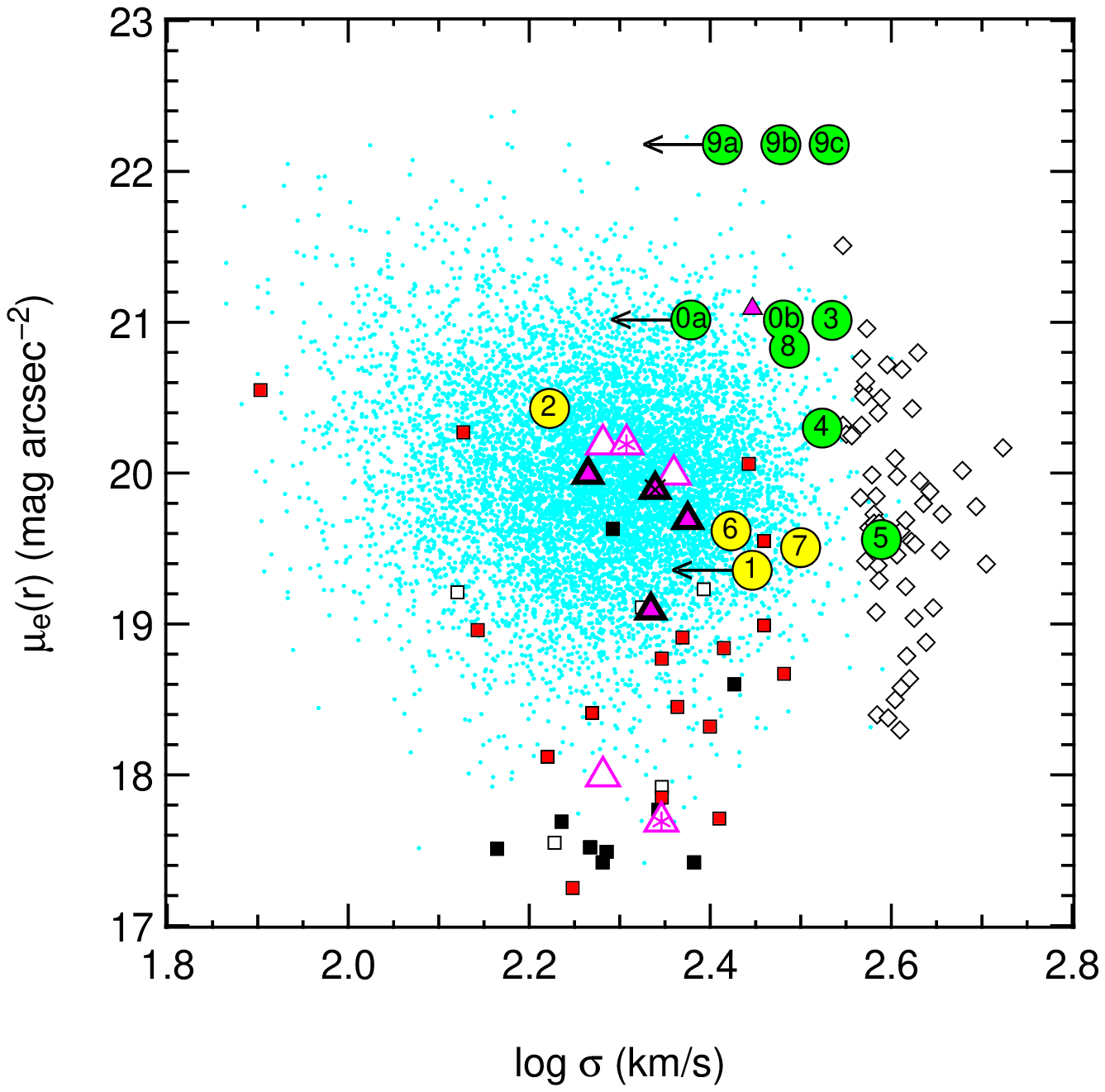}
 \includegraphics[scale=0.6]{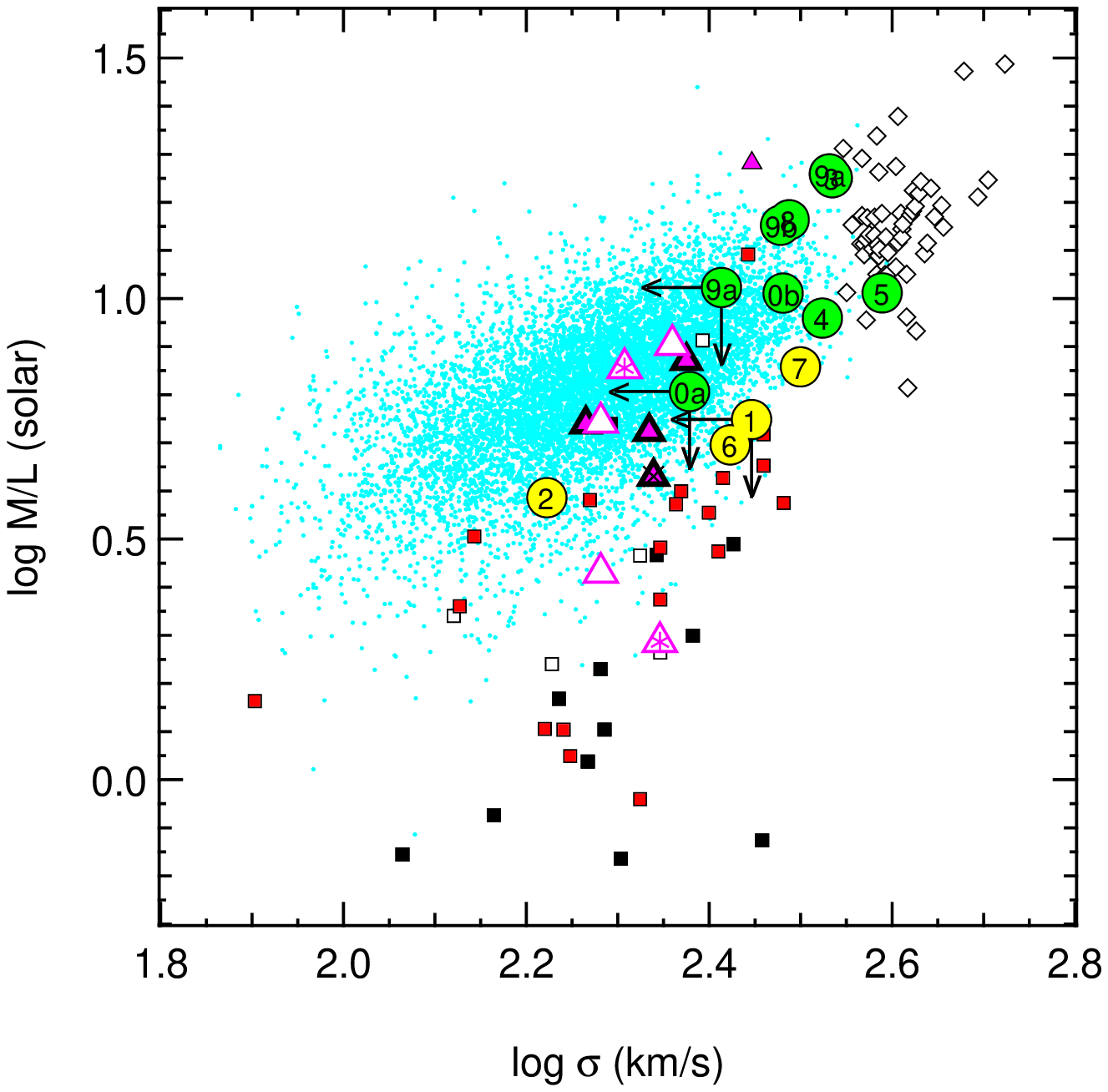}
 \caption{ Fundamental galaxy parameters. Symbols are the same as in
   Figure \ref{fp_fig}. Note that (1) is upper limits on $\sigma_{*}$
   and M.  \label{parameters_fig} }
\end{figure}

\begin{figure}
 \epsscale{0.7}
 \plotone{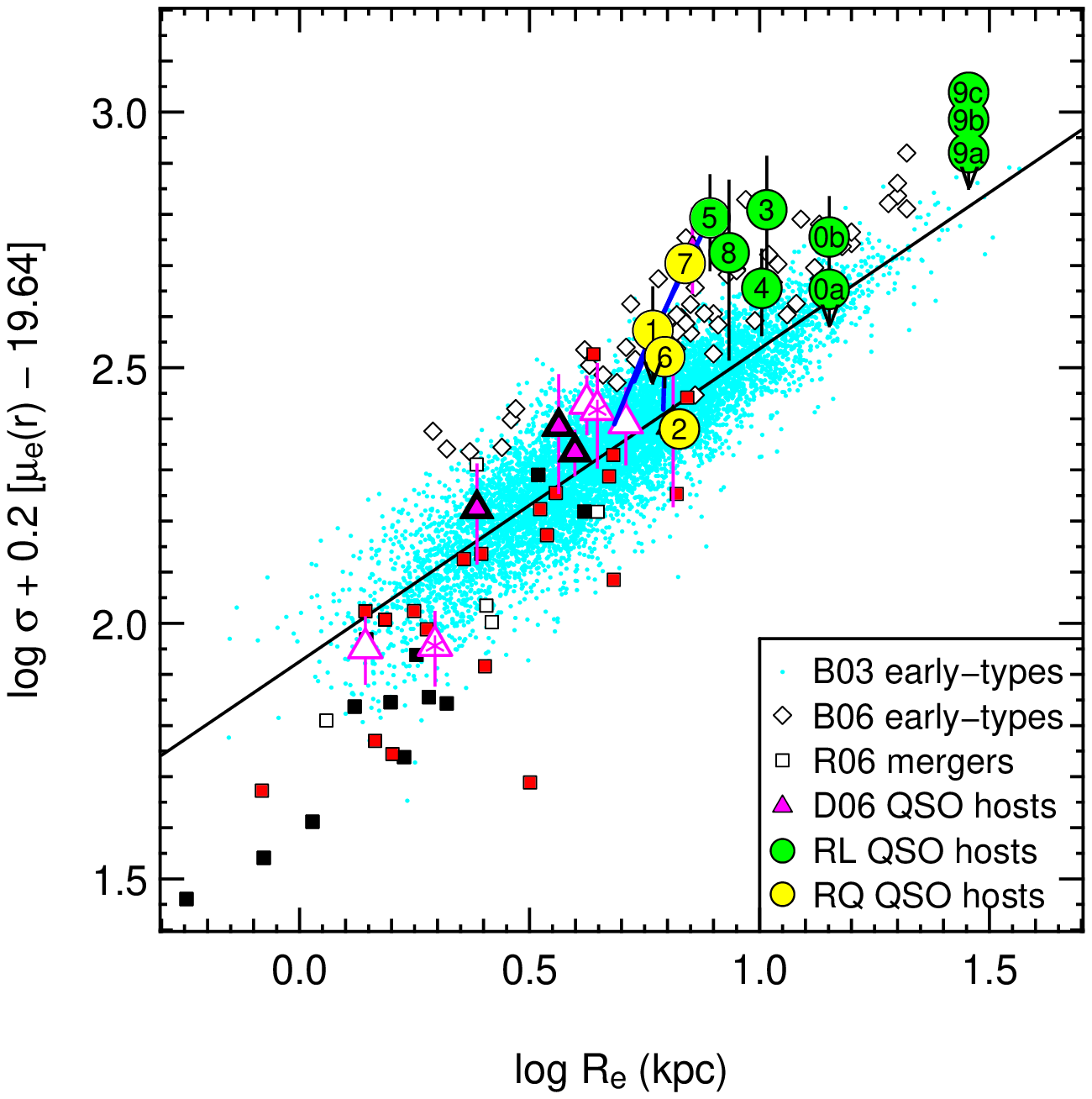}
 \caption{ The Fundamental Plane using galaxy parameters derived from
   exponential disk profile fits for PG~0052+251 (1), PKS~1302-102
   (5), PG~1309+355 (6), and    PG~1444+407 (7). Lines connected to these
   points show the spread between the exponential disk positions and
   the r$^{1/4}$ positions. Otherwise, symbols are the same as in
   Figure \ref{fp_fig}.
   \label{fpexp_fig} }
\end{figure}

\begin{figure}
 \includegraphics[scale=0.6]{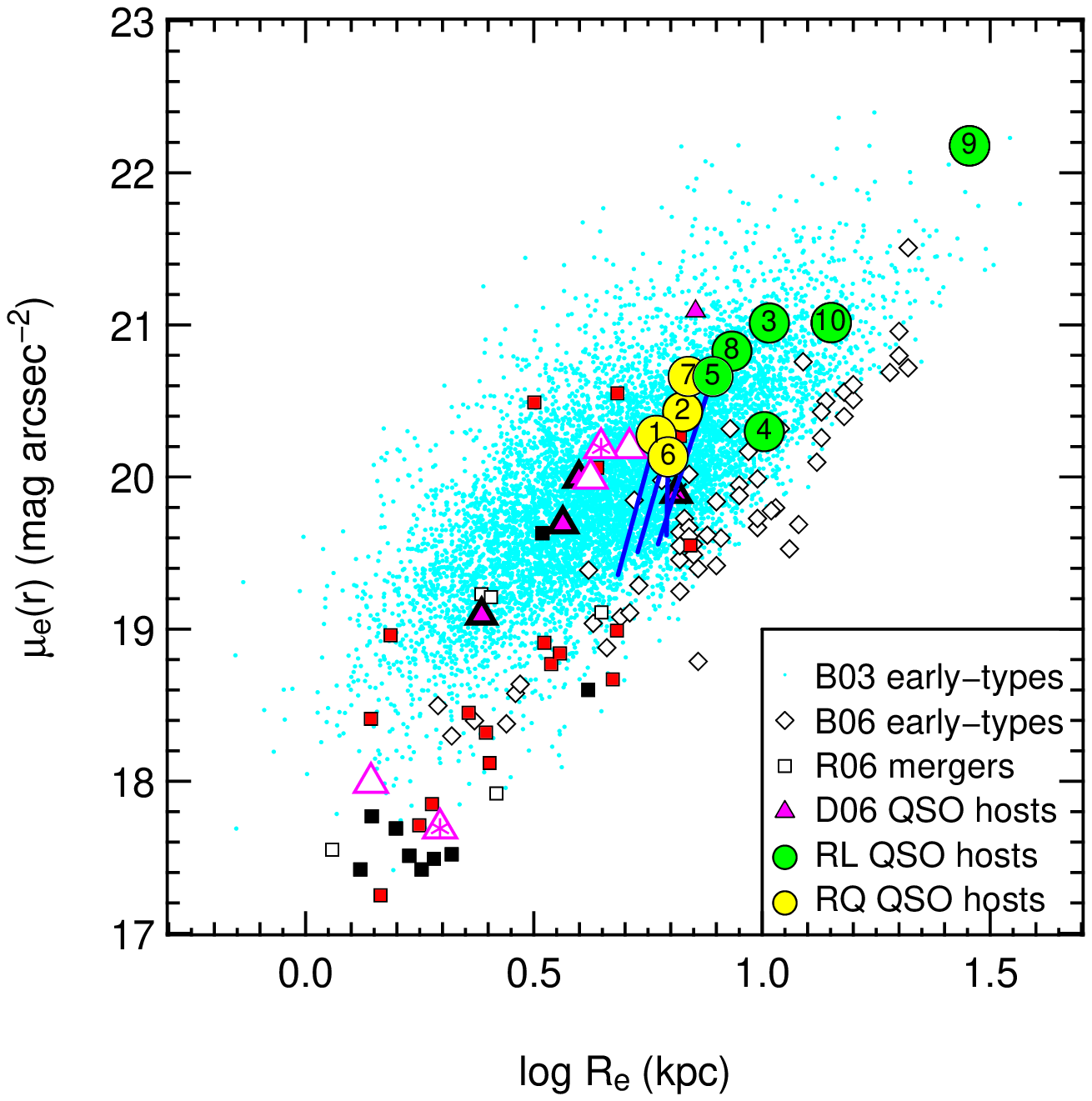}
 \includegraphics[scale=0.6]{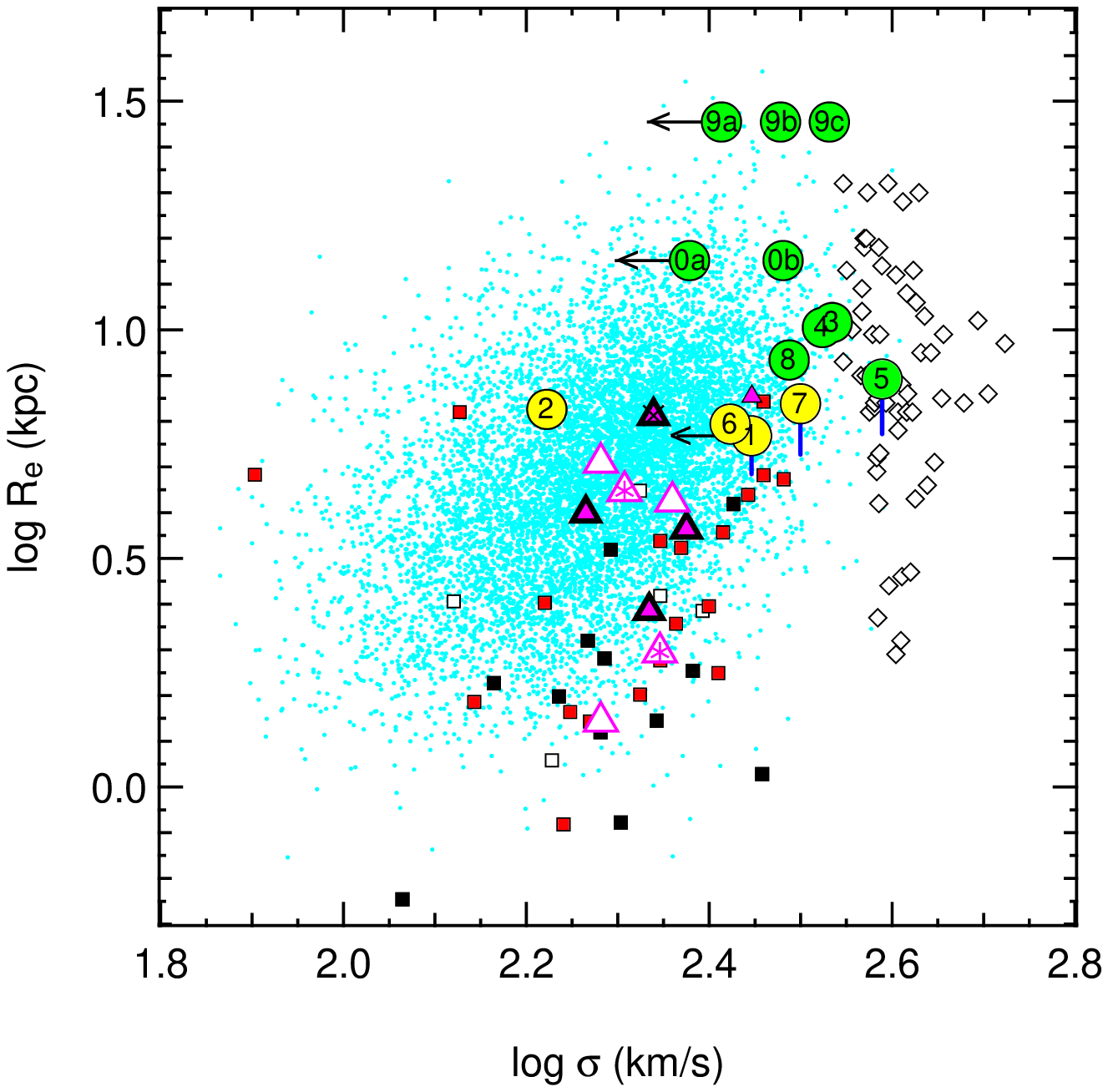}
 \includegraphics[scale=0.6]{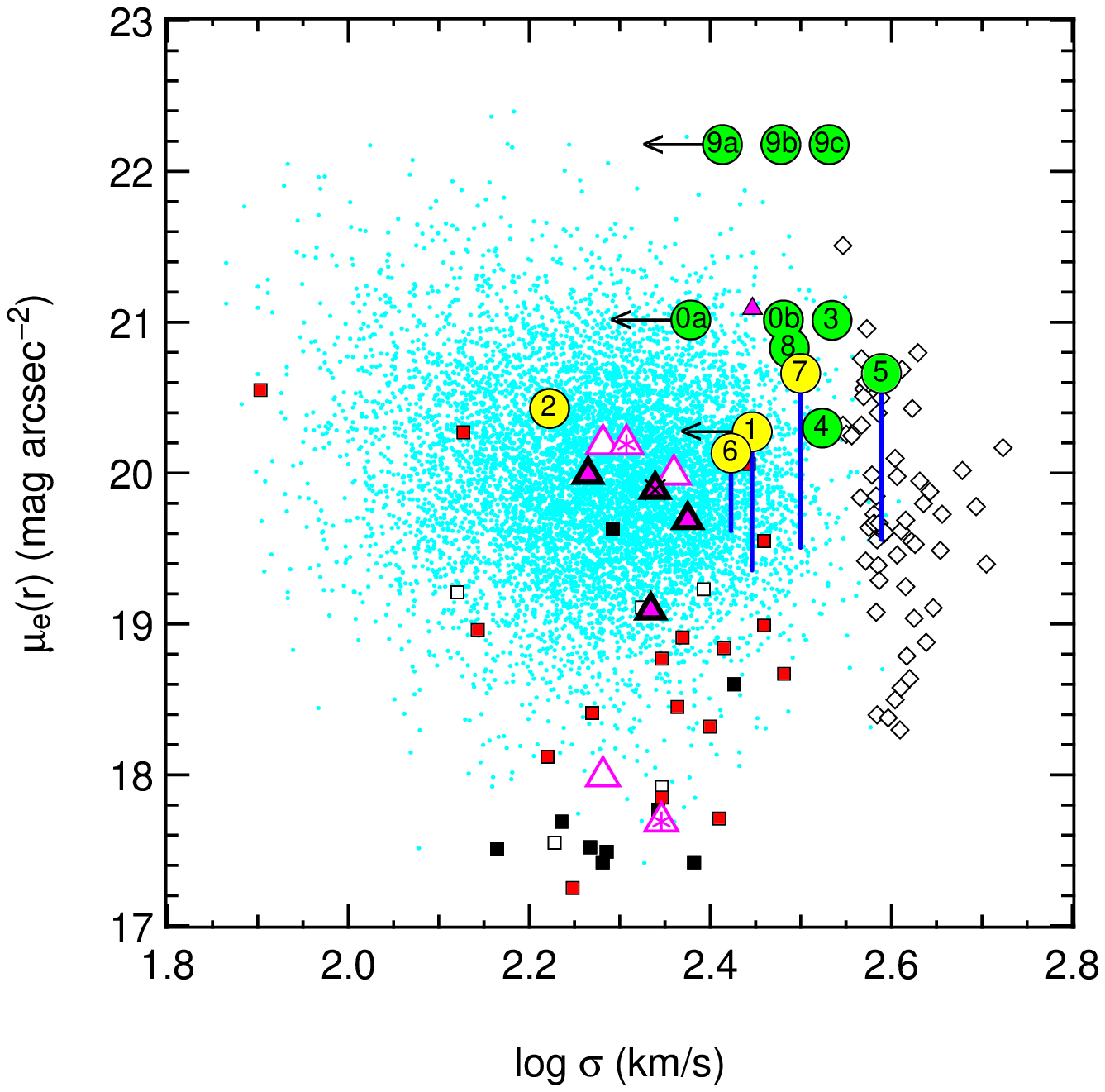}
 \includegraphics[scale=0.6]{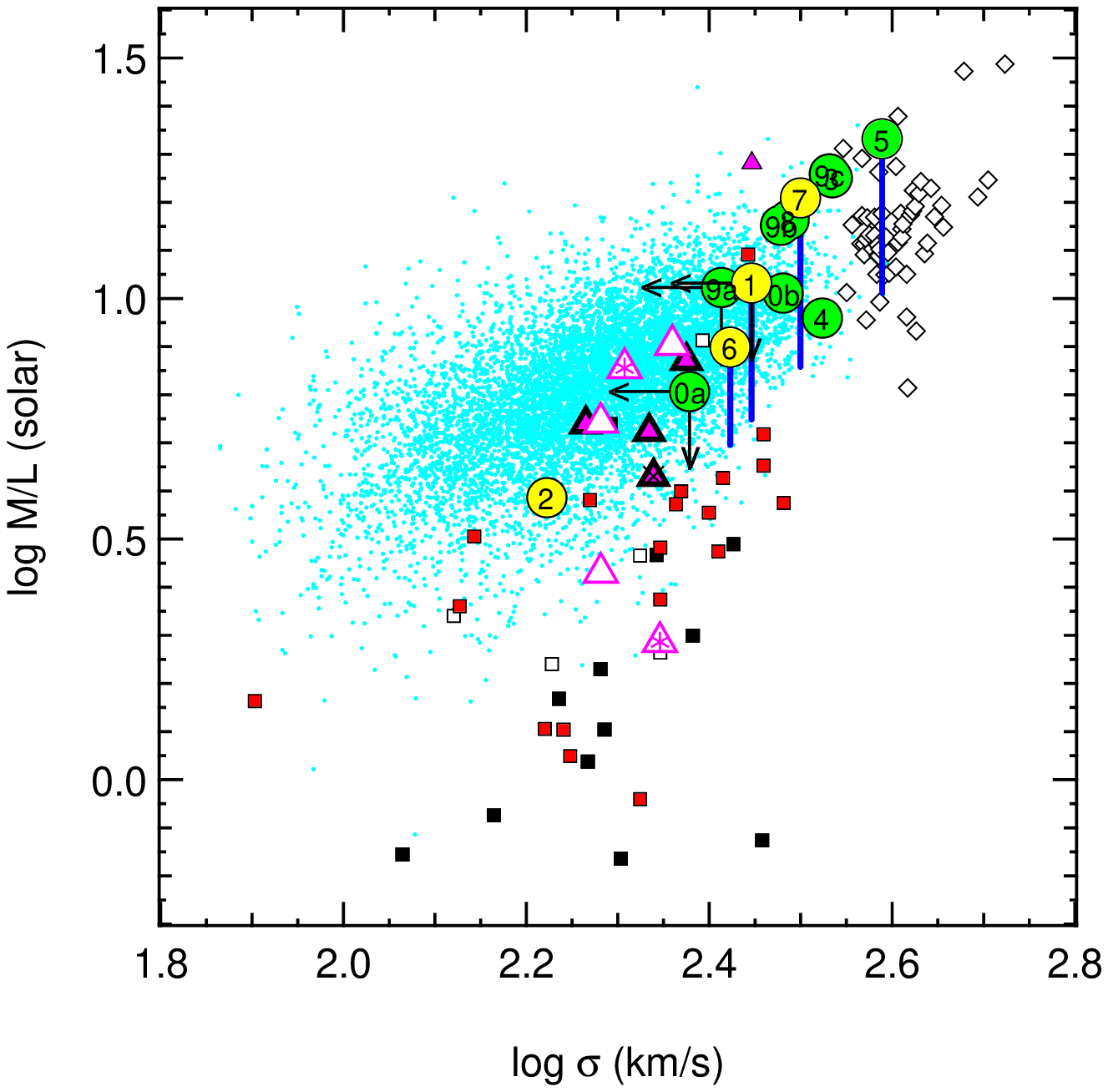}
 \caption{ Fundamental galaxy parameters derived from exponential disk
   profile fits for PG~0052+251 (1), PKS~1302-102 (5), PG~1309+355
   (6), and PG~1444+407 (7). Symbols are the same as in Figures
   \ref{fp_fig} and \ref{fpexp_fig}.  \label{expparam_fig} }
\end{figure}

If we look at the different groups of objects in the plots of Figures
\ref{fp_fig} and \ref{parameters_fig} we see trends suggesting
different classes of galaxies. First we address the general
trends. All objects -- quiescent early-type galaxies, massive
early-type galaxies, quasar host galaxies, and galaxy merger remnants
-- lie nearly on the Fundamental Plane (or are expected to lie there
once faded, in the case of the merger remnants). Of course, we expect
the early-type galaxies, including QSO hosts of that morphology, to be
there. The radio-loud hosts lie among the massive early-type galaxies
at the upper edge of the plane. The merger remnants were
morphologically chosen to be near the end of the merger sequence with
the stipulation that the cores have already coalesced into a single
nucleus. Since we believe that early-type galaxies form from mergers,
it is not surprising that these also occupy the Fundamental Plane. The
LIRG/ULIRGs have a slight offset below the plane due to their
currently high luminosity, but are expected to fade up to the FP after
the starbursts end. The radio-loud quasar hosts are morphologically
either early-type or interacting and on their way to forming
early-types, therefore they would be expected on the FP. Perhaps the
hardest to explain are the radio-quiet QSO hosts that show definite
signs of spiral structure (three of our radio-quiet objects and two of
the Dasyra sample).

The upper left plot of Figure \ref{parameters_fig} shows surface
brightness and effective radius. The merger remnants tend to be small
and bright, suggesting that nuclear starbursts are in progress. The
massive early-type galaxies from SDSS are bright and mostly large,
though a few stretch down to small radii. The Dasyra et al. QSO hosts
tend to have lower surface brightness for a given radius than either
the massive galaxies or the merger remnants, residing near the center
of the distribution of normal early-type galaxies from SDSS. Our three
radio-quiet hosts that are spirals lie in the transition region
between normal and massive early-type quiescent galaxies, while the
elliptical host resides up among the normal quiescent early-types.

In the upper right plot of Figure \ref{parameters_fig}, the stellar
velocity dispersions of the merger remnants and the lower luminosity
radio-quiet QSO hosts overlap and fall well within the distribution of
normal early-type galaxies, as concluded by \citet{dasyra07}. It has
been suggested that LIRG/ULIRGs form intermediate-sized elliptical
galaxies from the merger of two gas-rich spiral galaxies and will not
evolve into the brightest QSOs
\citep{genzel01,tacconi02,rothberg06,dasyra06}. This idea is supported
by the similar velocity dispersions of these two classes of objects,
and by their mass-to-light ratios (lower right plot of Figure
\ref{parameters_fig}; note that most of our radio-quiet QSO hosts are
on the high end of the other PG QSO hosts in $\sigma_{*}$ and
M$_{*}$). The merger remnants have lower M/L due to their currently
high luminosities, but after fading should reside among the lower
luminosity radio-quiet QSO hosts in Figures \ref{fp_fig} and
\ref{parameters_fig}. There is a clear distinction between these
objects and our massive radio-loud QSO hosts in $\sigma_{*}$ and
R$_{e}$ (and thus stellar mass). The separation occurs at
$\sigma_{*}\sim$~300~km~s$^{-1}$, R$_{e}\sim$~6~kpc, and corresponding 
M$_{*}\sim$~7 x 10$^{11}$~M$_{\sun}$. Such a distinction between
radio-loud and radio-quiet QSOs has been observed as a function of
estimated black hole mass \citep{dunlop03}, however, here we have
deduced it here using directly measured host galaxy stellar velocity
dispersions.

Because our method of calculating galaxy mass assumes that the
velocity anisotropies are small and that the system is dynamically hot
\citep{bender92}, the masses calculated in this way for spiral
galaxies or those with strong interactions will be systematically
incorrect. We now consider only the galaxies that are known to have
elliptical morphologies. The radio-quiet elliptical QSO host sample
consists of PHL~909, PG~0007+106, PG~1617+175, and PG~2214+139 (the
last 3 are from \citet{dasyra07}), while the radio-loud sample
includes 3C~273, PKS~2135-147, and 4C~31.63. Our calculated masses for
the radio-quiet group of ellipticals are 1.3 - 2.4 x
10$^{11}$~M$_{\sun}$ with an average of 1.9 x
10$^{11}$~M$_{\sun}$. The radio-loud group has calculated masses of
9.4 - 29.8 x 10$^{11}$~M$_{\sun}$ with an average of 17.4 x
10$^{11}$~M$_{\sun}$. This gives a radio-loud/radio-quiet division
somewhere between 2.4 - 9.4 x 10$^{11}$~M$_{\sun}$, or M$_{*} \sim$
5.9 $\pm$ 3.5 x 10$^{11}$~M$_{\sun}$, for the elliptical QSO host
galaxies.

Mean galaxy parameters are given in Table \ref{avg_table} for the
different groups of galaxies. Overall, our radio-quiet hosts show
properties similar to normal early-type galaxies, though slightly
larger and more massive. A comparison of properties of our radio-quiet
hosts and normal early-type galaxies show R$_{e}$ = 5.7, 4.9 kpc,
respectively, M/L = 5.3, 7.0, $\sigma_{*}$ = 241, 190 km s$^{-1}$, and
M$_{*}$ = 4.4, 2.1 x 10$^{11}$~M$_{\sun}$. On the other hand, our
radio-loud hosts are more analogous to the massive early-type
galaxies. They are larger (11.4, 8.0 kpc, respectively) with lower
surface brightness (20.8, 19.8 mag~arcsec$^{-1}$), lower $\sigma_{*}$
(321, 403 km~s$^{-1}$) and M/L (12.4, 14.7), but higher M$_{*}$ (14.8,
12.2 x 10$^{11}$~M$_{\sun}$). It appears that we have two classes of
objects, regardless of the surface brightness model used for the disk
galaxies (refer also to Figures \ref{fpexp_fig} and
\ref{expparam_fig}): very massive galaxies with radio loud quasars and
intermediate mass galaxies with radio quiet quasars. This supports the
conclusions of \citet{malkan84} who found in an imaging study that
radio-quiet quasars have the color, size, surface brightness, and
scale length of normal early-to-intermediate type spiral galaxies,
while radio-loud quasars appear to reside in moderate to bright
ellipticals.

\subsection{Radio Properties}

\begin{figure}
 \epsscale{0.7}
 \plotone{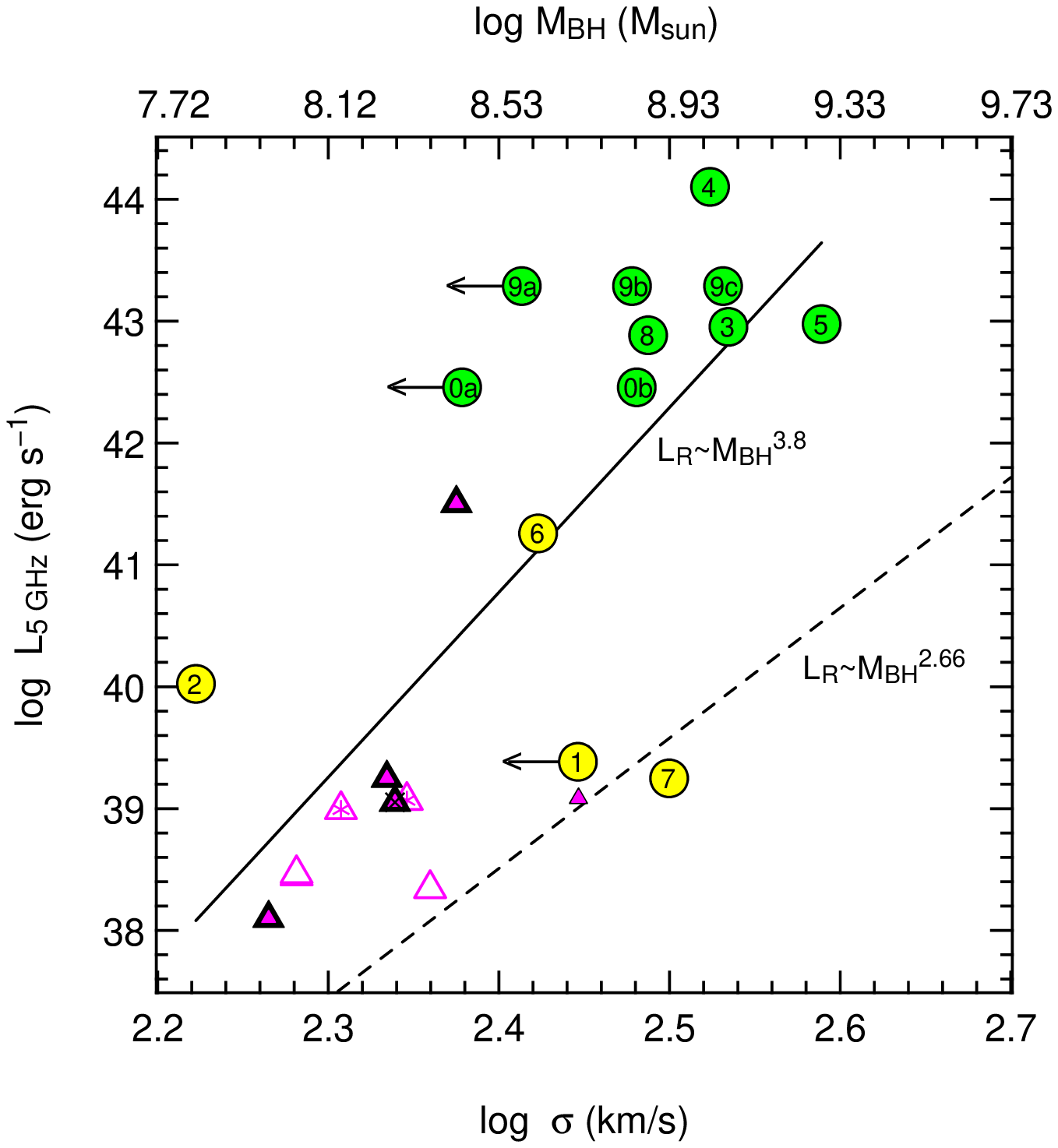}
 \caption{ Radio luminosity as a function of stellar velocity
   dispersion and black hole mass. Symbols are the same as in Figure
   \ref{fp_fig}. M$_{BH}$ was calculated from the \citet{tremaine02}
   M$_{BH}$-$\sigma$ relation using our measured $\sigma_{*}$. The
   solid line is the best linear regression fit to our objects and the
   lower luminosity \citet{dasyra07} QSOs (excluding PG 0052+251,
   \#1). The dashed line is the \citet{franceschini98}
   L$_{5~GHz}\varpropto$~M$_{BH}^{2.66}$ linear regression, derived
   for nonactive galaxies over log M$_{BH}\sim$6.2-9.5 M$_{\sun}$. 
   \label{radio_fig} }
\end{figure}

We see in Figures \ref{fp_fig} and \ref{parameters_fig} that all
radio-loud QSO hosts reside among the massive early-type galaxies. A
more detailed look at the radio properties of these objects reveals a
correlation between radio luminosity and $\sigma_{*}$, shown in Figure
\ref{radio_fig} with 5~GHz radio fluxes taken from
\citet{kellermann94} and the NASA/IPAC Extragalactic Database
(modified to our adopted cosmology of $\Omega_{M}$=0.3,
$\Omega_{\Lambda}$= 0.7, H$_{0}$=70~km~s$^{-1}$~Mpc$^{-1}$). A similar
correlation was found by \citet{nelson96} for radio-quiet Seyfert
galaxies, which extended to higher luminosity radio galaxies when
considering the core radio emission. They interpreted this correlation
as a dependence of radio emission on galaxy bulge mass.

\begin{figure}
 \epsscale{0.7}
 \plotone{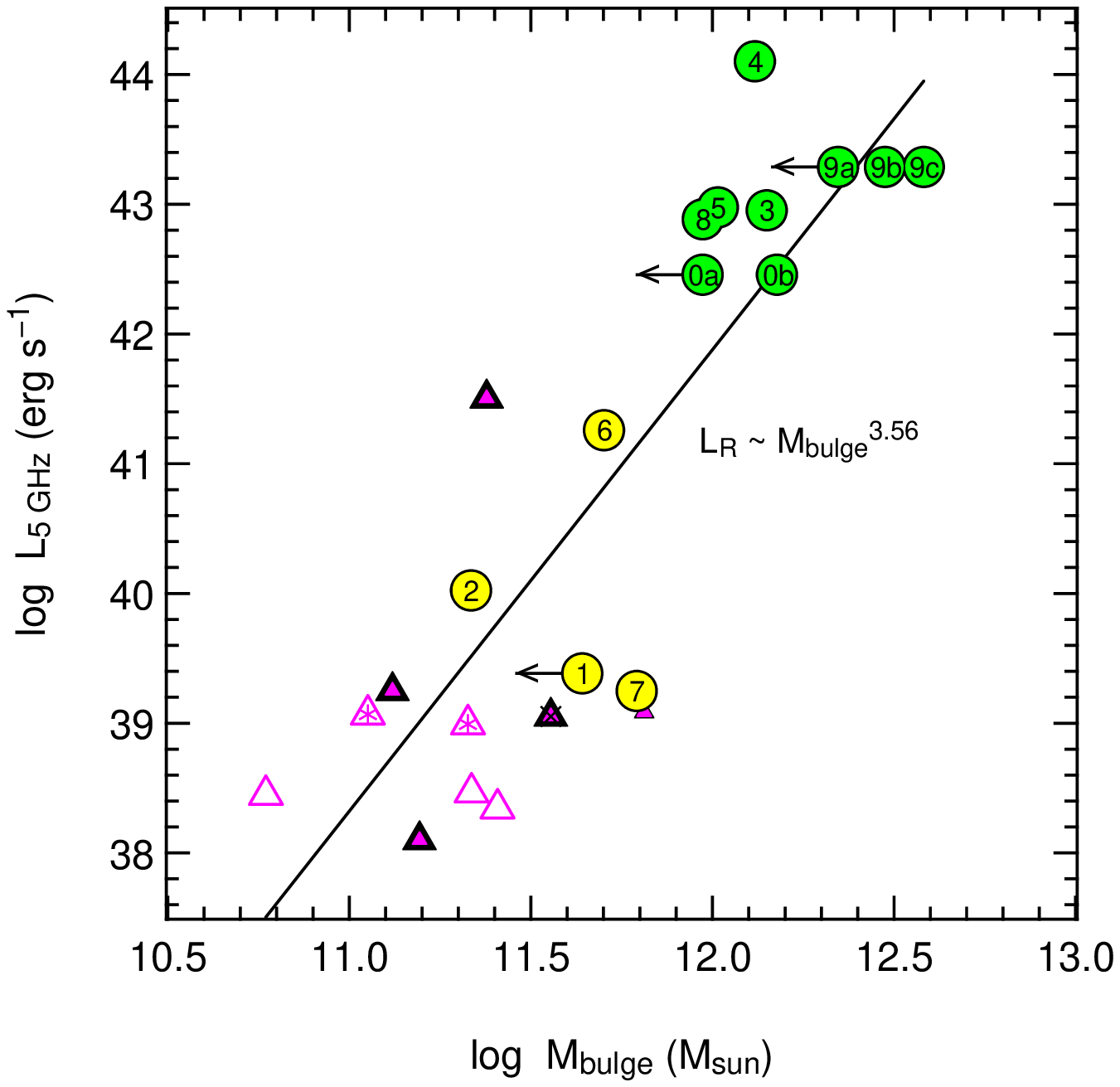}
 \caption{ Radio luminosity as a function of galaxy bulge mass
 (calculated from directly measured host galaxy $\sigma_{*}$ and
 R$_{eff}$). The solid line is a fit to the data, excluding PG 0052+251, \#1.
 \label{radio_mass} }
\end{figure}

Subsequent studies
\citep{franceschini98,MKDBOH99,nelson00,laor00,lacy01,ho02,dunlop03,snellen03,mclure04,woo05}
have looked at the dependence of radio luminosity on black hole mass
(M$_{BH}$), which we now know is related to $\sigma_{*}$ and galaxy
bulge mass \citep{gebhardt00a,ferrarese00}. \citet{franceschini98}
found a tight dependence of L$_{5~GHz} \varpropto$ M$_{BH}^{2.66}$ for
a sample of 13 nearby early-type galaxies with M$_{BH}$ estimated from
stellar or gas kinematics. There has been some disagreement on this
correlation as different types of objects were analyzed
\citep{MKDBOH99,lacy01,ho02,snellen03,woo05,mclure04} and scatter
about the relation grew to several orders of magnitude when AGN of 
different activity levels were added. For these larger samples a wide
range of radio luminosity is possible for a given M$_{BH}$ and it is
likely more dependent on BH accretion rate than on the mass
\citep{ho02}. \citet{ho02} suggested that the L$_{radio}$-M$_{BH}$
relation arises indirectly through more fundamental correlations
between radio luminosity and bulge mass and that between bulge mass
and black hole mass, similar to the original interpretation of
\citet{nelson96}. For strongly active quasars, we find a tighter
correlation of radio luminosity with host galaxy bulge mass
(calculated from directly measured host galaxy parameters) than with
black hole mass (inferred from the \citet{tremaine02}
M$_{BH}$-$\sigma$ relation). This relationship is shown in Figure
\ref{radio_mass}. Our best linear regression fit to the data gives the
relation $L_{radio} \sim M_{bulge}^{3.56}$ with rms scatter of 1.09
dex.

One consensus that has arisen from all of these studies is that
radio-loud AGN almost exclusively have
M$_{BH}\geq$~10$^{8}$~M$_{\sun}$. Indeed all of our radio-loud quasars
in Figure \ref{radio_fig} have M$_{BH}>$~10$^{8}$~M$_{\sun}$. The
best-fit line to the QSO hosts of this work and of Dasyra et
al. suggests a steeper slope of L$_{5~GHz}\varpropto$~M$_{BH}^{3.8}$
with rms scatter of 1.36 dex, however, given our limited sample size,
the small M$_{BH}$ range covered by the data, the large scatter
associated with this relation, and possible differences in black hole
mass estimation techniques (the black hole masses of the QSOs in this
study and their place on M$_{BH}$-$\sigma$ relation will be discussed
in future work), the QSOs in this study are consistent with the
previously determined L$_{5~GHz}$-M$_{BH}$ correlation. Radio
emission, like the host galaxy parameters in \S
\ref{evolution_section}, has a dependence on the mass of either the
host galaxy and/or the central black hole. Our data on active quasars
show a stronger dependence on the host galaxy mass.

\section{Summary and Conclusions \label{summary}}

We have for the first time directly measured host galaxy stellar
velocity dispersions for very luminous (M$_{V}<-$23) quasars,
including both radio-loud and radio-quiet objects, and analyzed their
structural properties. We compare the properties of these host
galaxies to those of normal early-type galaxies \citep{bernardi03a},
giant early-type galaxies \citep{bernardi06}, a sample of radio-quiet
PG QSO hosts \citep{dasyra07}, and galaxy merger remnants
\citep{rothberg06}. The following summarizes our main conclusions.

\begin{enumerate}
\item{ The six radio-loud QSO host galaxies lie at the upper extreme
    the Fundamental Plane of early-type galaxies. They occupy this
    location due to their large velocity dispersions (with an average
    of 321~km~s$^{-1}$), large effective radii (average of 11.4 kpc),
    low surface brightness (average of 20.8~mag~arcsec$^{-2}$), and
    high M/L (average of 12.4). The properties of these radio-loud
    host galaxies are similar to those of giant early-type galaxies in
    the SDSS, although only one has the spectrum of a purely old giant
    elliptical galaxy.  }
\item{ The four radio-quiet QSO host galaxies reside on the
    Fundamental Plane among normal early-type galaxies and at the high
    end of other PG QSO hosts, with a mean velocity dispersion of
    241~km~s$^{-1}$. Their surface brightness and effective radii are
    slightly higher than normal early-type galaxies. The M/L's are
    slightly below normal early-type galaxies.  }
\item{ The radio-loud hosts in our study are either elliptical
    galaxies or interacting, while the radio-quiet hosts show a
    mixture of spiral and elliptical structure. The distinction
    between the two groups is due to galaxy mass, inferred from
    measured structural parameters, rather than morphological galaxy 
    type. The separation occurs at galaxy velocity dispersions of
    $\sigma_{*}\sim$~300~km~s$^{-1}$, effective radii of R$_{e} \sim$
    6~kpc, and corresponding stellar masses of M$_{*} \sim$ 5.9 $\pm$
    3.5 x 10$^{11}$~M$_{\sun}$. }
\item{ We confirm a correlation between radio luminosity and stellar
    velocity dispersion, and thus implied black hole mass, of the host
    galaxies that suggests a higher slope (L$_{5
    GHz}\varpropto$~M$_{BH}^{3.8}$ with rms scatter of 1.36 dex) than
    found by Franceschini et al.~(L$_{5
    GHz}\varpropto$~M$_{BH}^{2.66}$), though it could be consistent
    with previous work given the large scatter in this relation
    \citep{mclure04}, our small sample size, the limited M$_{BH}$
    range of our data, and differences in M$_{BH}$ estimated from
    different techniques. We find a tighter correlation between radio
    luminosity and host galaxy bulge mass, $L_{radio} \sim
    M_{bulge}^{3.56}$ with rms scatter of 1.09 dex }

\end{enumerate}



\acknowledgments

MJW was supported for this work by the McKinney Postdoctoral
Fellowship at the University of Wisconsin. The authors wish to thank
Luis Ho for reading an early draft of this paper and providing very
helpful comments, Karl Gebhardt for the use of his velocity dispersion
code, and the following people for insightful conversations on various
aspects of this work: Eric Hooper, Joseph Miller, and Greg Shields. We
would also like to thank the referee, Matt Malkin, for very
constructive comments and suggestions that substantially improved the
paper.



{\it Facilities:} \facility{Keck}, \facility{WIYN}, \facility{HST}.



\appendix

\section{ Measurement Limits and Sigma Bias \label{bias} }

\begin{figure}
 \epsscale{0.7}
 \plotone{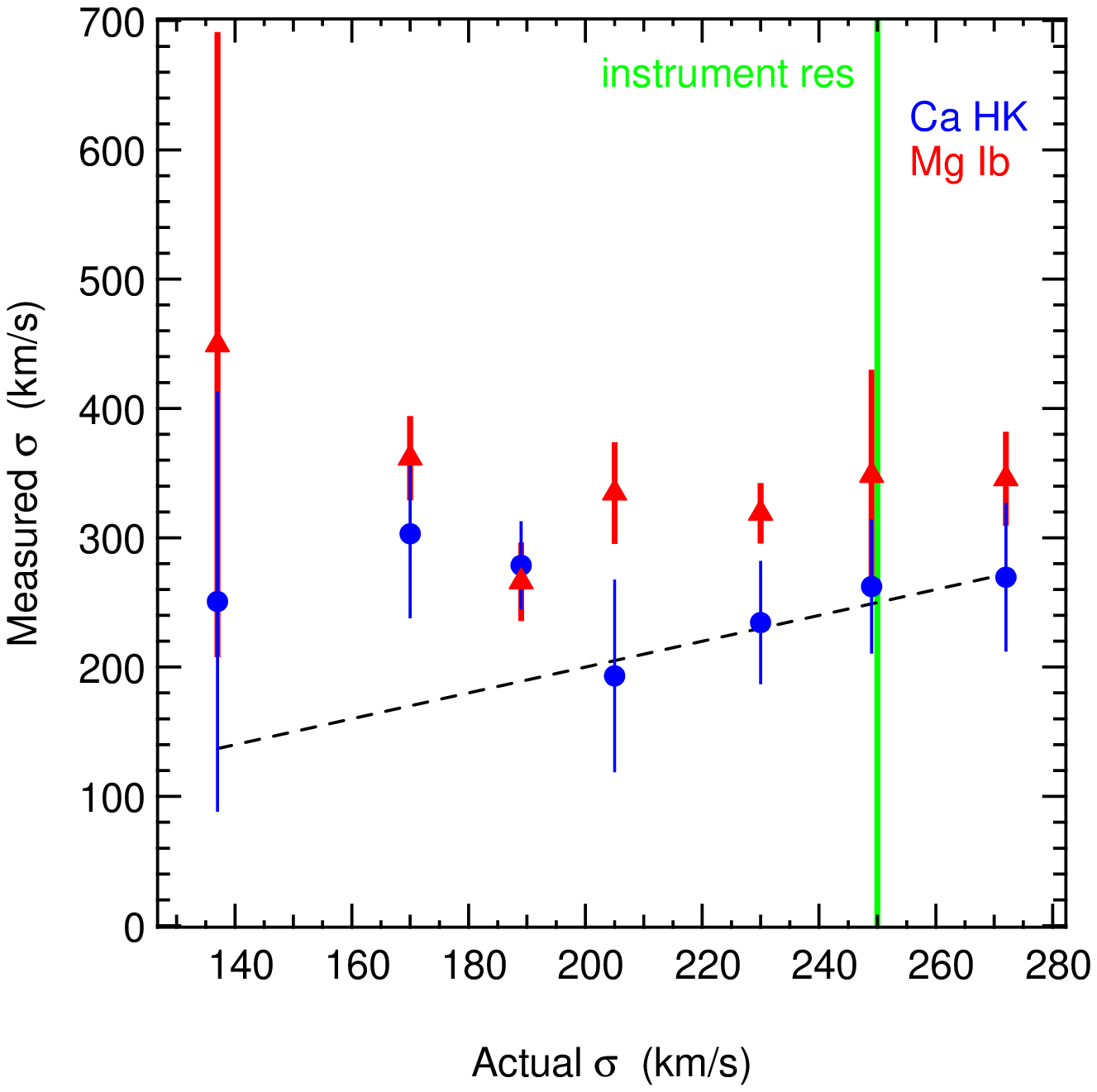}
 \caption{ Tests of velocity dispersion measurement limits. Actual
 $\sigma$'s are from \citet{mcelroy95} and measured $\sigma$'s are our
 results using the Gebhardt code. The dashed line marks a 1:1
 correlation between the two. Circles are measurements on Ca H\&K and
 triangles are on Mg Ib. Error bars were generated from Monte Carlo
 simulations. The instrument resolution of the data was
 250~km~s$^{-1}$. It is clear that $\sigma$ measured from the Ca H\&K
 lines is reliable down to 200~km~s$^{-1}$, 20\% below the instrument
 resolution. These tests also show that the Mg Ib triplet should not
 be used to measure $\sigma$ on these data.
 \label{sigma_tests} }
\end{figure}

To test our velocity dispersion measurement limitations we used seven
standard galaxies from \citet{mcelroy95}, a catalog of 86 galaxies
with at least three reliable central velocity dispersion measurements
in the literature. We used spectra of these galaxies from the STScI
database of UV-optical spectra of nearby galaxies (Storchi-Bergmann,
Calzetti, and Kinney) with instrumental resolution of approximately
250~km~s$^{-1}$ and S/N~$\sim$~4~\AA$^{-1}$. The galaxies we used are
listed in Table \ref{sigma_test_table} and fitting results are shown
in Figure \ref{sigma_tests}. In this plot the ``actual $\sigma$'s''
are from McElroy and ``measured $\sigma$'s'' are our results from the
Gebhardt code. The dashed line represents a 1:1 correlation between
the two. Circles are measurements using the Ca H\&K lines and
triangles use the Mg Ib triplet. It is clear that results using Ca
H\&K are good down to 200~km~s$^{-1}$, or 20\% below the instrument
resolution. These tests also show that the Mg Ib triplet cannot be
used to measure $\sigma_{*}$ on these data. For this reason we used
the Ca H\&K lines in this work.

Ideally, we should measure the same velocity dispersion of a galaxy
regardless of the input initial estimate.  However, in the presence of
noise the measurement can be biased. The Gebhardt code evaluates this
bias by examining the distribution of measured $\sigma_{*}$ values output
by the Monte Carlo simulations. If the bias in the distribution is much
smaller than the uncertainty then the measured $\sigma_{*}$ is valid. 

\begin{figure}
 \epsscale{0.5}
 \plotone{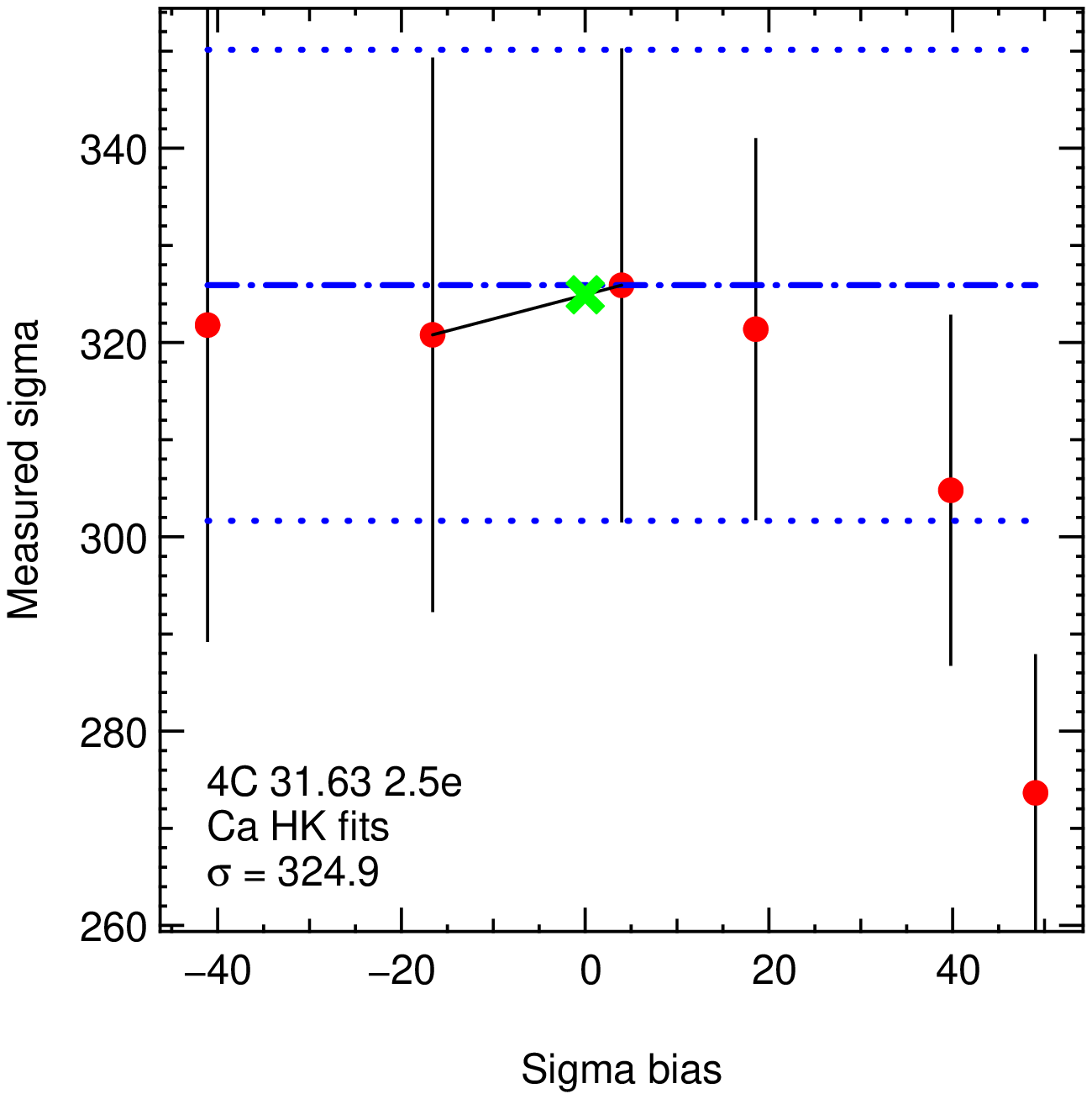}
 \caption{ Interpolation of measured $\sigma_{*}$ to zero bias.
   \label{interp_plot} }
\end{figure}

To investigate the problem of sigma bias, we made measurements on each
galaxy using a range of initial input $\sigma$'s. Plots in Figure
\ref{sigma_bias} show the results of these measurements for each of
our host galaxies. Each point and error bar represents 100 Monte Carlo
simulations of the measurement beginning with that input $\sigma$
value. Any values where the sigma bias was greater than 20-40\% of the
uncertainty are marked with crosses and not used. In all cases we are
left with 1-3 points that have relatively low sigma bias in the
measurement. If we plot measured $\sigma_{*}$ vs.~$\sigma$-bias, as
shown in Figure \ref{interp_plot} for 4C~31.63 2.5E, we can
interpolate between $\sigma$-bias values to obtain a measured
$\sigma_{*}$ that has zero $\sigma$-bias (marked by a green
cross). Velocity dispersions obtained in this manner are shown in
parentheses on the plots in Figure \ref{sigma_bias} and represent our
adopted $\sigma_{*}$ values for all galaxies except PG~0052+251. In
this case all measurements within our data resolution ($\sigma_{input}
>$ 250 km s$^{-1}$) were sigma biased. From the behavior of the
measurements, it appears that the $\sigma_{*}$ value with zero sigma
bias falls well below our measurement resolution. Therefore, we can
only set an upper limit of $\sigma_{*} <$ 250 km s$^{-1}$ for this
galaxy.

The black dashed lines in the plots mark the error bar overlap region
of the unbiased measurements and the blue dash-dot line marks the
center of this region. We adopt the average uncertainty of all good
measurements (marked by the blue dotted lines).

\begin{figure}
 \epsscale{0.5}
 \plotone{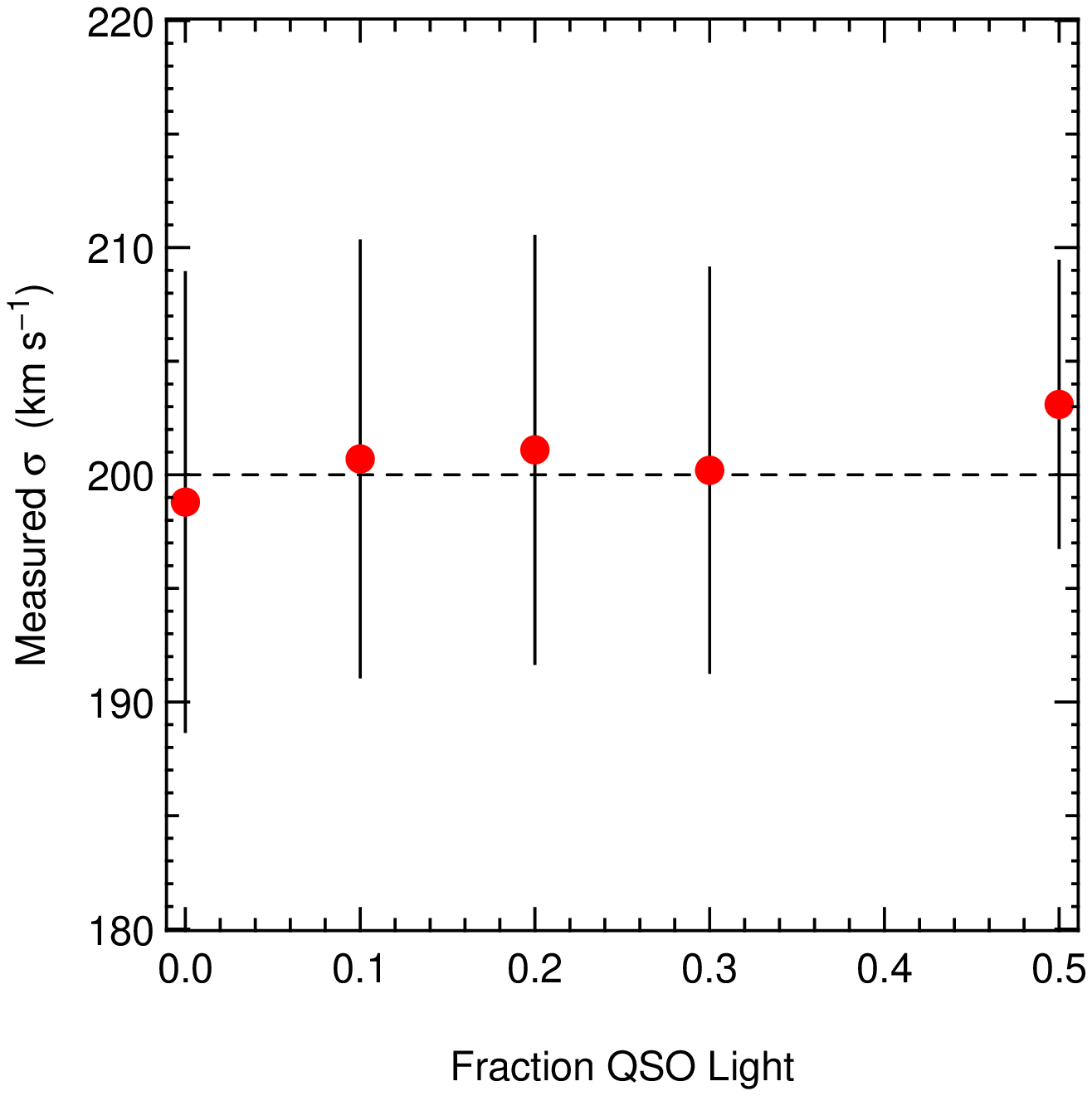}
 \epsscale{1.1}
 \plottwo{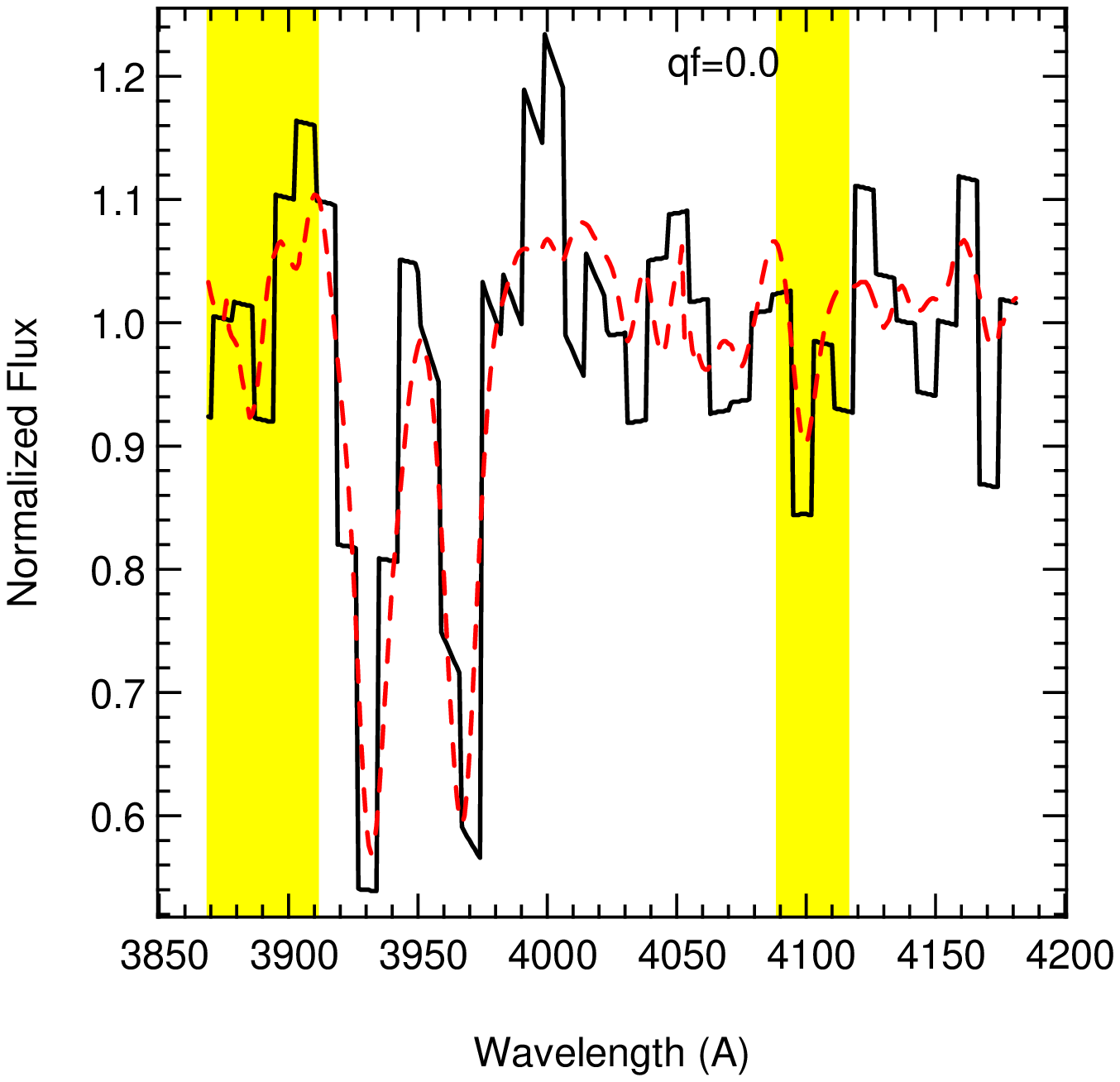}{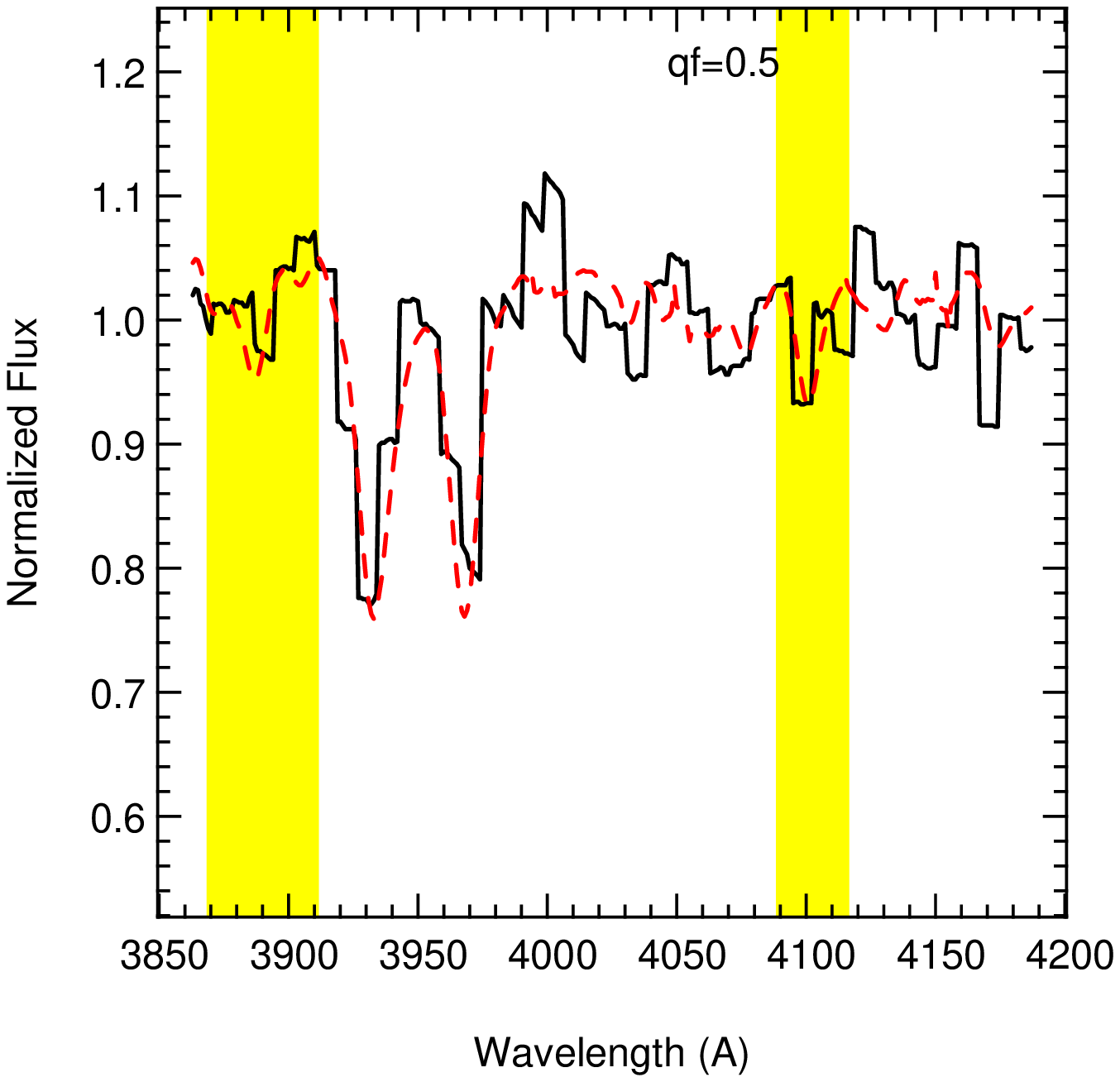}
 \caption{ Test of velocity dispersion measurement on galaxy spectra diluted
   by quasar light. We smoothed a model galaxy spectrum (age 5 Gyr,
   [Fe/H]=0.0) to $\sigma$~=~200~km~s$^{-1}$, added noise to reach
   S/N~=~5~\AA$^{-1}$, and added different fractions of a quasar
   spectrum (from PG 1309+355) to dilute the galaxy spectrum.  The
   top plot shows measured $\sigma$ as a function of quasar light
   fraction. Up to 50\% of the scatter-subtracted spectrum could be
   residual QSO light without adversely affecting the measurement of
   velocity dispersion for a galaxy with
   $\sigma$~=~200~km~s$^{-1}$. The bottom plots show the $\sigma$ fits
   for quasar fractions of 0.0 and 0.5.
   \label{dilute_test} }
\end{figure}

One last concern is the possibility of galaxy absorption line dilution
by residual scattered quasar light inflating the measured velocity
dispersion. Since many of the host galaxies with Keck data have
measured velocity dispersions near the instrumental resolution, we
wished to test whether a lower $\sigma$ actually could be measured in
the presence of line dilution. We tested this effect via simulations
in which we generated a model spectrum from \citet{bc03} for a 5~Gyr
old galaxy with [Fe/H]~=~0.0, smoothed it to $\sigma
=$~200~km~s$^{-1}$, added gaussian noise to achieve
S/N~=~5~\AA$^{-1}$, and added varying fractions of quasar light using
our observed spectrum of PG~1309+355 (which has instrument resolution
of 110~km~s$^{-1}$). We measured velocity dispersions in the same
manner as for the host galaxies for simulated quasar light fractions
of 0.1-0.5. The results are shown in Figure \ref{dilute_test}. The top
plot shows measured $\sigma$ as a function of quasar light
fraction. Our results show that up to 50\% of the spectrum could be
due to residual scattered quasar light without adversely affecting the
measurement of velocity dispersion for a galaxy with $\sigma
=$~200~km~s$^{-1}$. The bottom two plots show the velocity dispersion
fits for quasar light fractions of 0.0 and 0.5. Dilution of the lines
is obvious.





\clearpage

\begin{figure}
 \includegraphics[scale=0.39]{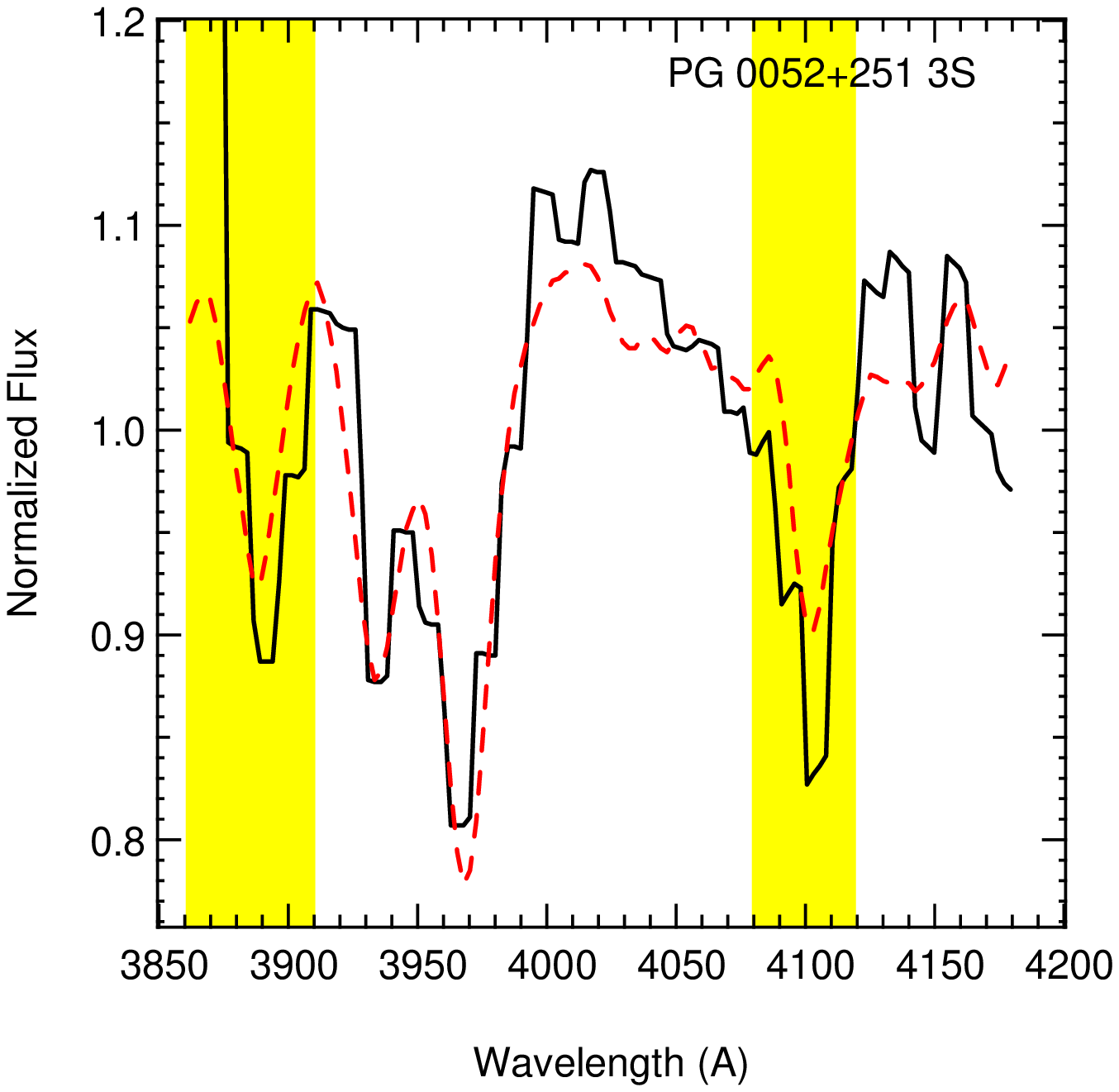}
 \includegraphics[scale=0.39]{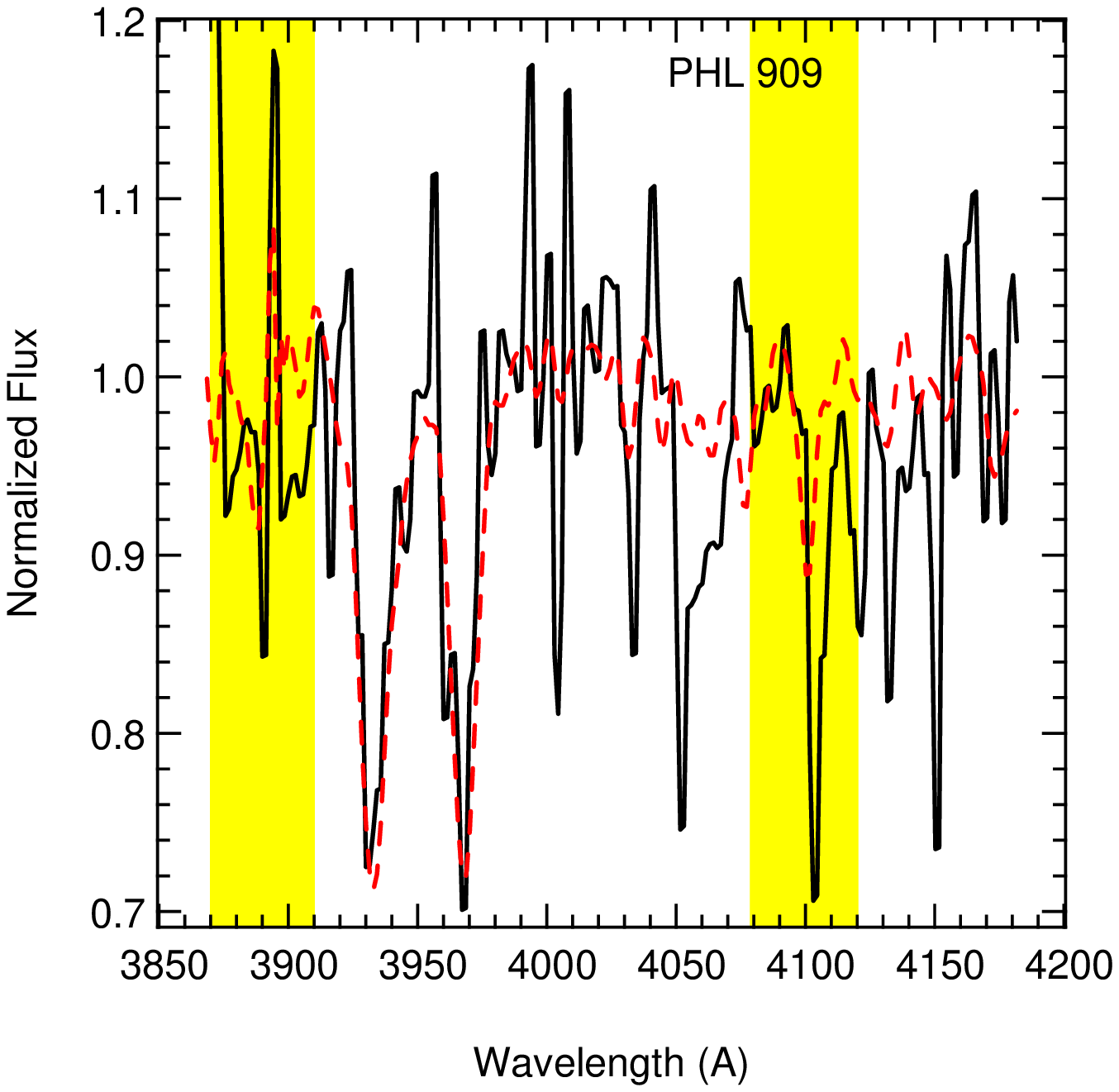}
 \includegraphics[scale=0.39]{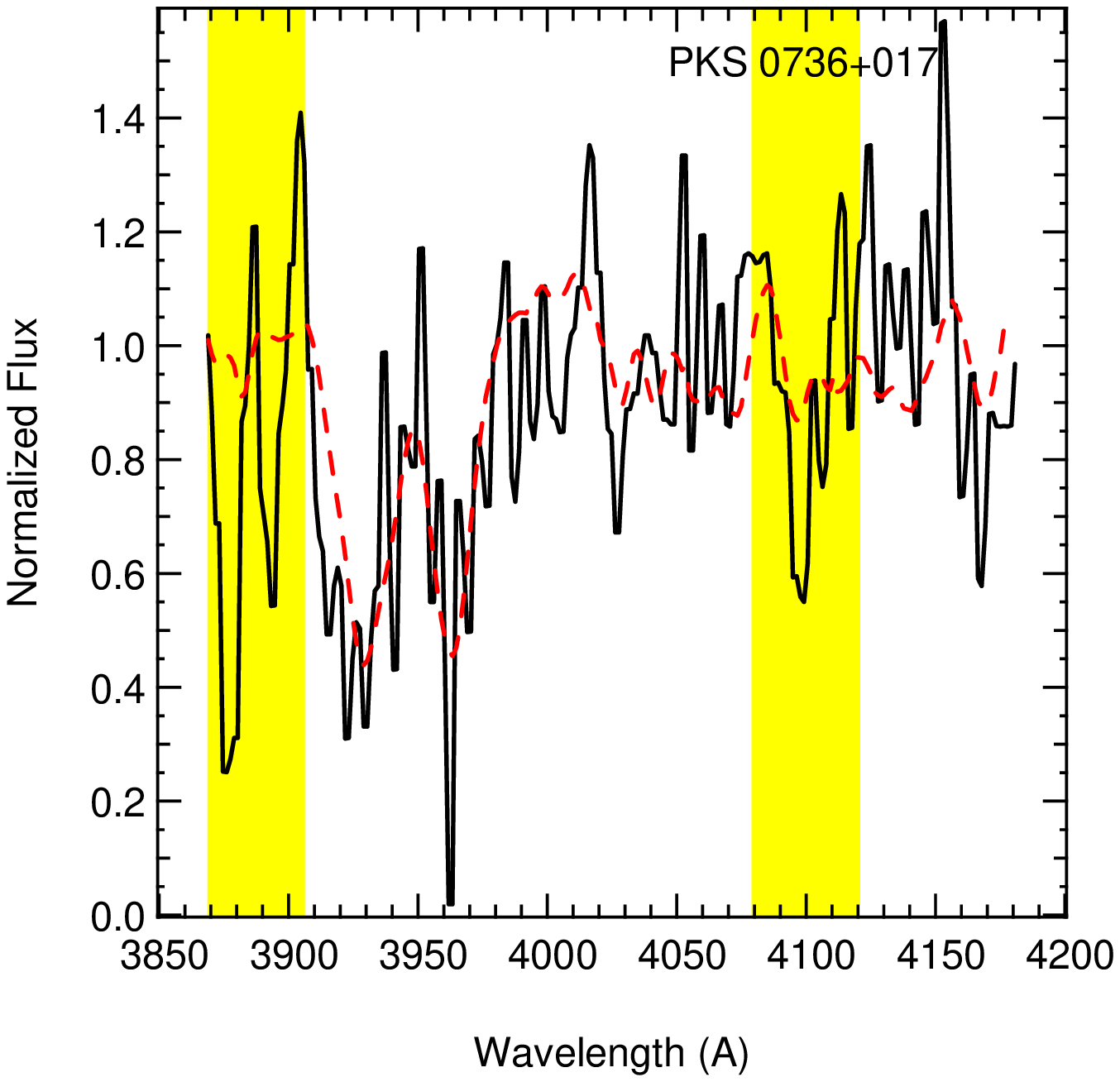}
 \includegraphics[scale=0.39]{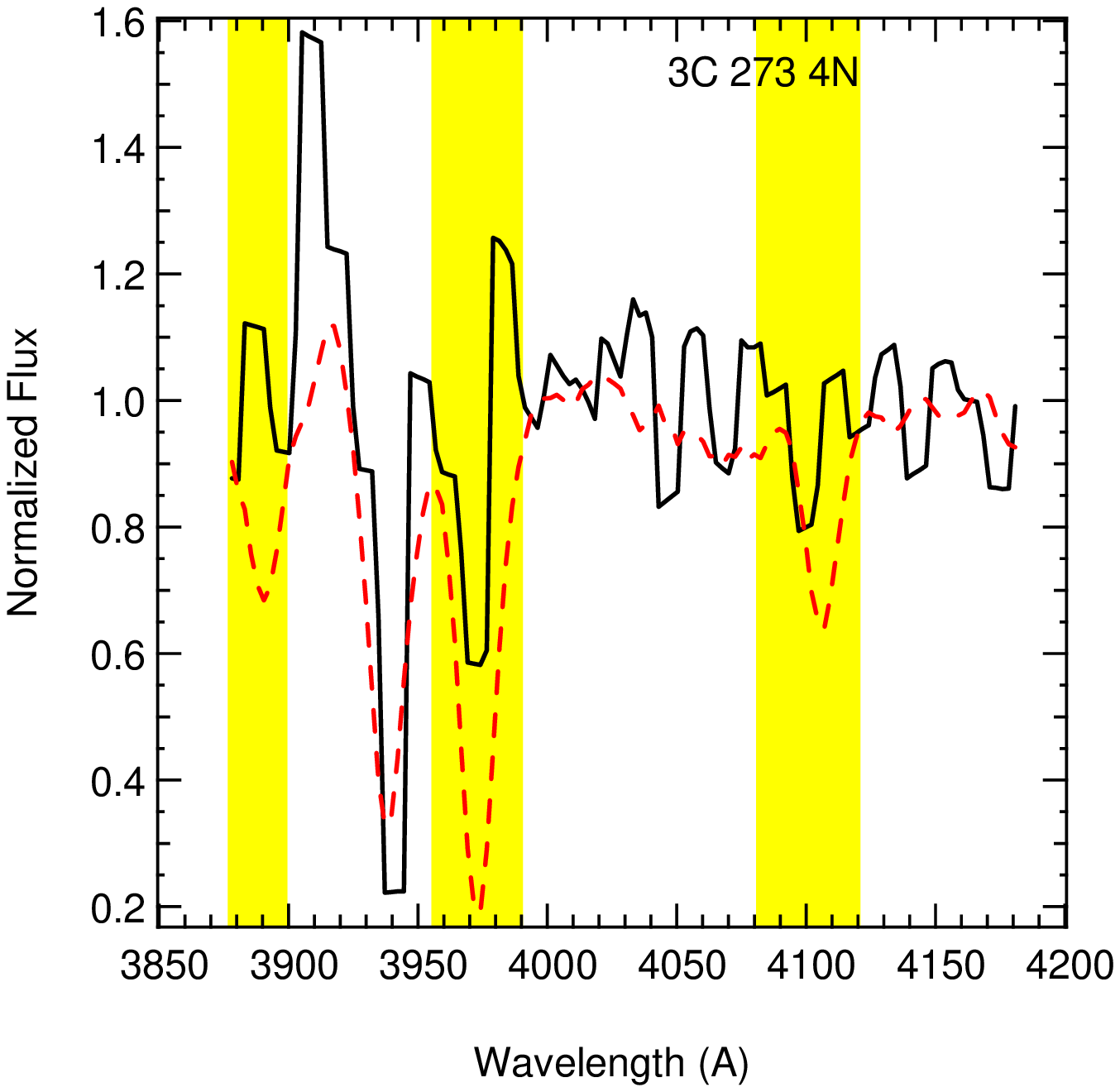}
 \includegraphics[scale=0.39]{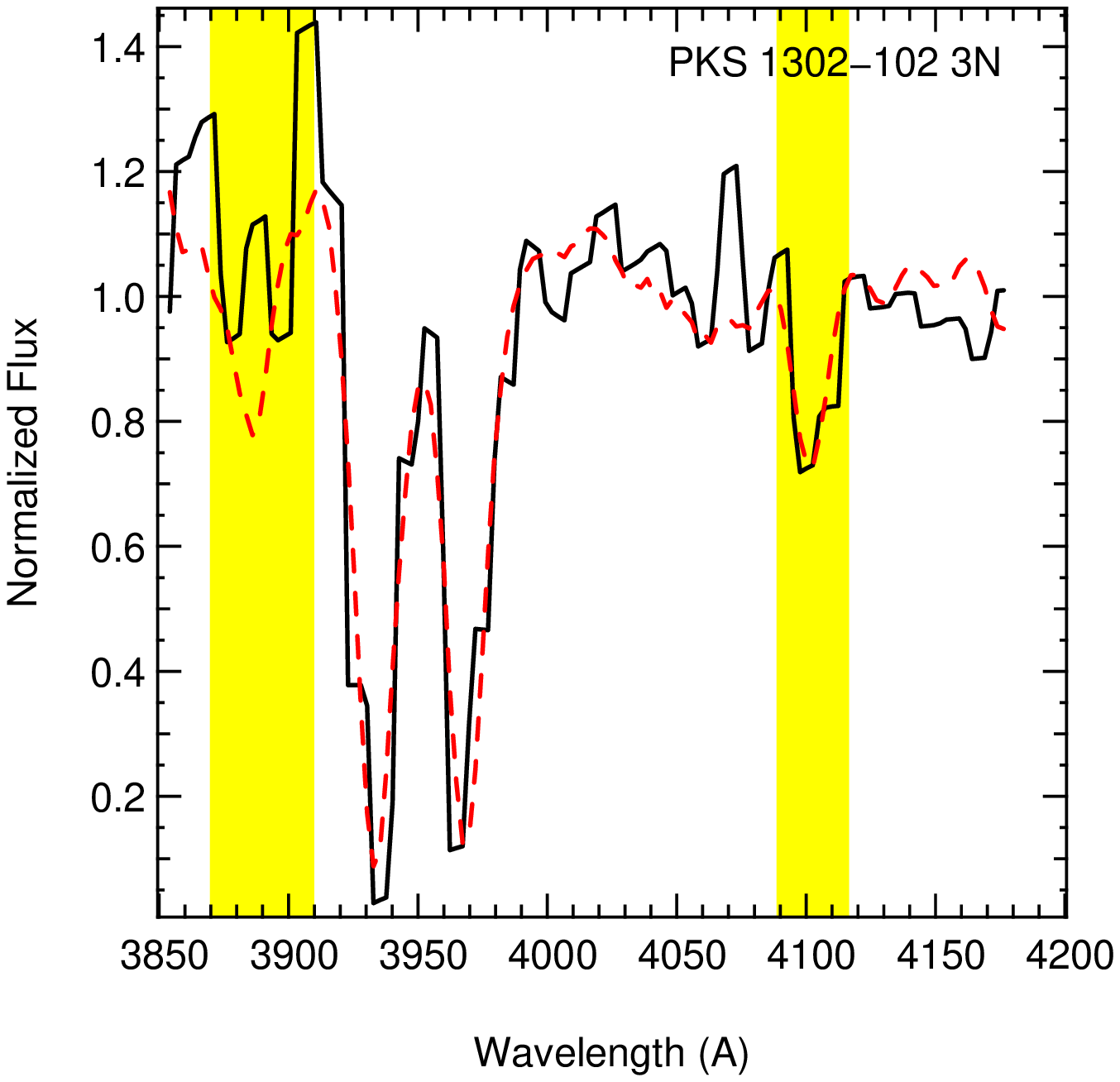}
 \includegraphics[scale=0.39]{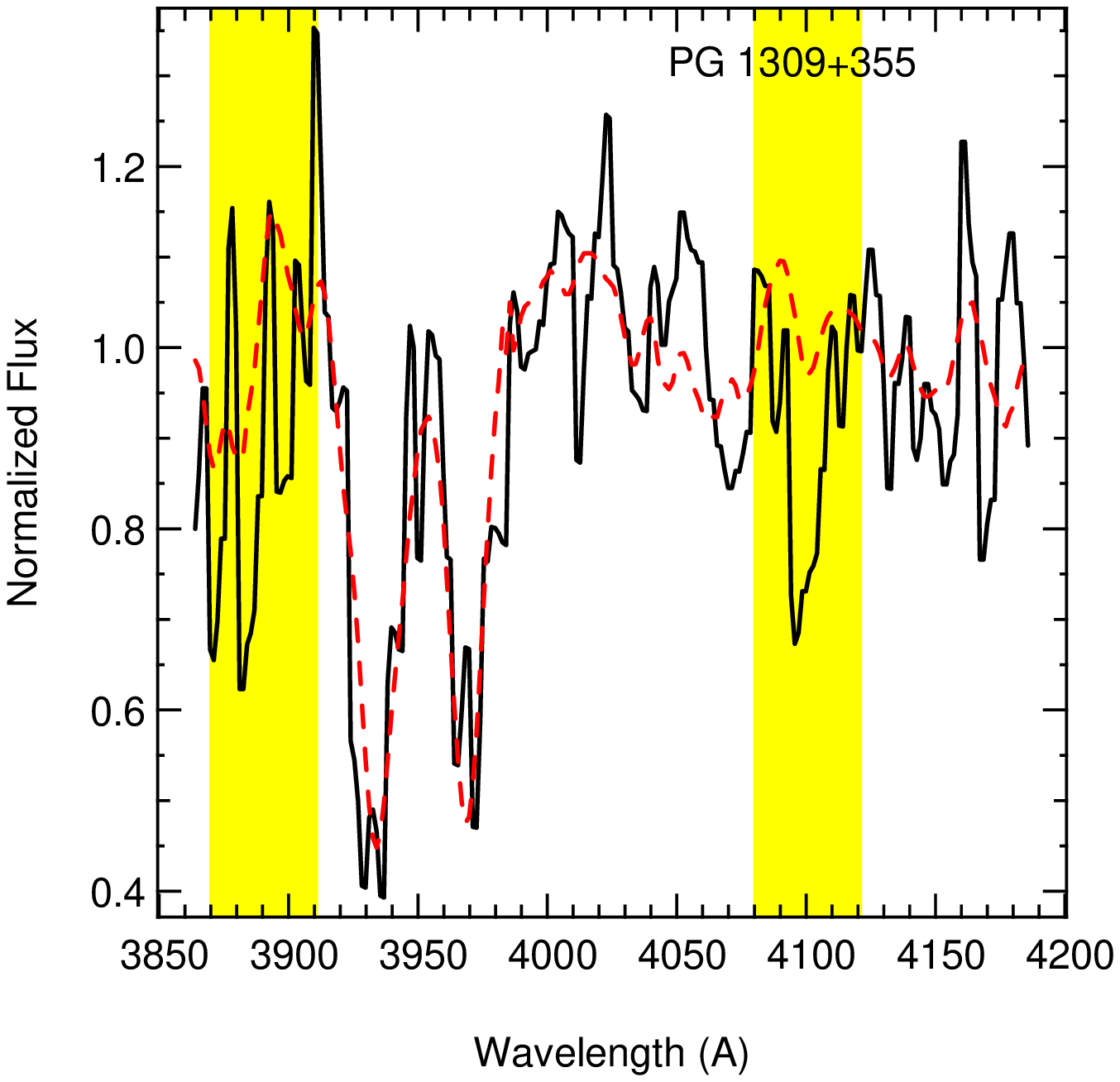}
 \includegraphics[scale=0.39]{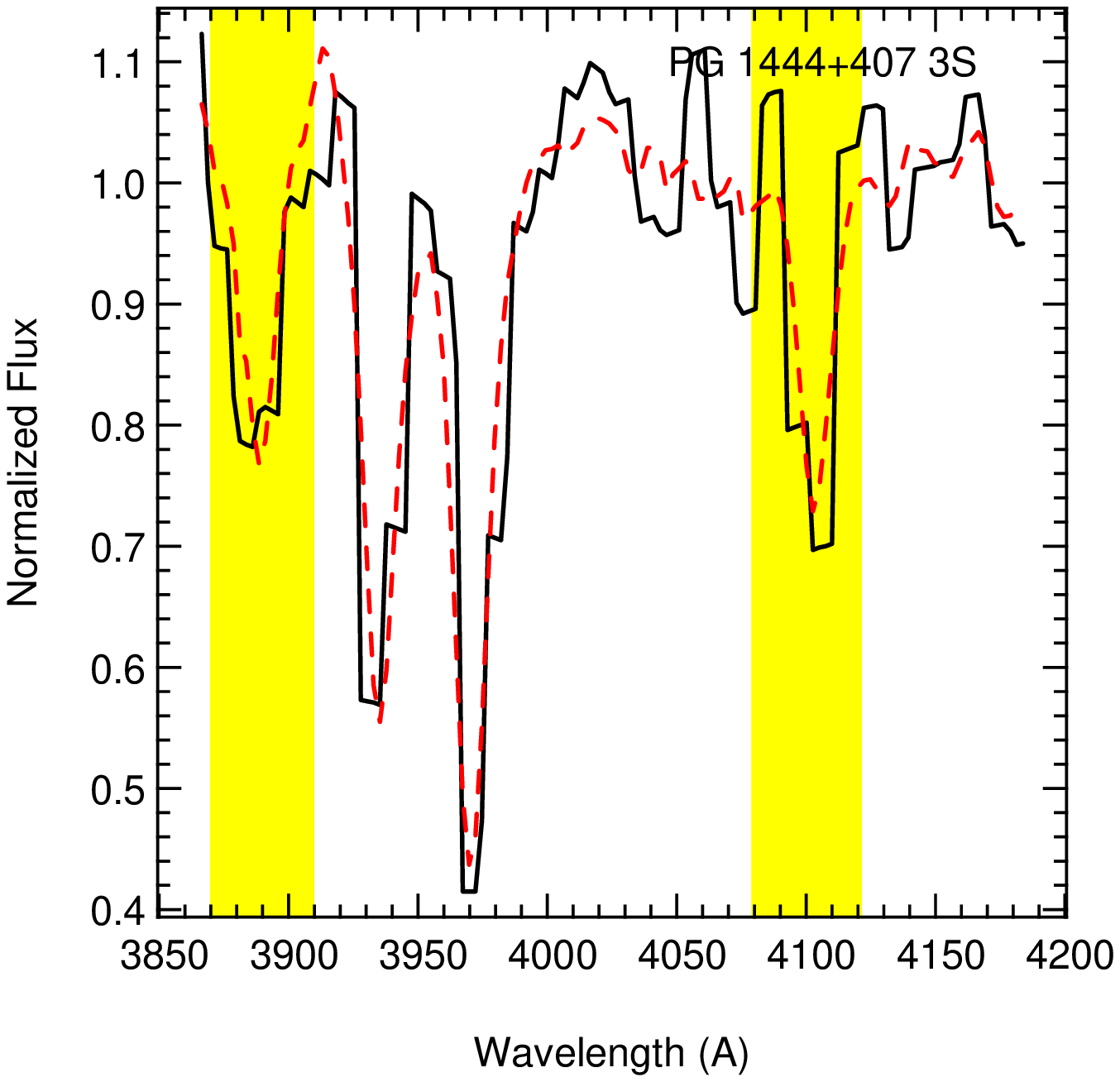}
 \includegraphics[scale=0.39]{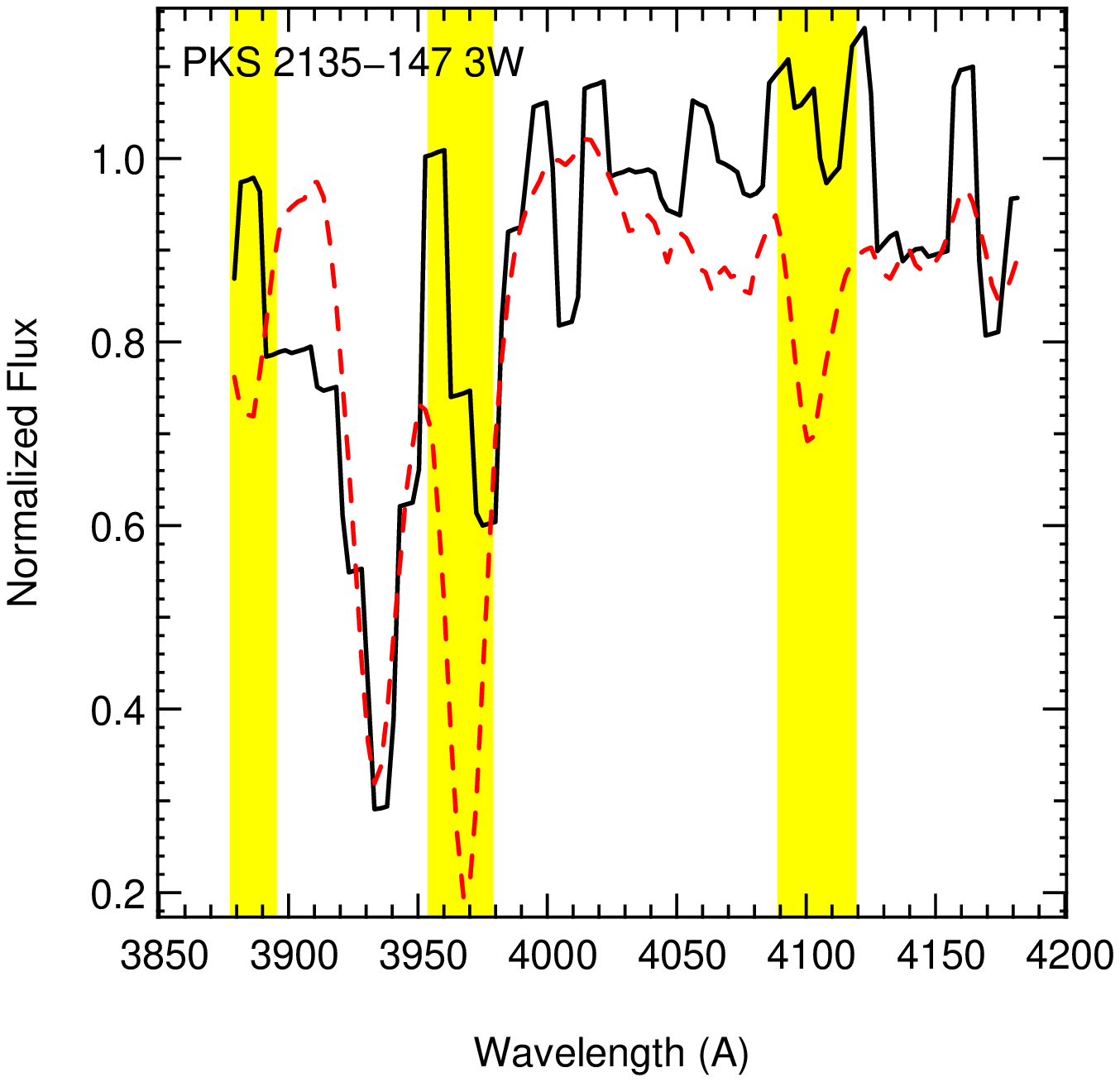}
 \includegraphics[scale=0.39]{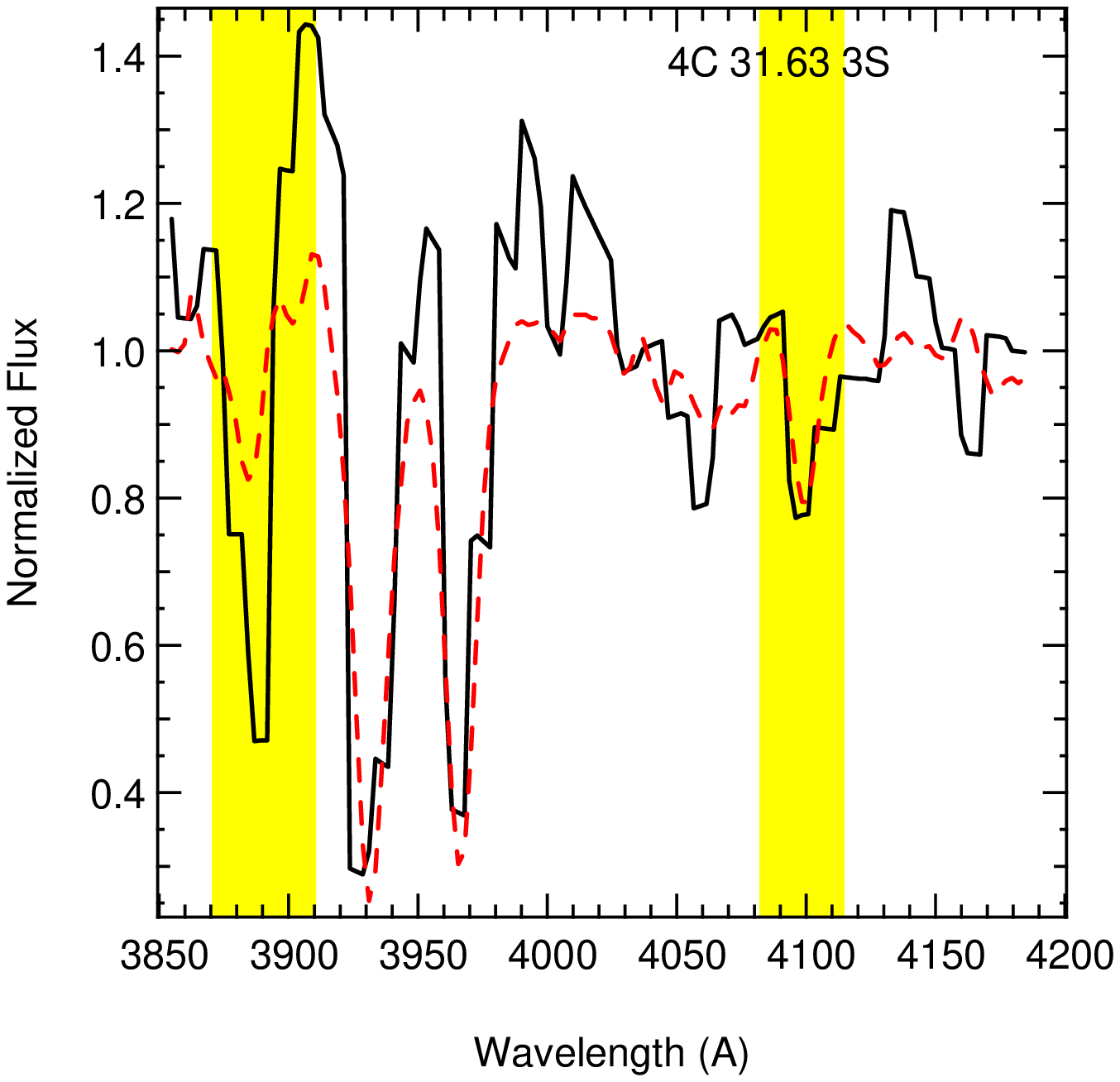}
 \caption{ Velocity dispersion fits to the Ca H\&K line region. Solid
 black lines are the host galaxy spectra, dashed red lines are the
 combined stellar templates convolved with the best-fitting gaussian
 line profiles, and yellow shading marks regions that were masked
 during the fit.  \label{vd_fig} }
\end{figure}

\begin{figure}
\figurenum{4}
 \includegraphics[scale=0.39]{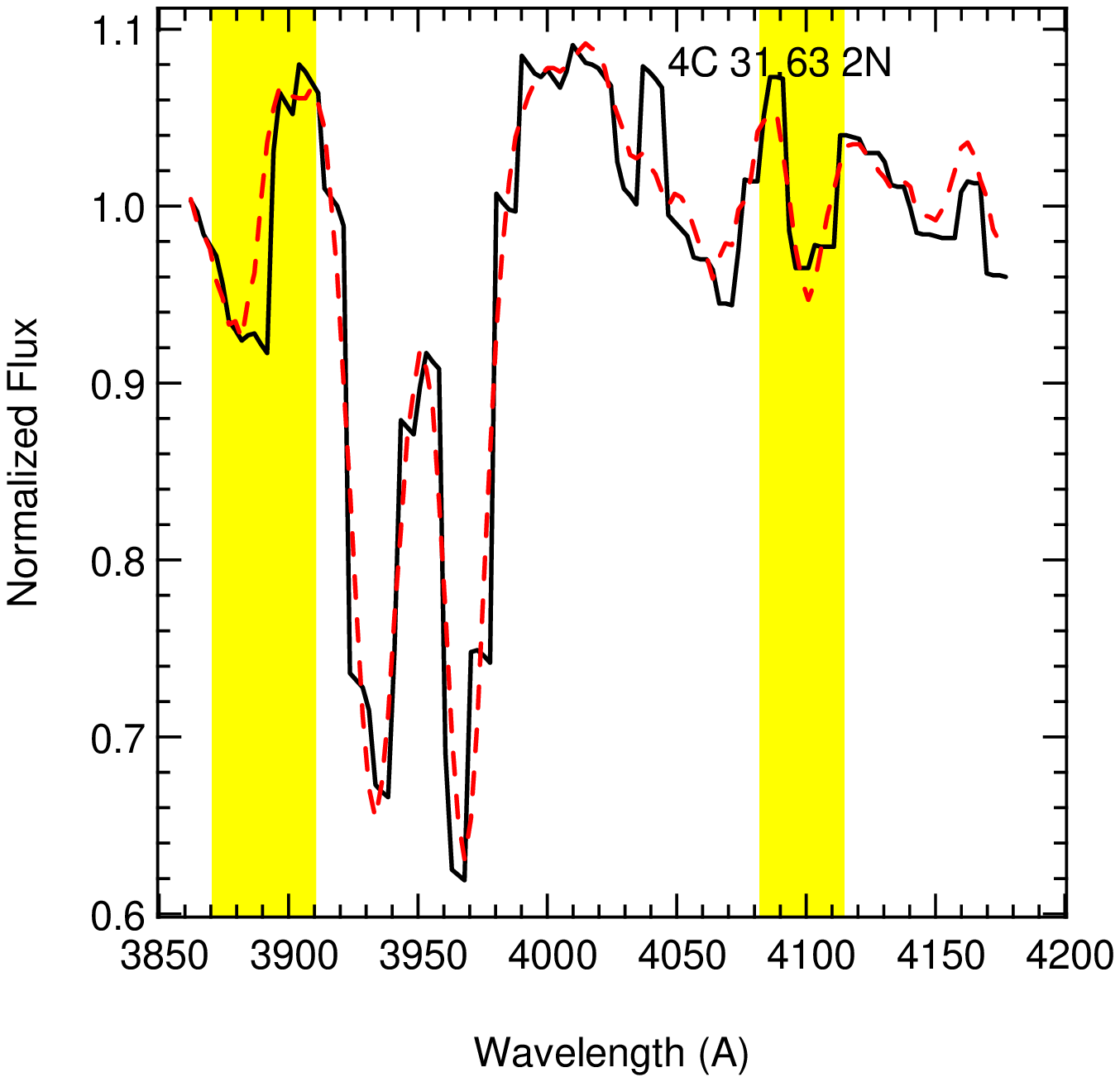}
 \includegraphics[scale=0.39]{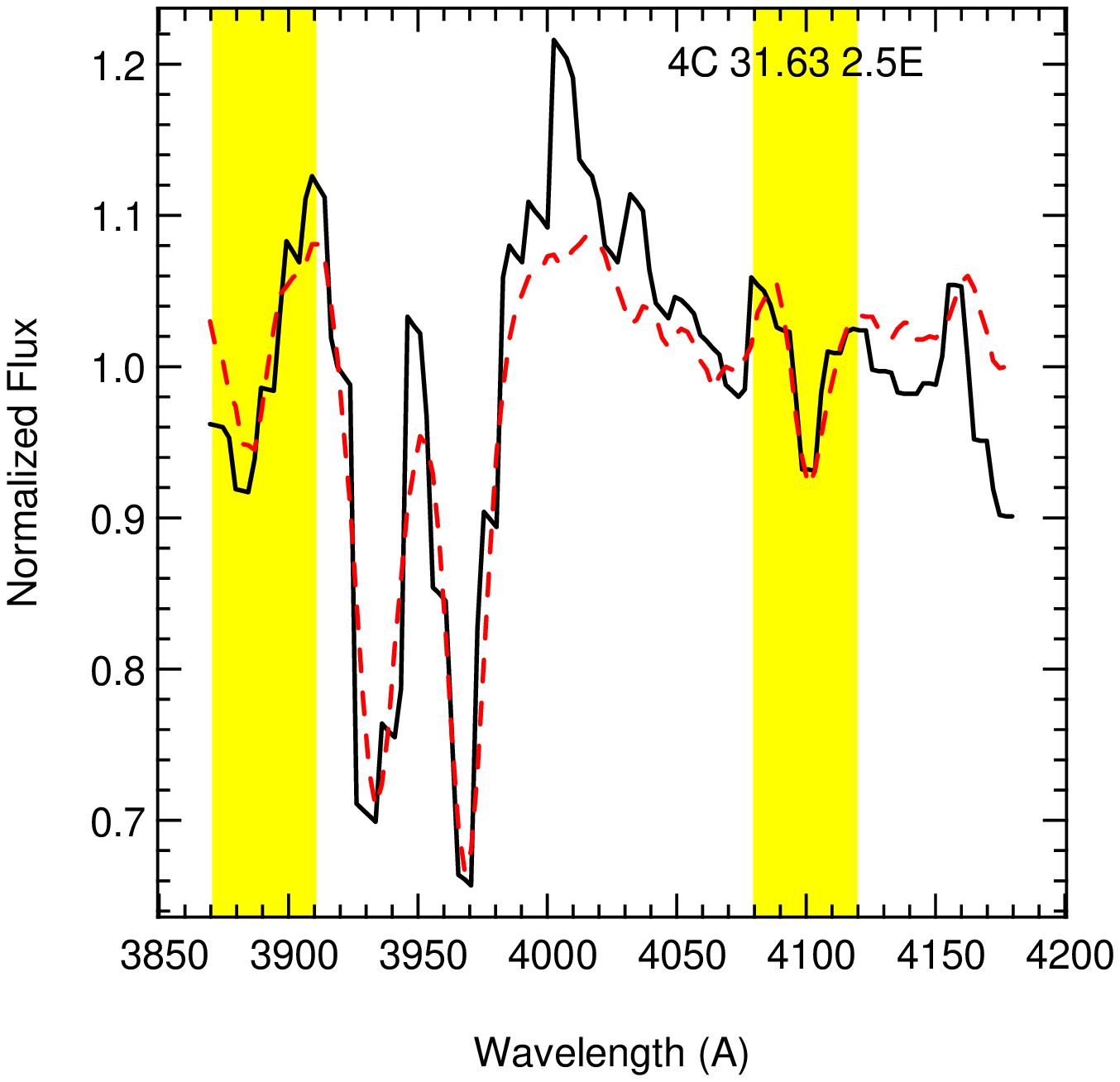}
 \includegraphics[scale=0.39]{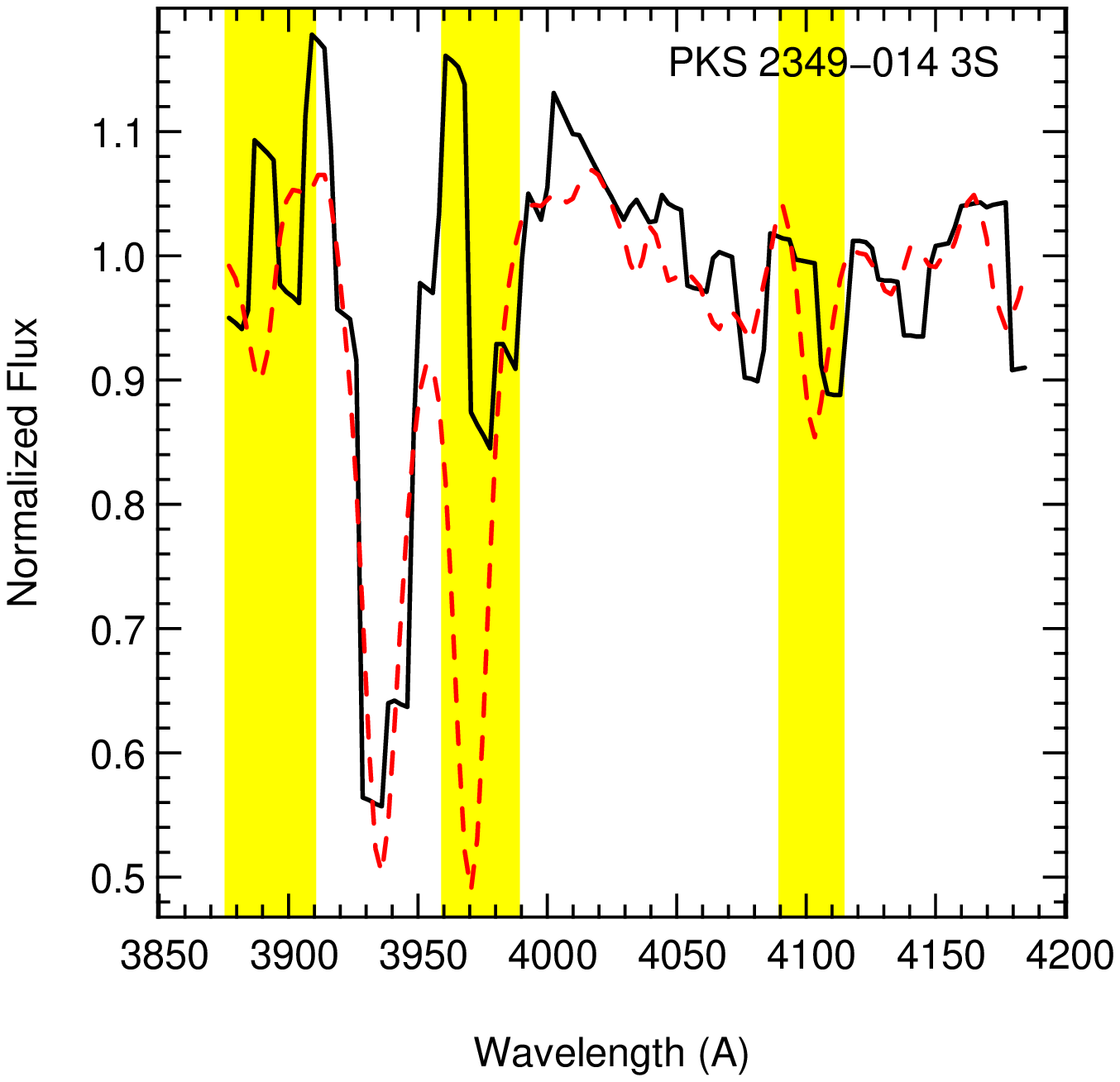}
 \includegraphics[scale=0.39]{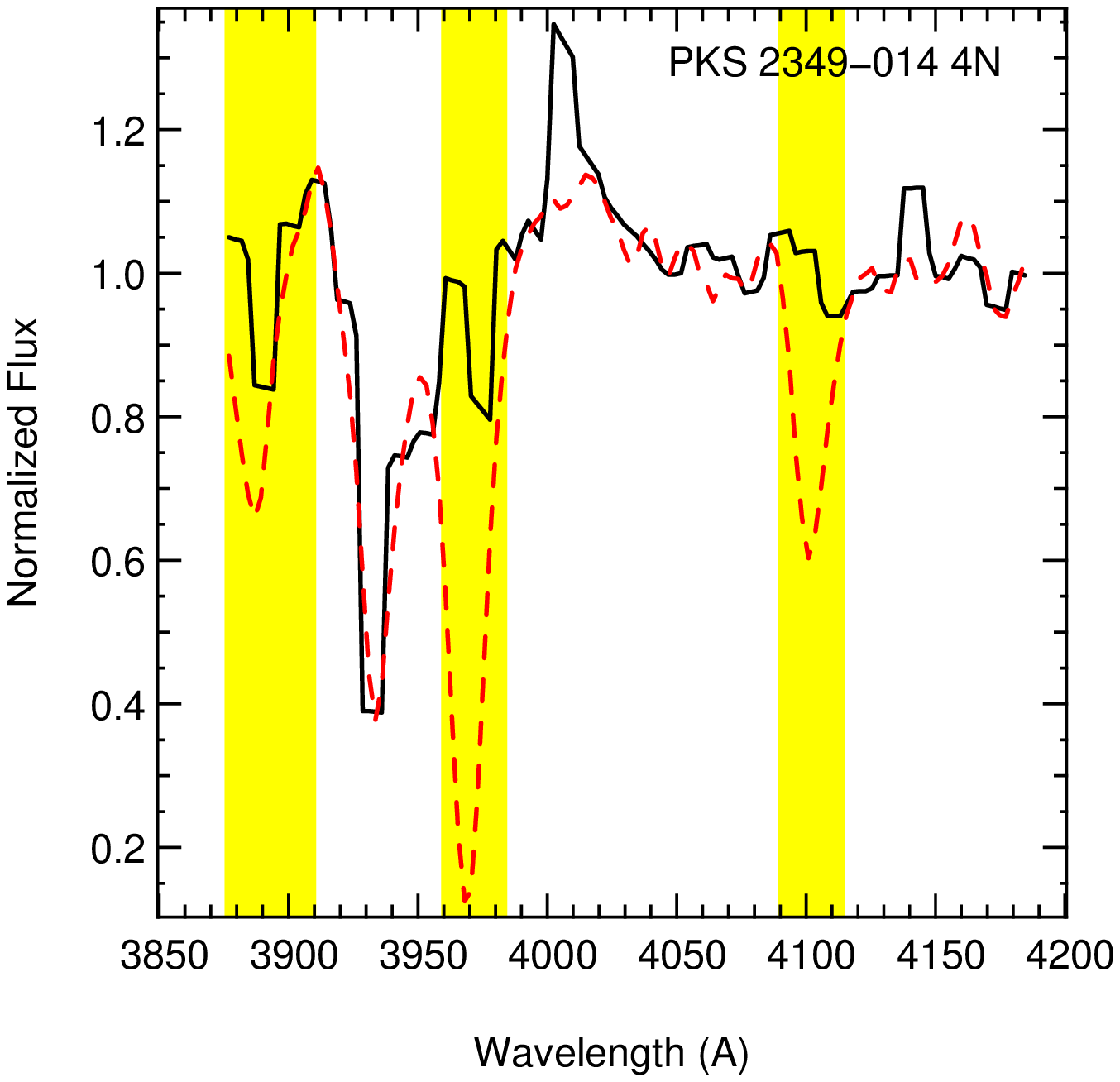}
 \caption{ cont. }
\end{figure}

\begin{figure}
 \includegraphics[scale=0.4]{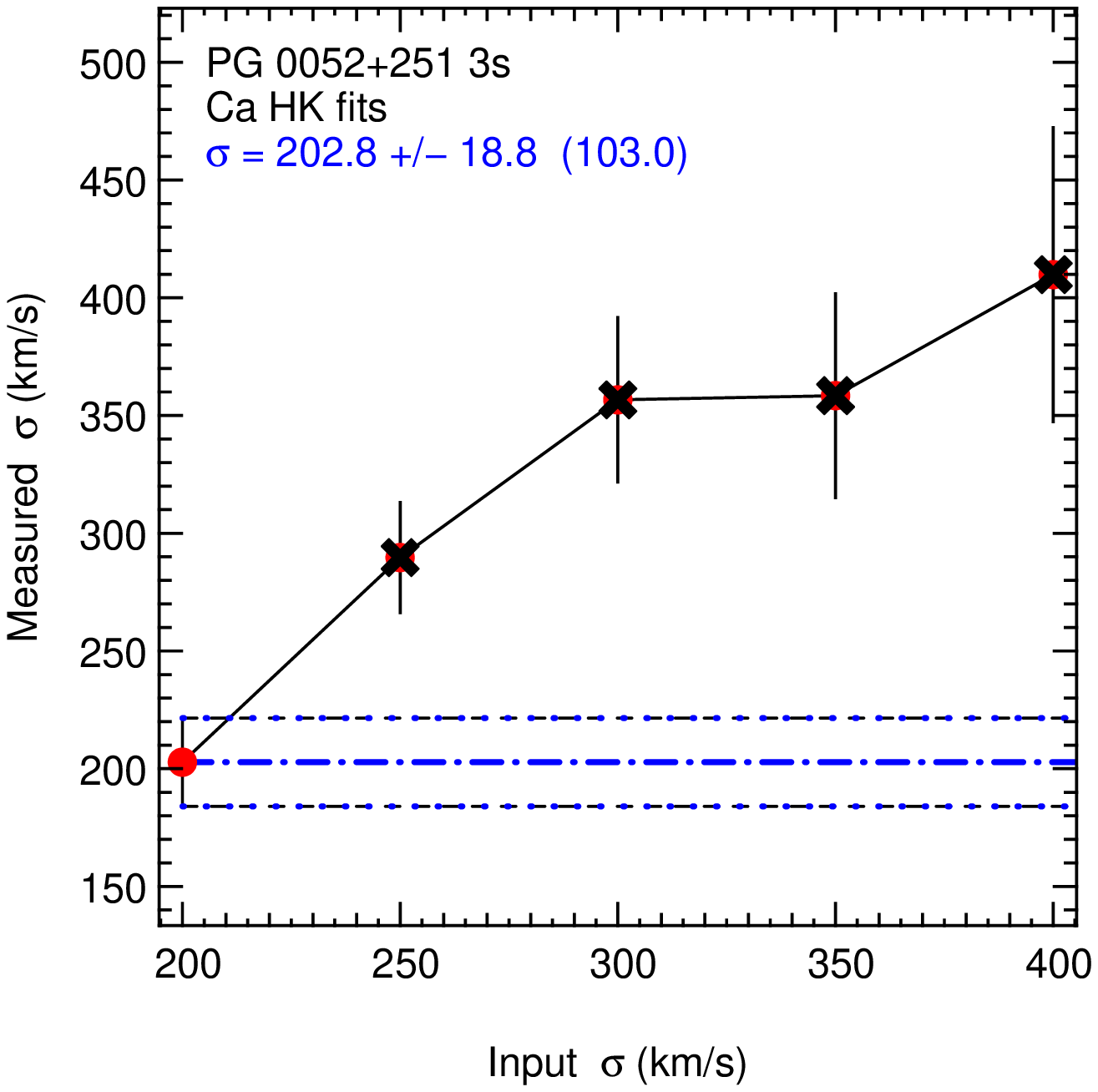}
 \includegraphics[scale=0.4]{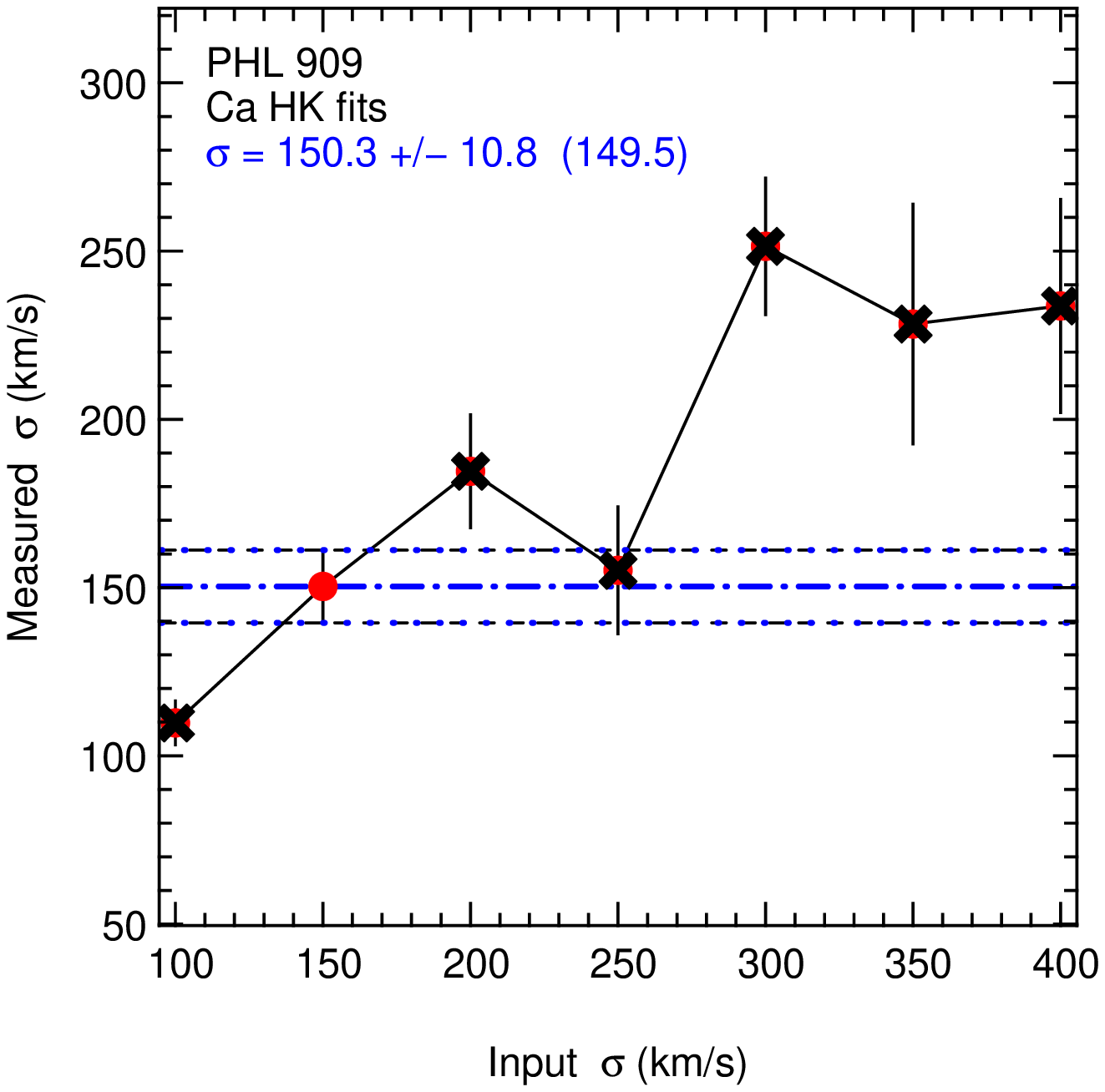} 
 \includegraphics[scale=0.4]{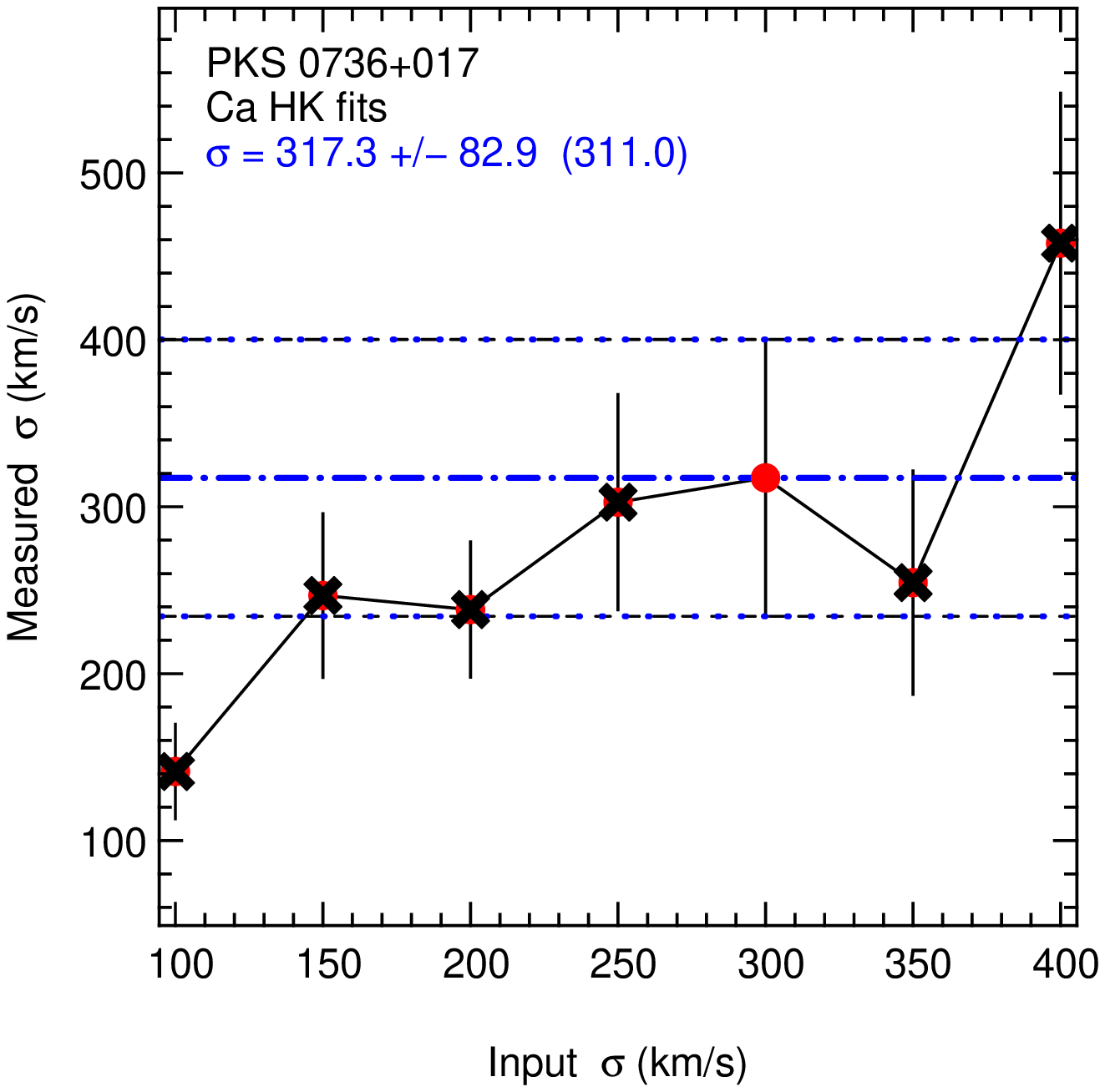}
 \includegraphics[scale=0.4]{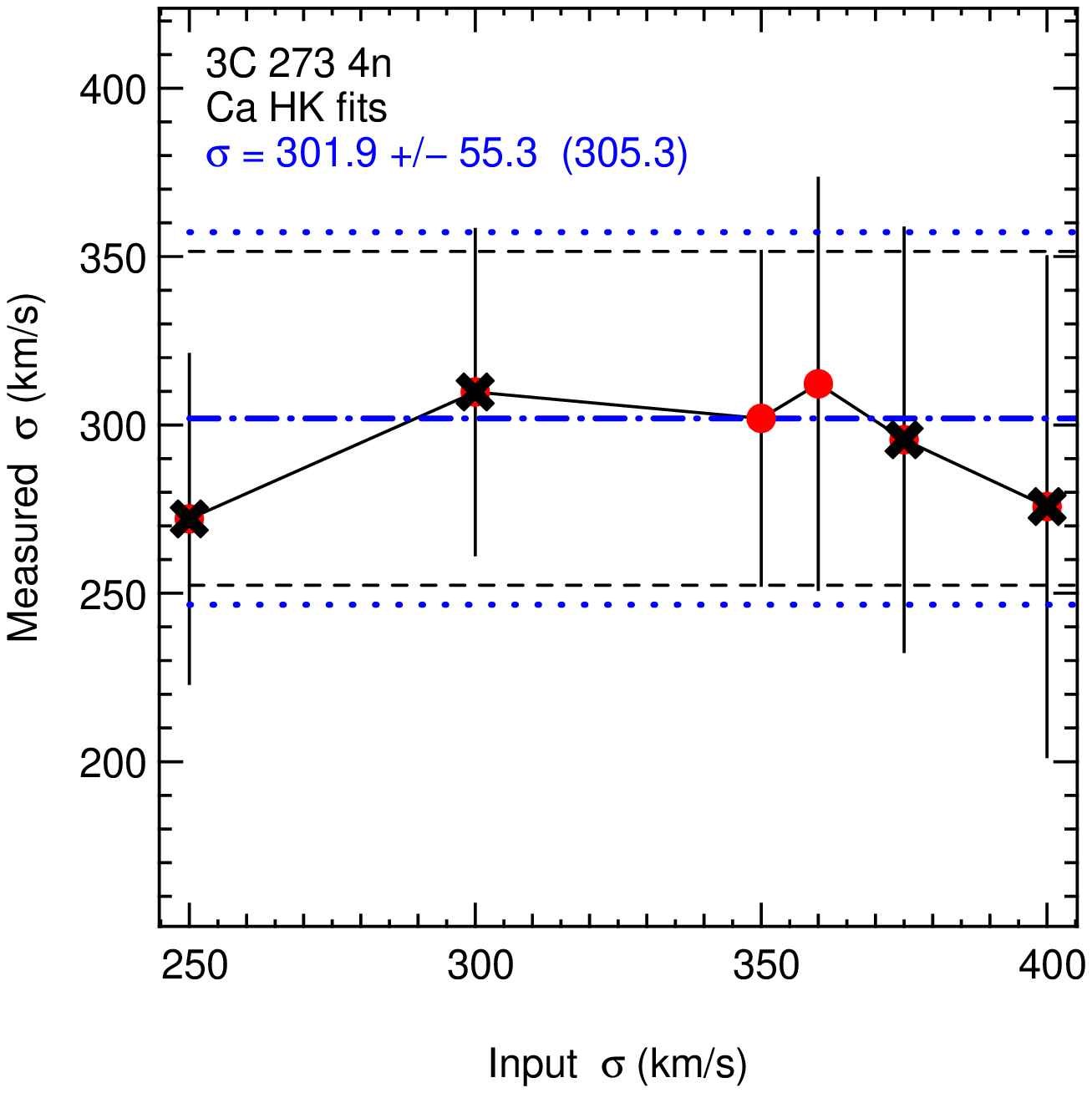}
 \includegraphics[scale=0.4]{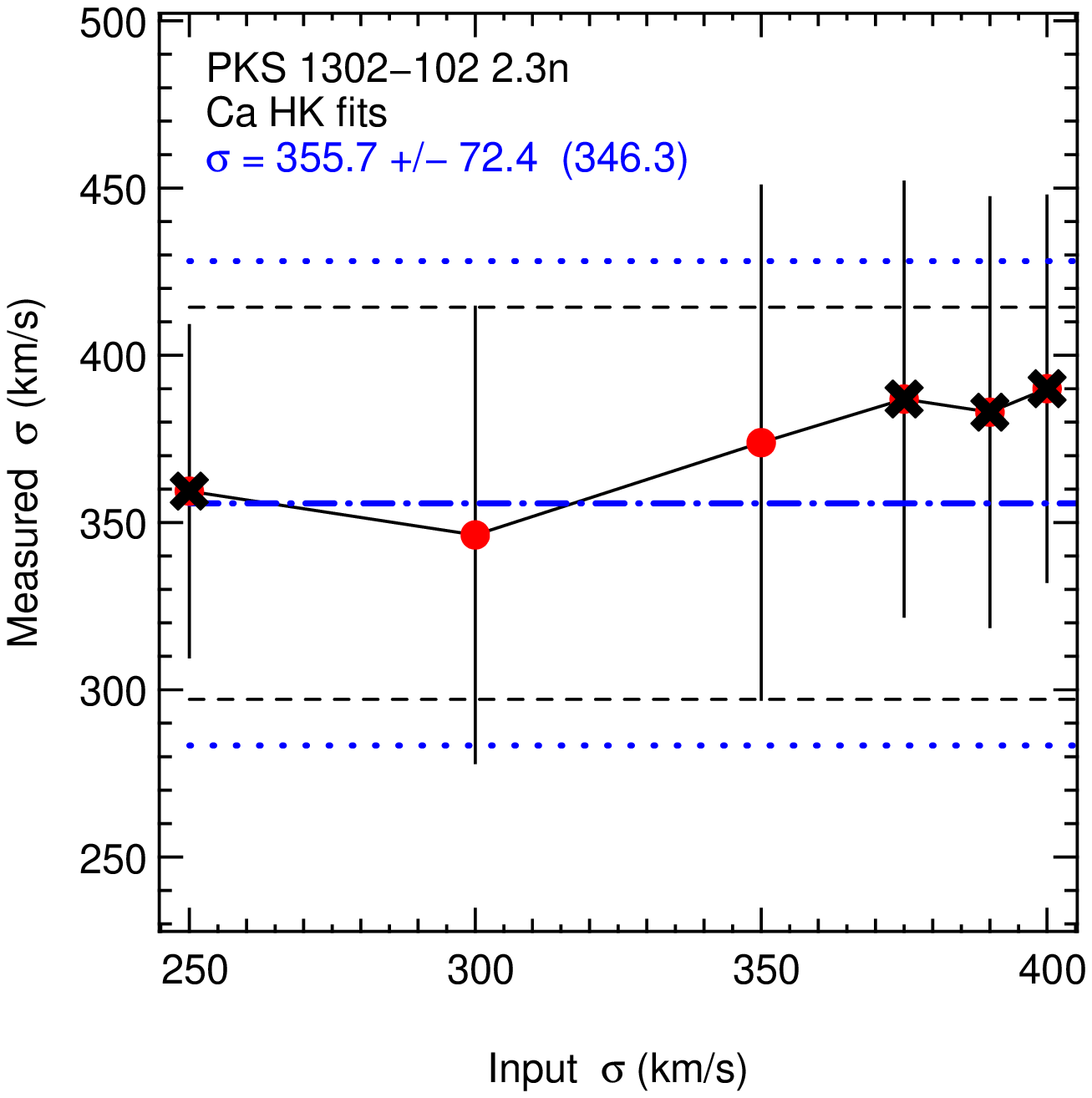}
 \includegraphics[scale=0.4]{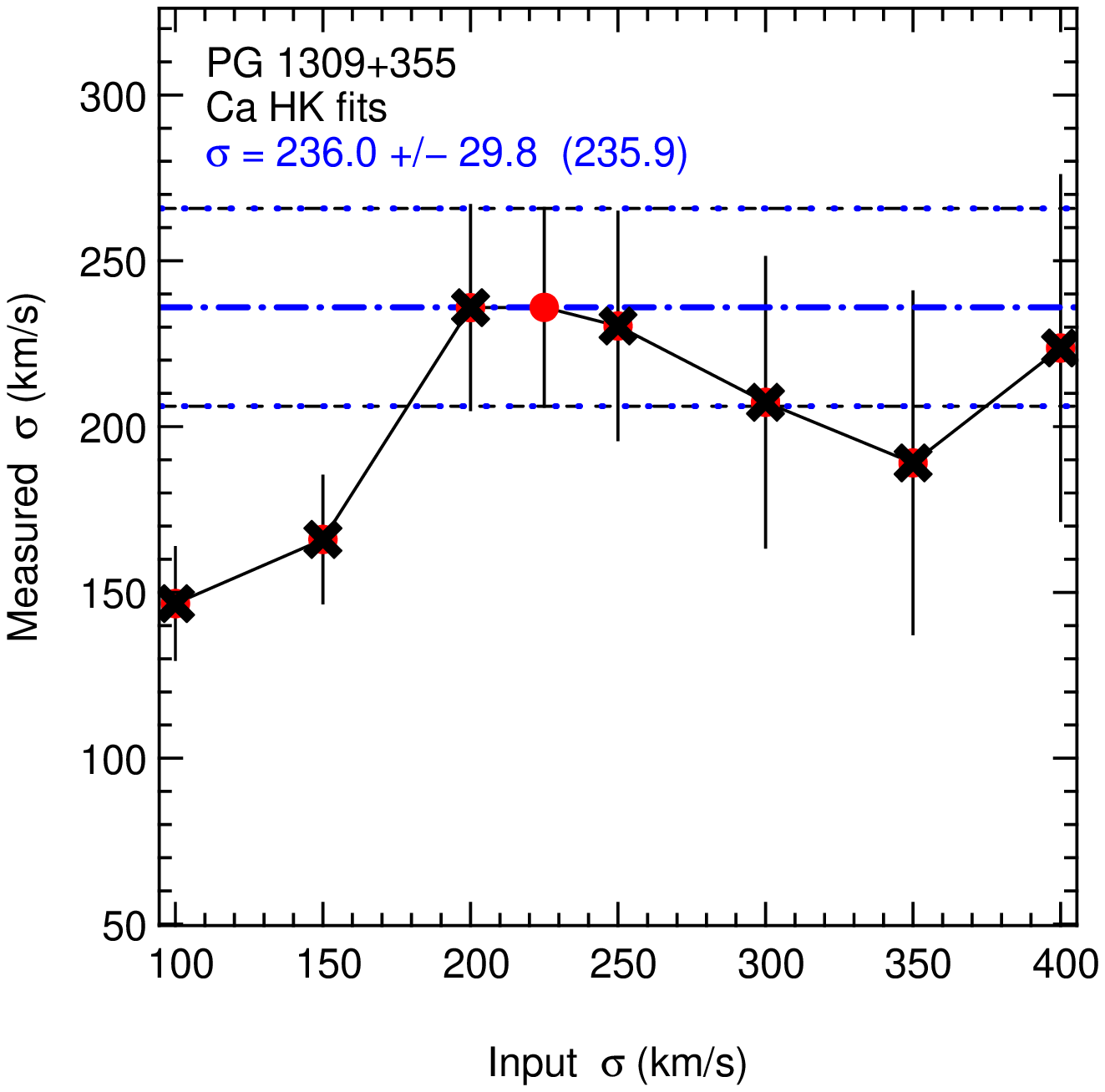}
 \includegraphics[scale=0.4]{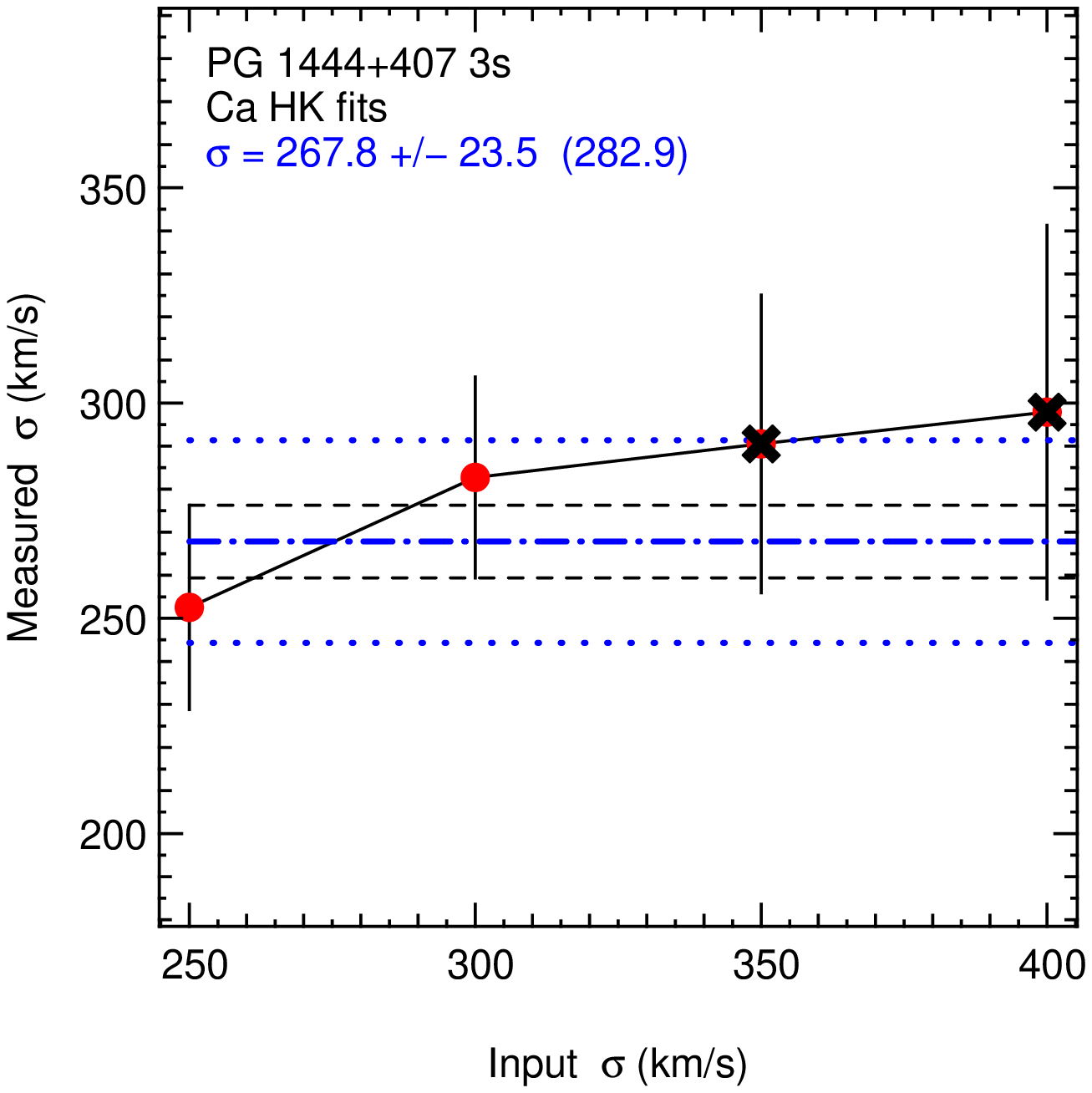}
 \includegraphics[scale=0.4]{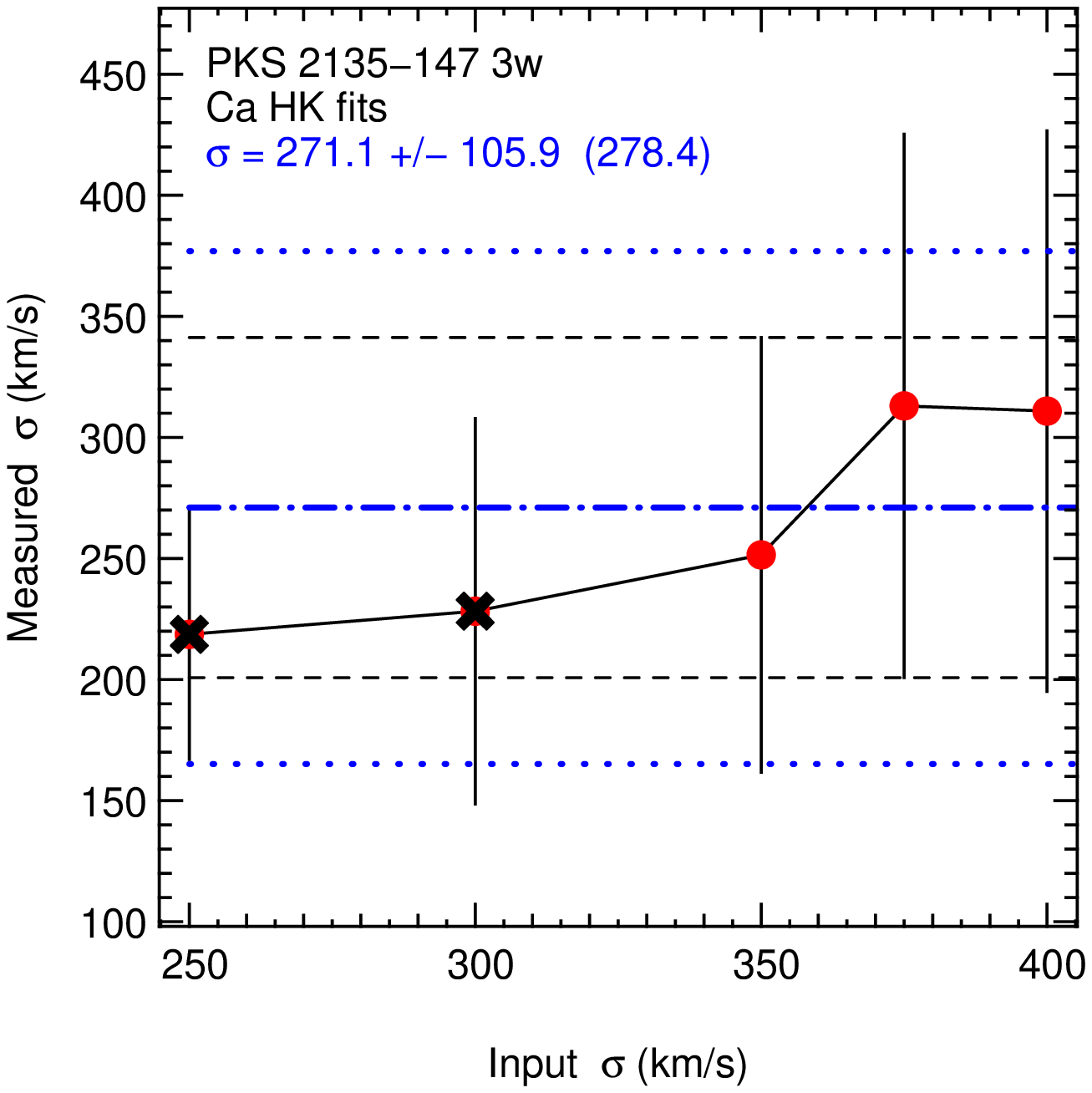}
 \includegraphics[scale=0.4]{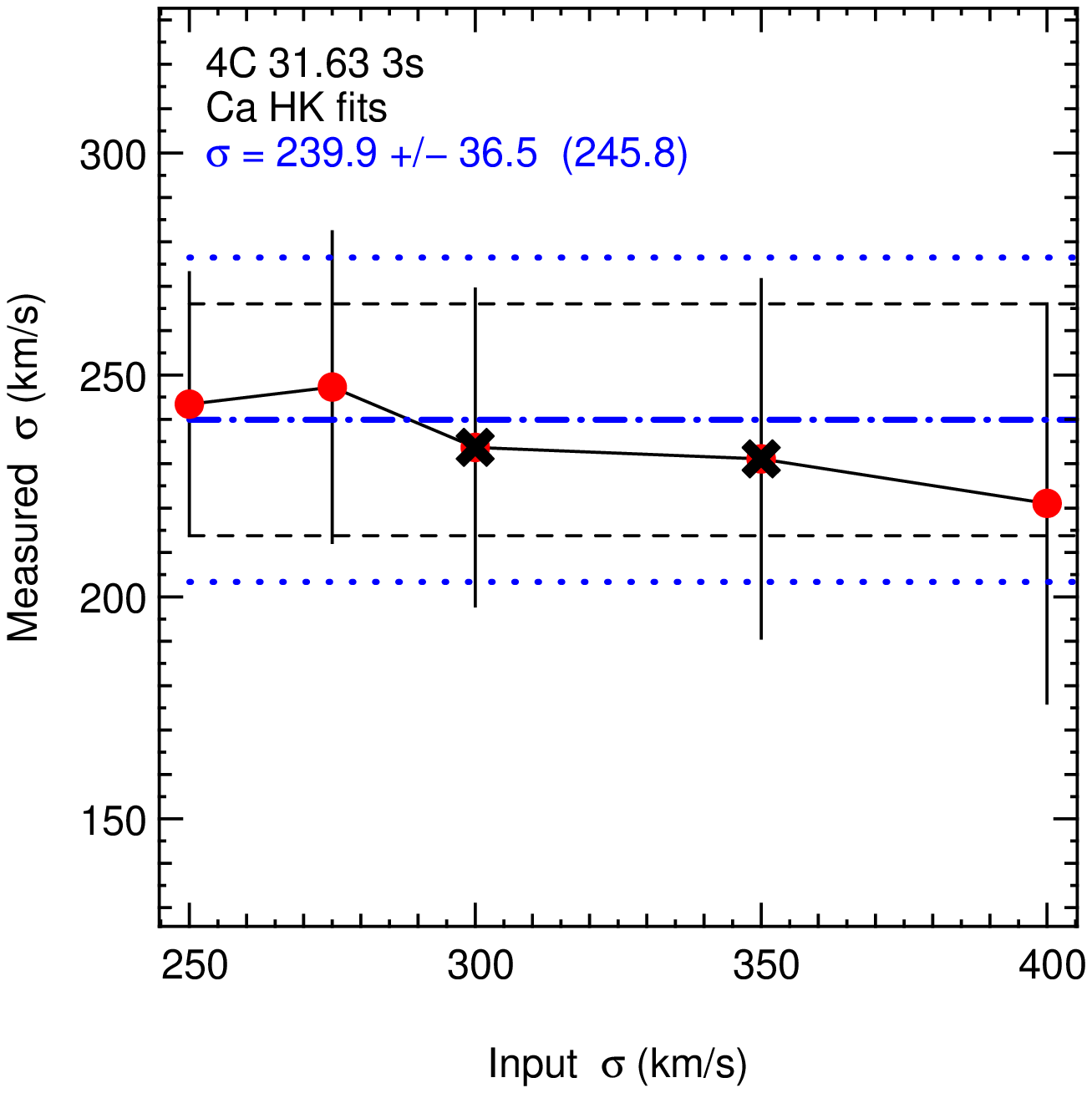}
 \caption{ Velocity dispersion measurement tests using a range of
   input initial $\sigma$'s. Each circle and its error bar is from 100
   Monte Carlo simulations at the given input $\sigma$. Black crosses mark
   points for which the sigma bias was $>$ 20-40\% of the uncertainty. The
   numbers in parentheses are the interpolated values for zero sigma 
   bias, as demonstrated in Figure \ref{interp_plot}, and represent 
   our adopted velocity dispersions, except for PG 0052+251 (see text).  
   \label{sigma_bias} }
\end{figure}

\begin{figure}
\figurenum{13}
 \includegraphics[scale=0.4]{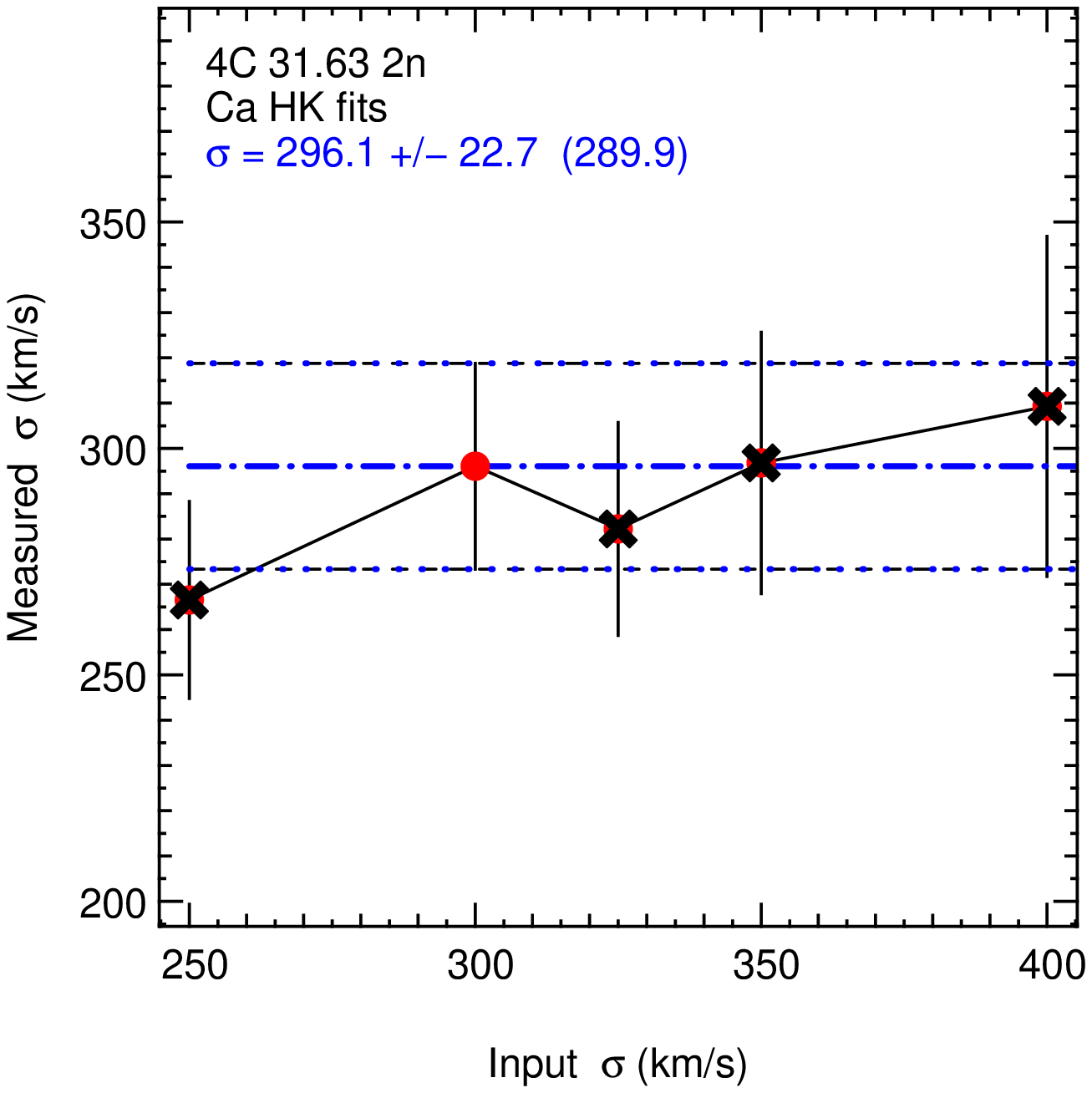}
 \includegraphics[scale=0.4]{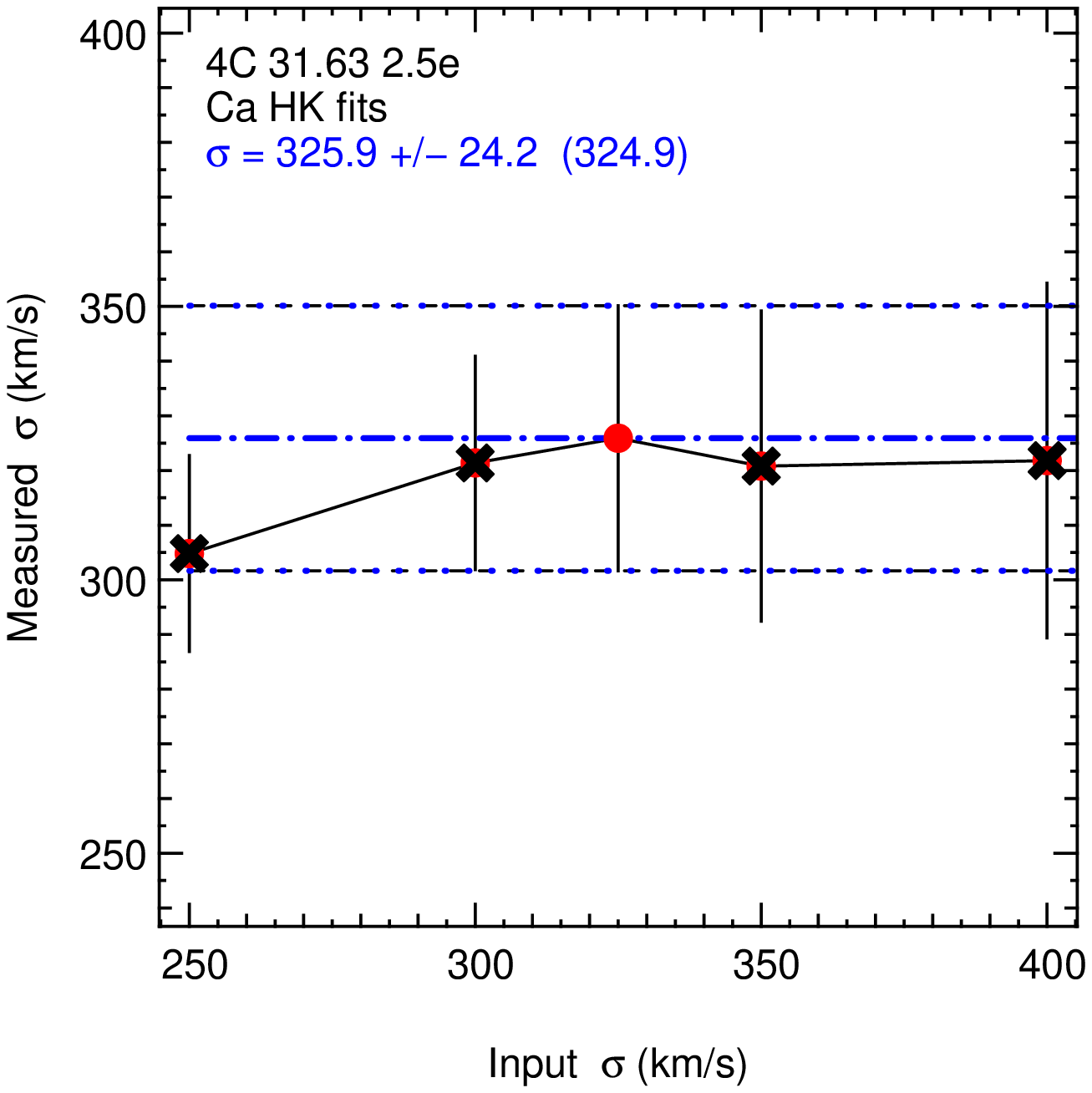}
 \includegraphics[scale=0.4]{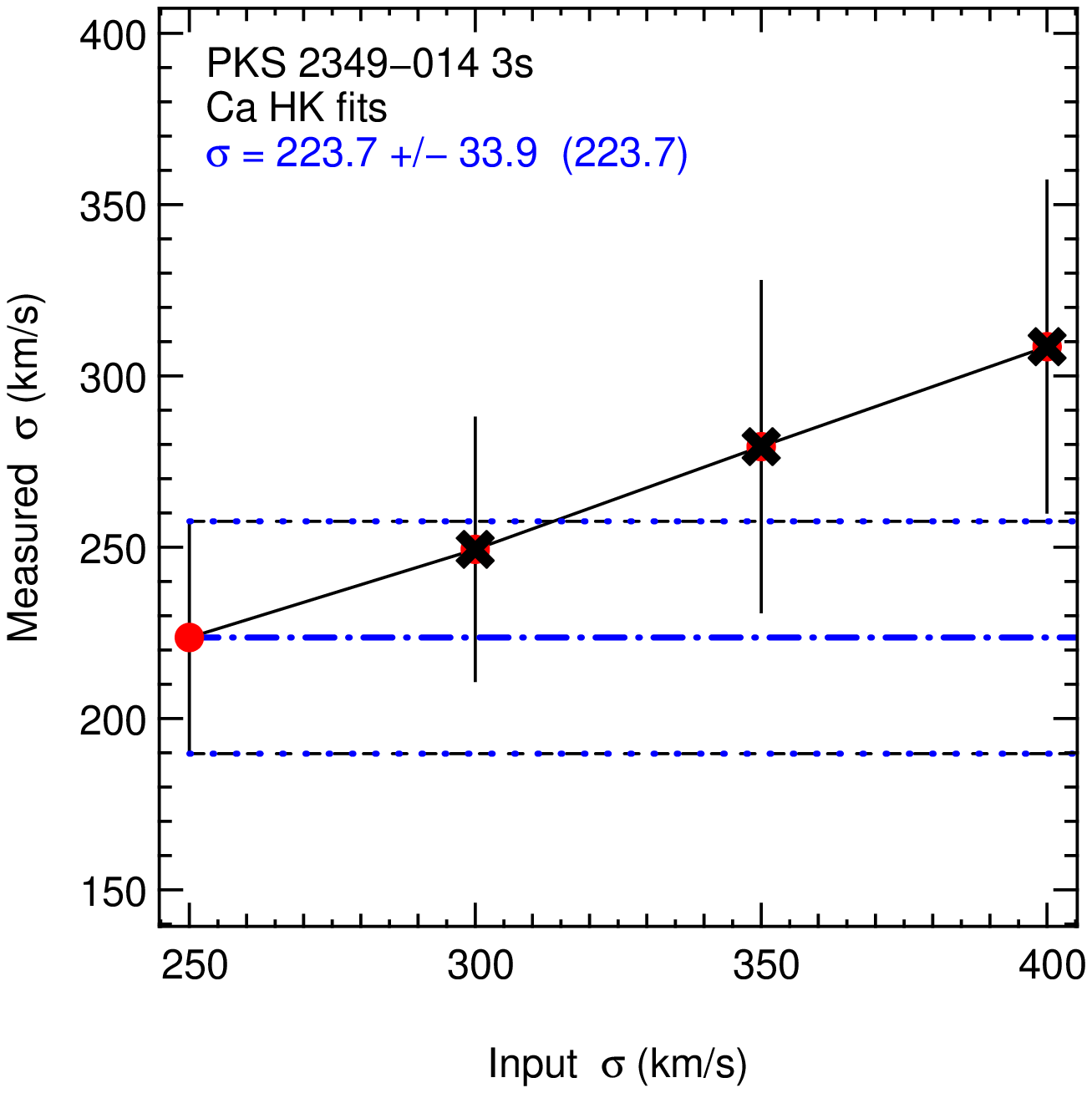}
 \includegraphics[scale=0.4]{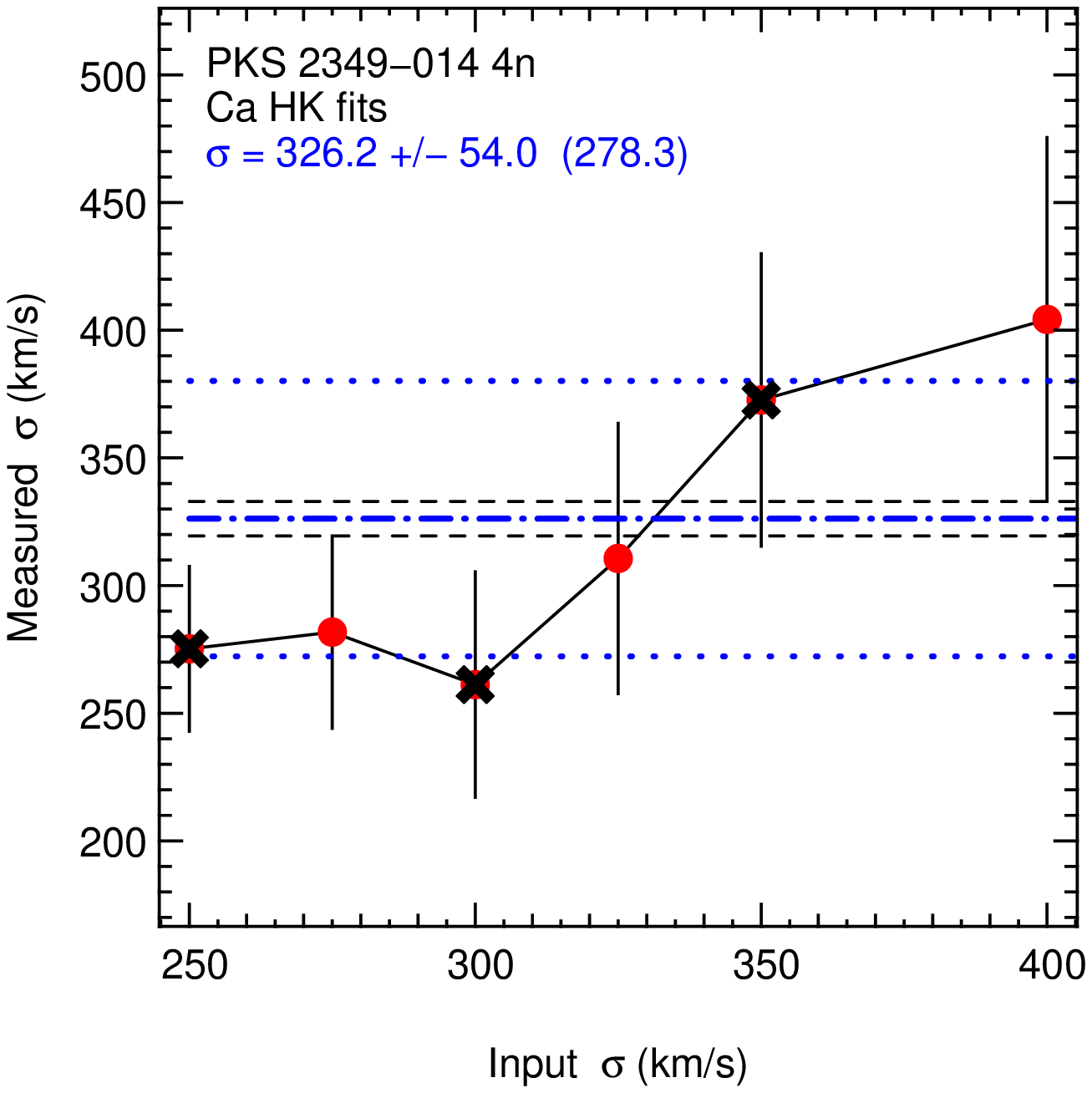}
 \caption{ cont. }
\end{figure}

\clearpage

\begin{deluxetable}{cccccccccccccc}
\tabletypesize{\scriptsize} \rotate \tablecaption{Object Data
\label{object_table}} \tablewidth{0pt} \tablehead{ \colhead{(1)} &
\colhead{(2)} & \colhead{(3)} & \colhead{(4)} & \colhead{(5)} &
\colhead{(6)} & \colhead{(7)} & \colhead{(8)} & \colhead{(9)} &
\colhead{(10)} & \colhead{(11)} & \colhead{(12)} & \colhead{(13)} &
\colhead{(14)} \\ \colhead{} & \colhead{} & \colhead{} & \colhead{B97}
& \colhead{H02} & \colhead{B97} & \colhead{H02} & \colhead{} &
\colhead{} & \colhead{} & \colhead{} & \colhead{}& \colhead{} &
\colhead{} \\ \colhead{\#} & \colhead{Object} & \colhead{Redshift} &
\colhead{M$_{V}$(QSO)} & \colhead{M$_{V}$(QSO)} &
\colhead{M$_{V}$(host)} & \colhead{M$_{V}$(host)} &
\colhead{m$_{V}$(host) \tablenotemark{a}} & \colhead{F606-V} &
\colhead{R$_{e}$} & \colhead{R$_{e}$} & \colhead{$\mu_{e}$(r)} &
\colhead{host} & \colhead{radio} \\ \colhead{} & \colhead{Name} &
\colhead{(z)} & \colhead{(mag)} & \colhead{(mag)} & \colhead{(mag)} &
\colhead{(mag)} & \colhead{(mag)} & \colhead{(mag)} &
\colhead{(as)} & \colhead{(kpc)} & \colhead{(mag as$^{-2}$)} &
\colhead{morphology} & \colhead{loudness} }

\startdata

1 & PG~0052+251 & 0.1550  & -24.1 & -23.2 & -22.5 & -22.4  & 17.01  
& -0.31  & 1.8 & 4.8  & 19.36 & Sb  & RQ  \\
2 & PHL 909 & 0.171 & -24.1 & -23.5 & -22.2 & -22.6 & 17.61 & -0.41 
& 2.3 & 6.7 & 20.43 & E4 & RQ \\
 & \it{(0054+023)} & & & & & & & & & & & & \\
3 & PKS 0736+017 & 0.191 & \nodata & -23.2 & -22.6 \tablenotemark{b} 
& -22.8 & 17.51 \tablenotemark{b} & \nodata & 3.3 & 10.4
\tablenotemark{b} & 21.01 & E, int. & RL \\
4 & 3C~273 & 0.1583  & -26.7 & -26.7 & -23.2 & -23.6 & 16.40  &
-0.40 & 3.7  & 10.1 & 20.30  & E4 & RL  \\
 & \it{(PG 1226+023)} & & & & & & & & & & & & \\
5 & PKS~1302-102 & 0.2784  & -25.9 & -25.6 & -22.9 & -23.4  & 18.20  
& -0.50  & 1.4 & 5.9  & 19.56 &  E4(?)  & RL  \\
6 & PG~1309+355 & 0.1840 & -24.4 & -24.1  & -22.8 & -23.1  & 17.15
& -0.35 & 2.0 & 6.2 & 19.62 & Sab & RQ \\
7 & PG~1444+407 & 0.2673  & -25.3 & -24.9 & -22.7 & -23.3 & 18.27  
& -0.47  & 1.3 & 5.3  & 19.51 & E1(?), S \tablenotemark{e}  & RQ  \\
8 & PKS~2135-147 & 0.2003  & -24.7 & -23.4  & -22.4 & -22.7  & 17.85  
& -0.45  & 2.6 & 8.6  & 20.83 & E1  & RL  \\
9 & 4C~31.63 & 0.2950  & \nodata  & -25.1 & \nodata & -23.8 & 17.55 
\tablenotemark{c} & \nodata  & 6.5 \tablenotemark{d}  & 28.5  & 22.18  
&  E \tablenotemark{f} & RL  \\
 & \it{(2201+315)} & & & & & & & & & & & & \\
10 & PKS~2349-014 & 0.1740  & -24.5 & -23.3 & -23.2 & -23.6 & 16.61  
& -0.41  & 4.8 & 14.2  & 21.02 &  E, int.  & RL  \\
                                       
\enddata

\tablecomments{ Absolute magnitudes are calculated from apparent
  magnitudes with k-corrections. Column (2) is the object name, column
  (3) is the redshift, column (4) is the V-band absolute magnitude of
  the QSO from \citet{bahcall97}, column (5) is the V-band absolute
  magnitude of the QSO from \citet{hamilton02}, column (6) is the
  V-band absolute magnitude of the host galaxy from the best fit 
  of \citet{bahcall97}, column (7) is the V-band absolute magnitude of
  the host galaxy from the best fit of \citet{hamilton02}, column (8)
  is the V-band apparent magnitude of the host galaxy from B97
  r$^{1/4}$ 2-D fits except as noted, column (9) is the F606W-V color
  from B97, column (10) is the effective radius in arcsec from B97 2-D
  r$^{1/4}$ fits to F606W HST images, column (11) is the effective
  radius in kpc, column (12) is the r-band average surface brightness
  within R$_{e}$, column (13) is the host galaxy morphology from B97
  except as otherwise noted, and column (14) is the radio loudness of
  the QSO: radio-loud (RL) or radio-quiet (RQ), divided at L$_{5
    GHz}=$~10$^{26}$~W~Hz$^{-1}$ as in \citet{kellermann94}.}

\tablenotetext{a}{These magnitudes are from r$^{1/4}$ law fits,
  converted to r-band using average values of B-V=0.95 and B-r=1.25
  from \citet{faber89} and \citet{gebhardt03}. These values also
  correspond to the average of E and S0 galaxy color corrections from
  \citet{fukugita95}. }

\tablenotetext{b}{ m$_{host}$(R) from \citet{dunlop03} and converted
  to m$_{V}$ with color corrections from \citet{fukugita95}, R$_{e}$
  from \citet{dunlop03}. }

\tablenotetext{c}{ m$_{host}$(F702W) 
from \citet{hamilton02} converted to m$_{V}$ with color corrections 
from \citet{fukugita95}. }

\tablenotetext{d}{Effective radius derived from 2-D r$^{1/4}$ fit to H-band
AO image by \citet{guyon06}. }

\tablenotetext{e}{Morphology from \citet{hamilton02}}

\tablenotetext{f}{Morphology from \citet{hamilton02} and
  \citet{guyon06}. }

\end{deluxetable}

\clearpage

\begin{deluxetable}{ccccc}
\tabletypesize{\scriptsize}
\tablecaption{Stars Used in Template Spectra \label{star_table}}
\tablewidth{0pt}
\tablehead{
\colhead{(1)} & \colhead{(2)} & \colhead{(3)} & \colhead{(4)} &
\colhead{(5)}  \\
\colhead{Name} & \colhead{Type} & \colhead{T$_{eff}$ (K)} & 
\colhead{log \it{g}} & \colhead{[Fe/H]} 
}

\startdata

HD 39283 & A2V & 4000 & 1.3 & -0.12 \\
HD 60179 & A1V & 10286 & 4 & 0.98 \\
HD 85235 & A3IV & 11200 & 3.55 & -0.4 \\
HD 70110 & F9V & 5955 & 4.07 & 0.07 \\
HD 10307 & G1.5V & 5898 & 4.31 & -0.02 \\
HD 199960 & G1V & 5813 & 4.2 & 0.11 \\
HD 52711 & G4V & 5890 & 4.31 & -0.16 \\
HD 111812 & G0IIIP & \nodata & \nodata & 0.01 \\
HD 161797 & G5IV & 5411 & 3.87 & 0.16 \\
HD 107950 & G6III & 5030 & 2.61 & -0.16 \\
HD 10761 & G8III & 4980 & 2.82 & -0.11 \\
HD 181276 & G9III & 5000 & 2.95 & -0.08 \\
HD 219449 & K0III & 4575 & 2.1 & -0.03 \\
HD 220954 & K1III & 4625 & 1.95 & -0.08 \\
HD 76291 & K1IV & 4536 & 2.74 & 0.08 \\
HD 81146 & K2III & 4370 & 2.34 & 0.01 \\
HD 173780 & K3III & 4400 & 2.57 & -0.12 \\
HD 124547 & K3III & 4130 & 2.04 & 0.17 \\
HD 136726 & K4III & 4120 & 2.03 & 0.07 \\
HD 120933 & K5III & 3820 & 1.52 & 0.5 \\
HD 112300 & M3III & 3700 & 1.3 & -0.16 \\

\enddata

\end{deluxetable}

\clearpage

\begin{deluxetable}{cccccccccccccc}
\tabletypesize{\scriptsize}
\rotate
\tablecaption{Host Galaxy Velocity Dispersions \label{vd_table}}
\tablewidth{0pt}
\tablehead{
\colhead{(1)} & \colhead{(2)} & \colhead{(3)} & \colhead{(4)} &
\colhead{(5)} & \colhead{(6)} & \colhead{(7)} & \colhead{(8)} & 
\colhead{(9)} & \colhead{(10)} & \colhead{(11)}  \\
\colhead{Object} & \colhead{Object} & \colhead{Measured $\sigma_{*}$} & 
\colhead{$\sigma_{*}$ Error} & \colhead{R$_{avg}$} & 
\colhead{f$_{AV}$} & \colhead{f$_{F-GV}$} & \colhead{f$_{G-KIII}$} & 
\colhead{S/N~\AA$^{-1}$ } & \colhead{f$_{QSO}$} 
& \colhead{Inst. Res.} \\
\colhead{Number} & \colhead{Name} & \colhead{(km~s$^{-1}$)} & 
\colhead{(km~s$^{-1}$)} & \colhead{(arcsec)} & \colhead{} & \colhead{} 
& \colhead{} & \colhead{(3850-4200~\AA)} & \colhead{} & 
\colhead{(km~sec$^{-1}$)}     
}
\startdata

1 & PG~0052+251 3S & $<$250.0 & 52.9 & 3.63 & 0.46 & 0.00 & 0.54 &
13.9 & 0.67 & 300 \\ 
2 & PHL 909 4.5N & 149.5 & 10.8 & 4.5 & 0.0 & 0.0 & 1.0 & 5.8 & 0.67 &
110 \\
3 & PKS 0736+017 4.5NW & 311.0 & 82.9 & 4.5 & 0.0 & 0.0 & 1.0 & 2.4 &
0.46 & 110 \\
4 & 3C~273 4N & 305.3 & 57.5 & 4.36 & 0.18 & 0.82 & 0.00 & 4.5 & 0.40 & 300 \\ 
5 & PKS~1302-102 2.3N & 346.3 & 72.4 & 3.05 & 0.07 & 0.93 & 0.00 & 2.2 
& 0.78 & 300 \\ 
6 & PG~1309+355 4.5SW & 235.9 & 29.8 & 4.5 & 0.06 & 0.09 & 0.85 & 8.9
& 0.39 & 110 \\ 
7 & PG~1444+407 3S & 278.9 & 22.4 & 3.70 & 0.27 & 0.73 & 0.00 & 4.4 
& 0.74 & 300 \\ 
8 & PKS~2135-147 3W & 278.4 & 105.9 & 3.81 & 0.39 & 0.00 & 0.61 & 3.0 
& 0.69 & 300 \\ 
9a & 4C~31.63 3S & $<$245.8 & 36.5 & 3.0 & 0.00 & 1.00 & 0.00 & 2.0 & 0.27 & 
300 \\ 
9b & 4C~31.63 2N & 289.9 & 22.7 & 2.0 & 0.06 & 0.94 & 0.00 & 5.6 & 0.45 & 300
\\ 
9c & 4C~31.63 2.5E & 324.9 & 24.2 & 2.5 & 0.14 & 0.00 & 0.86 & 13.4 & 
0.08 & 300 \\ 
10a & PKS~2349-014 3S & $<$223.7 & 33.9 & 3.24 & 0.00 & 0.00 & 1.00 &
1.8 & 0.27 & 300 \\ 
10b & PKS~2349-014 4N & 278.3 & 54.0 & 4.83 & 0.00 & 0.00 & 1.00 & 2.0
& 0.30 & 300 \\

\tablecomments{The velocity dispersions listed here are not
  aperture-corrected. Aperture corrections range from 6-13\% with an
  average of 9\%. Any measured $\sigma$'s less than 250 km
  s$^{-1}$ are marked as upper limits. The measurement on PG 0052+251 has other
  problems described in the text.}

\enddata

\end{deluxetable}

\clearpage

\begin{deluxetable}{cccc}
\tabletypesize{\scriptsize}
\tablecaption{Host Galaxy M/L and Mass \label{mass_table}}
\tablewidth{0pt}
\tablehead{
\colhead{(1)} & \colhead{(2)} & \colhead{(3)} & \colhead{(4)}\\
\colhead{Object} & \colhead{Object} & \colhead{M/L} & 
\colhead{M$_{*}$} \\
\colhead{Number} & \colhead{Name} & \colhead{(M$_{\sun}$/L$_{\sun}$)} & 
\colhead{10$^{11}$~M$_{\sun}$} \\
}
\startdata

1 & PG~0052+251 &  $<$5.6 & $<$4.4 \\
2 & PHL 909 & 3.8 & 2.2 \\
3 & PKS 0736+017 & 17.8 & 14.1 \\
4 & 3C~273 & 9.1 & 13.1 \\
5 & PKS~1302-102 & 10.3 & 10.3 \\
6 & PG~1309+355 & 5.0 & 5.0 \\
7 & PG~1444+407 & 7.2 & 6.2 \\
8 & PKS~2135-147 & 14.6 & 9.4 \\
9 & 4C~31.63 & 14.3 & 30.1 \\
10 & PKS~2349-014 & 8.3 & 12.2 \\

\enddata

\end{deluxetable}

\clearpage

\begin{deluxetable}{ccccccc}
  \tabletypesize{\scriptsize} \rotate \tablecaption{Mean Galaxy
    Parameters \label{avg_table}} \tablewidth{0pt} \tablehead{
    \colhead{(1)} & \colhead{(2)} & \colhead{(3)} & \colhead{(4)} &
    \colhead{(5)} & \colhead{(6)}& \colhead{(7)} \\
    \colhead{Group} & \colhead{R$_{e}$} & \colhead{$\sigma_{*}$
      \tablenotemark{a}} & \colhead{$\mu_{e}(r)$} & \colhead{M/L} &
    \colhead{M$_{*}$} &
    \colhead{references} \\
    \colhead{} & \colhead{(kpc)} & \colhead{(km s$^{-1}$)} &
    \colhead{(mag~arcsec$^{-1}$)} & \colhead{(M$_{\sun}$/L$_{\sun}$)}
    & \colhead{(10$^{11}$ M$_{\sun}$)} & \colhead{} }

\startdata

 Our QSO hosts &  9.2 & 291.6 \tablenotemark{b} & 20.5 & 10.0 
\tablenotemark{b} & 11.3 \tablenotemark{b} & this work \\
 RL hosts & 11.4 & 320.9 & 20.8 & 12.4 & 14.8 & this work \\
 RQ hosts & 6.0 & 240.8 \tablenotemark{b} & 19.8 & 5.3
 \tablenotemark{b} & 4.4 \tablenotemark{b} & this work \\
 PG QSO hosts & 3.6 & 183 & 19.6 & 4.8 & 1.7 & \citet{dasyra07} \\
 Early-type galaxies & 4.9 & 190 & 20.0 & 7.0 & 2.1 & \citet{bernardi03a} \\
 High $\sigma$ early-type galaxies & 8.0 & 403 & 19.8 & 14.7 & 12.2
 &  \citet{bernardi06} \\
 Merger remnants & 2.2 & 190 & 18.2 & 2.8 & 1.5 & \citet{rothberg06} \\

\enddata

\tablenotetext{a}{Aperture-corrected values.}

\tablenotetext{b}{Excluding PG 0052+251, for which we only have upper limits.}

\end{deluxetable}

\clearpage

\begin{deluxetable}{cccccccc}
\tabletypesize{\scriptsize}
\tablecaption{Host Galaxy Properties from Exponential Disk Fits
  \label{expfit_table} } 
\tablewidth{0pt} 
\tablehead{ 
\colhead{(1)} & \colhead{(2)} & \colhead{(3)} & \colhead{(4)} & 
\colhead{(5)} & \colhead{(6)} & \colhead{(7)} & \colhead{(8)} \\
\colhead{Object} & \colhead{Object} & \colhead{m$_{V}$(host)} & 
\colhead{R$_{e}$} & \colhead{R$_{e}$} & \colhead{$\mu_{e}$(r)} &
\colhead{M/L} & \colhead{M$_{*}$} \\ 
\colhead{\#} & \colhead{Name} & \colhead{(mag)} & \colhead{(arcsec)} & 
\colhead{(kpc)} & \colhead{(mag arcsec$^{-1}$)} &
\colhead{(M$_{\sun}$/L$_{\sun}$)} & \colhead{(10$^{11}$ M$_{\sun}$)}
}

\startdata

1 & PG 0052+251 & 17.51 & 2.2 & 5.9 & 20.27 & $<$10.8 & $<$5.3 \\
5 & PKS 1302-102 & 18.70 & 1.8 & 7.8 & 20.66 & 21.5 & 13.6 \\
6 & PG 1309+355 & 17.65 & 2.0 & 6.2 & 20.13 & 7.9 & 5.1 \\
7 & PG 1444+407 & 18.87 & 1.7 & 6.9 & 20.66 & 16.2 & 8.0 \\

\enddata

\end{deluxetable}

\clearpage

\begin{deluxetable}{cccccc}
\tabletypesize{\scriptsize}
\tablecaption{ Measurements on Sigma Standard Galaxies 
  \label{sigma_test_table} } 
\tablewidth{0pt} 
\tablehead{ 
\colhead{(1)} & \colhead{(2)} & \colhead{(3)} & \colhead{(4)}  &
\colhead{(4)} & \colhead{(5)} \\
\colhead{Galaxy} & \colhead{Actual $\sigma$} & \colhead{Measured
  $\sigma$} & \colhead{Error} & \colhead{S/N~\AA$^{-1}$} & 
\colhead{Galaxy Type} \\
\colhead{} & \colhead{(km~s$^{-1}$)} & \colhead{ (km~s$^{-1}$)} & 
\colhead{ (km~s$^{-1}$)} & \colhead{(3850-4200~\AA)} & \colhead{}   }

\startdata

NGC 1407 & 272 & 269.5 & 56.4 & 2.9 & E0 \\
NGC 4594 & 249 & 262.3 & 50.8 & 6.9 & Sa \\
NGC 584 & 230 & 234.4 & 46.5 & 4.2 & E4 \\
NGC 2768 & 205 & 193.2 & 73.3 & 4.0 & E6 \\
NGC 4579 & 189 & 278.7 & 32.9 & 5.4 & Sb \\
NGC 3031 & 170 & 303.1 & 64.2 & 3.5 & Sab \\
NGC 6340 &137 & 250.7 & 161.5 & 2.0 & S0/a \\

\enddata

\end{deluxetable}


\end{document}